\documentclass[12pt]{article}

\usepackage[top=1in, bottom=1in, left=0.9in, right=0.9in]{geometry}

\usepackage{latexsym, multirow} 
\usepackage{amsmath}
\usepackage{amsfonts}
\usepackage{graphicx} 
\usepackage{enumerate} 
\usepackage{authblk} 
\usepackage{subfigure} 

\usepackage[authoryear]{natbib} 

\usepackage{amsthm}
\theoremstyle{plain}
\newtheorem{proposition}{Proposition}
\newtheorem{theorem}{Theorem}
\newtheorem{lemma}{Lemma}

\def\T{{ \mathrm{\scriptscriptstyle T} }}
\DeclareMathOperator*{\argmin}{arg\,min}
\DeclareMathOperator*{\argmax}{arg\,max}
\DeclareMathOperator*{\logit}{logit}

\def\ind{\begin{picture}(9,8)
         \put(0,0){\line(1,0){9}}
         \put(3,0){\line(0,1){8}}
         \put(6,0){\line(0,1){8}}
         \end{picture}
        }

\def\pr{\textup{pr}}
\def\diag{\textup{diag}}

\begin{document}
\title{\bf \Large Inverse regression for causal inference with multiple outcomes}

\author[1]{Wei Zhang}
\author[1]{Qizhai Li}
\author[2]{Peng Ding\thanks{pengdingpku@berkeley.edu}}
\affil[1]{State Key Laboratory of Mathematical Sciences, Academy of Mathematics and Systems Science, Chinese Academy of Sciences, Beijing 100190, China}
\affil[2]{Department of Statistics, University of California, 425 Evans Hall, Berkeley, California 94720, U.S.A.}

\date{\today}
\maketitle

\begin{abstract}
With multiple outcomes in empirical research, a common strategy is to define a composite outcome as a weighted average of the original outcomes. However, the choices of weights are often subjective and can be controversial. We propose an inverse regression strategy for causal inference with multiple outcomes. The key idea is to regress the treatment on the outcomes, which is the inverse of the standard regression of the outcomes on the treatment. Although this strategy is simple and even counterintuitive, it has several advantages. First, testing for zero coefficients of the outcomes is equivalent to testing for zero treatment effects, even though the inverse regression is deemed misspecified. Second, the coefficients of the outcomes provide data-driven weights for defining a composite outcome. Interestingly, these weights maximize the standardized effect size, a measure commonly used for univariate outcomes in empirical studies. We also discuss the associated inference issues. Third, this strategy is applicable to general study designs. We illustrate the theory in both randomized experiments and observational studies. We implement the proposed method in our R package \texttt{invreg}, available at \texttt{https://github.com/zhangwei0125/invreg}.

\end{abstract}

\section{Introduction}\label{sec-intro}

Causal studies involving multiple outcomes are common in medical and social sciences. A conventional strategy is to analyse each outcome separately, yielding outcome-specific treatment effect estimates. However, this approach ignores the joint impact of treatment on multiple outcomes and may become difficult to interpret when the number of outcomes is large. Recent work on outcome-wide longitudinal designs emphasizes the importance of assessing treatment effects on multiple outcomes simultaneously to obtain comprehensive causal inferences \citep{vanderweele2020outcome}. To construct a summary measure of treatment effects across multiple outcomes, \cite{Kling2007} combined standardized outcomes with equal weights, while \cite{kennedy2019estimating} introduced a weighted average of scaled treatment effects that allows different weights.  \cite{Lupparelli2020} proposed aggregating multiple binary outcomes through their product, which is limited to binary data but not applicable to other outcome types.

Composite outcomes, defined as weighted averages of single outcomes, are widely used as primary endpoints in clinical studies (e.g., \citealt{Kling2007}; \citealt{Cordoba2010}; \citealt{Armstrong2017}; \citealt{kennedy2019estimating}). Their main advantage lies in improving statistical efficiency and power by increasing the overall event rate through the aggregation of related outcomes. However, the choice of weights often relies on clinical expertise or prior research, making it subjective and potentially controversial, as clinicians and researchers may differ in their perspectives on the relative importance of each outcome \citep{Hong2011}. To address multiple testing, \cite{anderson2008multiple} developed a summary index that combines multiple standardized outcomes using inverse-covariance weighing. While the index accounts for correlations among outcomes, it assumes that treatment effects share a common direction and reverses signs when this assumption fails, which limits its applicability because effect directions are often unclear. To our knowledge, no objective, data-adaptive, and universally applicable weighting method has been proposed.

We fill this gap by proposing an inverse regression approach that uses the treatment as the response and multiple outcomes as the regressors to determine the weights based on observed data.  Inverse regression, also called reverse regression, has been applied in diverse research areas. In genetic association studies, it is used to test pleiotropy, where a single genetic variant influences multiple phenotypic traits \citep{OReilly2012,Banerjee2021,ZhangSun2022}. However, most existing tests require the inverse regression model be correctly specified, an assumption that is often deemed invalid. In randomized trials, \cite{Zhang2014} applied an inverse logistic regression framework to jointly analyse multiple outcomes, formulating it as a semiparametric density ratio model for the outcome distributions in the two treatment groups. In randomized experiments, \cite{Ludwig2017} employed inverse regression to develop a test for treatment effects on the joint distribution of multiple outcomes, leveraging machine learning techniques. \cite{Bind2023} used inverse regression of treatment on high-dimensional outcomes to construct the test statistics for Fisher randomization tests in randomized experiments.  Other earlier work includes sliced inverse regression for dimension reduction \citep{li1991} and reverse regression to assess fairness in employment \citep{Conway1983,goldberger1984reverse}.

Although the idea of inverse regression is not entirely new, we further the literature in the following directions. First, we establish an equivalence relationship between the inverse regression coefficients and outcome-specific treatment effects, which ensures that testing for zero coefficients is equivalent to testing for zero treatment effects. Based on the inverse regression coefficients, we develop a test for the null hypothesis of zero effects without assuming the inverse regression model is correctly specified. Second, we use the inverse regression coefficients to define a composite outcome as a weighted average of the original outcomes, yielding a data-driven measure of the overall treatment effect. This weighting strategy enjoys six advantages: (i) it is free from subjective judgment, ensuring uniform and generalizable results across studies; (ii) it naturally accounts for correlations among outcomes, providing a comprehensive assessment of treatment effects; (iii) it admits an optimality interpretation by maximizing the standardized effect size \citep{hedgesstatistical}; (iv) it is flexible to incorporate additional covariates within the regression framework; (v) it is simple to implement based on standard least squares software routines; and (vi) it is applicable to general study designs. We illustrate the proposed approach in both randomized experiments and observational studies. In each setting, we develop an estimator for the treatment effect on the composite outcome and establish its asymptotic properties.

\section{Completely randomized experiments with multiple outcomes}\label{sec-CREnoX}
\subsection{Notation: potential outcomes and average treatment effect} \label{subsec-CREnoX-ATE}

To introduce the key idea, we begin with the simplest setting of a completely randomized experiment. Consider a randomized experiment with $n$ units sampled from an infinite population and with $L\geq1$ outcomes of interest.  Each unit is randomly assigned to either an experimental treatment $(Z=1)$ or a control treatment $(Z=0)$. Let $Z$ denote the binary indicator for the treatment, with treatment probability $p=\pr(Z=1)$. Let $Y = (Y_1, \ldots, Y_L)^{\T}$ denote the outcome vector. For each $\ell=1,\ldots,L$, let $Y_{\ell}(1)$ and $Y_{\ell}(0)$ denote the potential outcomes under treatment and control, respectively, vectorized as $Y(1)=(Y_1(1), \ldots, Y_L(1))^{\T}$ and $Y(0)=(Y_1(0), \ldots, Y_L(0))^{\T}$. The observed outcome is $Y=ZY(1)+(1-Z)Y(0)$. Define the average treatment effect  of $Z$ on $Y_{\ell}$ as $\tau_{\ell} = E\{Y_{\ell}(1)\} - E\{Y_{\ell}(0)\}$, vectorized as $\tau=(\tau_1,\ldots, \tau_L)^\T$.

Under randomized treatment assignment, $Z\ind\{Y(1),Y(0)\}$, so $E(ZY/p)=E\{Y(1)\}$ and $E\{(1-Z)Y/(1-p)\}=E\{Y(0)\}$. Combining the two identities yields
\begin{equation}\label{CRE-eq-invdiff}
\tau=E\{Y(1)-Y(0)\}=E\left\{\frac{ZY}{p}-\frac{(1-Z)Y}{1-p}\right\},
\end{equation}
which corresponds to the difference of weighted means between the treatment and control groups, with weights $1/p$ for treated units and  $1/(1-p)$ for control units. Moreover,  we can show that $\tau$ equals the coefficient of $Z$ from the element-wise weighted least squares regression of $Y$ on $(1, Z)$ with weights
\begin{equation}\label{CRE-eq-weight}
W=\frac{Z}{p}+\frac{1-Z}{1-p}.
\end{equation}
Specifically, for each outcome $Y_{\ell}$ , the treatment effect $\tau_{\ell}$ equals the coefficient of $Z$ from
\begin{equation}\label{CRE-eq-wls}
(\tau_{0\ell}, \tau_{\ell}) = \argmin_{a_0,a_{\ell}} E\{W(Y_{\ell}-a_0-a_{\ell}Z)^2\},
\end{equation}
and stacking these coefficients yields $\tau=(\tau_1,\ldots,\tau_L)^{\T}$.

To define an overall treatment effect for $Y$, we propose to regress the scalar $Z$ on vector $Y$ with an intercept, using the weights defined in \eqref{CRE-eq-weight}. Let $(\beta_0, \beta) = \argmin_{b_0,b} E\{W(Z-b_0-b^\T Y)^2\}$ denote the population weighted least squares coefficients,
where $\beta=(\beta_1,\ldots,\beta_L)^\T$.
Defining $\Phi_{\textsc{yy}}^{\textsc{w}}=E(WYY^{\T})-E(WY)E(W)^{-1}E(WY^{\T})$ and $\Phi_{\textsc{yz}}^{\textsc{w}}=E(WZY)-E(WY)E(W)^{-1}E(WZ)$, we write $\beta=(\Phi_{\textsc{yy}}^{\textsc{w}})^{-1}\Phi_{\textsc{yz}}^{\textsc{w}}$.
Using the inverse regression coefficient $\beta$, we construct the composite outcome
\begin{equation*}\label{compY-CR-noX}
Y^{\textup{c}} =\beta^{\T}Y= \sum_{\ell = 1}^L \beta_{\ell} Y_{\ell}.
\end{equation*}
The average treatment effect of $Z$ on $Y^{\textup{c}}$ is  $\tau^{\textup{c}}=\beta^\T\tau$.
The coefficient vector $\beta$ ensures that the composite outcome $Y^{\textup{c}}$ is invariant under any nonsingular linear transformation of $Y$. That is, if the outcomes are transformed as $\Pi Y$ with a nonsingular $L\times L$ matrix $\Pi$, $Y^{\textup{c}}$ is unchanged.

For each unit $i$, let $y_i=(y_{i1},\ldots,y_{iL})^\T$ denote the observed outcome vector and $z_i$ the assigned treatment. Define  $y=(y_1,\ldots,y_n)^\T$ and $z=(z_1,\ldots,z_n)^\T$. Assume that $(z_i, y_i(1), y_i(0))$ $(i=1,\cdots,n)$ are independent and identically distributed. We estimate $\tau^{\textup{c}}$ by the inner product of the estimators of $\beta$ and $\tau$. Since both $\beta$ and $\tau$ are population weighted least squares coefficients, we estimate them using the corresponding sample analogues. Specifically, we estimate $\beta$ as the coefficient of $y_i$ from the weighted regression of $z_i$ on $(1, y_i)$ with weights $w_i=z_i/p+(1-z_i)/(1-p)$, and denote the resulting estimator by $\hat{\beta}$. Similarly, we estimate $\tau$ as the coefficient of $z_i$ from the component-wise weighted regression of $y_i$ on $(1, z_i)$ with the same weights $w_i$, and denote the estimator by $\hat{\tau}$. We estimate the treatment effect $\tau^{\textup{c}}$ by
$$\hat{\tau}^{\textup{c}}=\hat{\beta}^{\T}\hat{\tau}.$$


\subsection{Equivalence between inverse regression coefficients and average treatment effects}\label{subsec-CREnoX-equiv}

We first establish an equivalence relationship between the inverse weighted least squares coefficient vector $\hat{\beta}$ and the treatment effect estimator  $\hat{\tau}$, as well as that between their respective estimands, $\beta$ and $\tau$. For generic vectors $a_i$ and $b_i$ $(i=1,\ldots,n)$, we define the weighted sample moment $S^{w}_{\smash{ab}}=n^{-1}\sum_{i=1}^{n}w_ia_ib^{\T}_i$. Using this notation, we write
$\hat{\beta}=(\phi_{yy}^{w})^{-1}\phi_{yz}^{w}$
and
$\hat{\tau}=(\phi_{zz}^{w})^{-1}\phi_{yz}^{w}$, where $\phi_{yy}^{w}=S^{w}_{\smash{yy}}-S^{w}_{\smash{y1}}(S^{w}_{\smash{11}})^{-1}S^{w}_{\smash{1y}}$, $\phi_{zz}^{w}=S^{w}_{\smash{zz}}-S^{w}_{\smash{z1}}(S^{w}_{\smash{11}})^{-1}S^{w}_{\smash{1z}}$, and $\phi_{yz}^{w}=S^{w}_{\smash{yz}}-S^{w}_{\smash{y1}}(S^{w}_{\smash{11}})^{-1}S^{w}_{\smash{1z}}$.

\begin{proposition}\label{CREnoX-equiv}
(i) (Sample version) We have
$$\hat{\beta}=\phi_{zz}^{w}\big(\phi_{yy}^{w}\big)^{-1}
\hat{\tau}=\frac{\hat\Sigma^{-1}\hat\tau}{1+\hat\tau^\T \hat\Sigma^{-1}\hat\tau},$$
where $\hat\Sigma=(\phi_{zz}^{w})^{-1}[\hat{\textup{cov}}\{Y(1)\}+\hat{\textup{cov}}\{Y(0)\}]$ with  $\hat{\textup{cov}}\{Y(1)\}=(np)^{-1}\sum_{i=1}^{n}z_i\{y_i-\bar{y}(1)\}\{y_i$ $-\bar{y}(1)\}^{\T}$,  $\hat{\textup{cov}}\{Y(0)\}=\{n(1-p)\}^{-1}\sum_{i=1}^{n}(1-z_i)\{y_i-\bar{y}(0)\}\{y_i-\bar{y}(0)\}^{\T}$, $\bar{y}(1)=\sum_{i=1}^{n}z_iy_{i}/$ $(\sum_{i=1}^{n}z_i)$, and $\bar{y}(0)=\sum_{i=1}^{n}(1-z_i)y_{i}/\{\sum_{i=1}^{n}(1-z_i)\}$.

(ii) (Population version) Under  complete randomization with $Z\ind\{Y(1),$ $Y(0)\}$, we have
\begin{equation*}
\begin{aligned}
\beta&=\Phi_{\textsc{zz}}^{\textsc{w}}\big(\Phi_{\textsc{yy}}^{\textsc{w}}\big)^{-1}\tau
=\frac{\Sigma^{-1}\tau}{1+\tau^\T\Sigma^{-1}\tau},
\end{aligned}
\end{equation*}
where  $\Phi_{\textsc{zz}}^{\textsc{w}}=E(WZ^2)-E(WZ)E(W)^{-1}E(WZ)=1/2$ and $\Sigma=2[\textup{cov}\{Y(1)\}+\textup{cov}\{Y(0)\}]$.
\end{proposition}

Proposition \ref{CREnoX-equiv} characterizes a nonlinear mapping that defines equivalence relationships between $\hat{\beta}$ and $\hat{\tau}$ and between $\beta$ and $\tau$. In particular, $\hat{\beta}=0$  if and only if $\hat{\tau}=0$, and $\beta=0$ if and only if $\tau=0$. However, this equivalence does not imply that each $\beta_{\ell}$ has the same zero status or sign as $\tau_{\ell}$. Because $\Sigma$ is not necessarily diagonal, $\beta_{\ell}$ may be nonzero when $\tau_{\ell}=0$, and $\beta_{\ell}$ and $\tau_{\ell}$ may have different signs.
Consider a stylized example with two outcomes. Suppose $Y_1=Z+\epsilon_1$ and $Y_2=\gamma Z+\epsilon_2$, where $0\leq\gamma<1$, and let $\epsilon=(\epsilon_1, \epsilon_2)^{\T}$ have mean zero and covariance matrix $\Sigma_{\epsilon}$, with unit diagonal elements and off-diagonal element $\delta$.  Then $\textup{cov}\{Y(1)\}=\textup{cov}\{Y(0)\}=\Sigma_{\epsilon}$, $\Sigma=4\Sigma_{\epsilon}$, and $\tau=(1, \gamma)^{\T}$.  Both outcomes therefore have nonnegative treatment effects.  For $\delta\in(-1,1)$, $\Sigma^{-1}\tau=(1-\gamma\delta, \gamma-\delta)^{\T}/\{4(1-\delta^2)\}$. Because $\beta$ is a positive scalar multiple of $\Sigma^{-1}\tau$, the coefficient on $Y_1$ is always positive, whereas the coefficient on $Y_2$ is positive when $\gamma>\delta$, zero when $\gamma=\delta$, and negative when $\gamma<\delta$. A negative coefficient on $Y_2$ therefore does not indicate a negative treatment effect on $Y_2$. Rather, the negative coefficient arises because $Y_2$ is highly correlated with $Y_1$ but has a relatively weaker treatment effect.

This possible sign discrepancy is related to, but distinct from, the regression paradox in the classical reverse regression literature on employment discrimination \citep{Conway1983,goldberger1984reverse,chen2009regression}. In employment discrimination analyses,  forward regression typically regresses salary on job qualifications and a group indicator, such as gender or race. Reverse regression interchanges salary and job qualifications while continuing to focus on the coefficient of the group indicator. Forward and reverse regressions may yield apparently conflicting conclusions about group effects because they answer different conditional questions. By contrast, the proposed inverse regression has a different formulation and purpose: it interchanges the treatment indicator and the outcomes, and uses the outcome coefficients to construct a composite outcome. Consequently, sign discrepancies do not undermine the intended role of $\beta$.

We interpret $\beta$ as a treatment-aligned direction that accounts for both the treatment effects
and the covariance structure of the outcomes. By Proposition \ref{CREnoX-equiv}, $\beta$ is proportional to $\Sigma^{-1}\tau$, which maximizes $b^{\T}\tau/(b^{\T}\Sigma b)^{1/2}$ over all nonzero $L$-dimensional vectors $b$. The objective function is the standardized mean difference between the composite potential outcomes $b^{\T}Y(1)$ and $b^{\T}Y(0)$, with standardization determined by $\Sigma$. Thus, $\beta$ identifies  the linear combination of outcomes with the largest standardized mean difference. By basic linear algebra, $\max_{b\neq0}b^{\T}\tau/(b^{\T}\Sigma b)^{1/2}=(\tau^{\T}\Sigma^{-1}\tau)^{1/2}$, so $\tau^{\textup{c}}$ is a bounded and normalized version of the squared maximum standardized mean difference. The standardized mean difference is closely related to common effect size measures in social sciences, including Cohen's $d$ \citep{cohen1988statistical} and Glass's $\Delta$ \citep{glass1976primary}. For a scalar outcome, Cohen's $d$ standardizes the difference in group means by the pooled standard deviation, while Glass's $\Delta$ uses the standard deviation of the control group. The standardized mean difference $b^{\T}\tau/(b^{\T}\Sigma b)^{1/2}$ extends the same standardization principle to multiple outcomes by using a constant multiple of the sum of the covariance matrices under treatment and control for standardization \citep{hedgesstatistical}.

The optimality interpretation also connects $\beta$ to Fisher's linear discriminant analysis \citep{fisher1936multiple}, which identifies the projection with the largest separation between group means relative to within-group variability. The optimal projection is  proportional to $\Sigma^{-1}_{\textsc{w}}(\mu_1-\mu_0)$, where $\mu_1$ and $\mu_0$ are the group mean vectors of the outcomes and $\Sigma_{\textsc{w}}$ is the sum of the within-group covariance matrices. Therefore, $\beta$ can be viewed as a causal analogue of the Fisher linear discriminant direction, with $\tau$ replacing $\mu_1-\mu_0$ and $\Sigma$ replacing  $\Sigma_{\textsc{w}}$. The resulting composite outcome best separates the potential outcome means under treatment and control. When $\textup{cov}\{Y(1)\}=\textup{cov}\{Y(0)\}$, $\beta$ coincides with the usual Fisher linear discriminant direction up to scale.

Proposition \ref{CREnoX-equiv} further implies $\hat{\tau}^{\textup{c}}=\hat\tau^{\T}\hat\Sigma^{-1}\hat\tau/(1+\hat\tau^\T \hat\Sigma^{-1}\hat\tau)\in[0,1)$ and $\tau^{\textup{c}}=\tau^{\T}\Sigma^{-1}\tau/(1+\tau^\T \Sigma^{-1}\tau)\in[0,1)$, because $\hat{\Sigma}$ and $\Sigma$ are positive semi-definite, ensuring $\hat{\tau}^\T\hat{\Sigma}^{-1}\hat{\tau}\geq0$ and $\tau^\T\Sigma^{-1}\tau\geq0$. Hence, $\tau^{\textup{c}}$ is a bounded monotone transformation of $\tau^\T\Sigma^{-1}\tau$. For example, if $\tau=\theta(1,\ldots,1)^{\T}$,  then $\tau^{\textup{c}}$ increases toward 1 as $\theta\rightarrow\infty$ but remains strictly below 1 for every finite $\theta$. Because $\Sigma$ depends only on the moments of $(Y(1), Y(0))$ and not on the randomization mechanism,  the summary measure $\tau^{\textup{c}}$ is design-independent. Since $\tau$ also equals the coefficient of $Z$ from the element-wise ordinary least squares regression of $Y$ on $(1, Z)$, we could alternatively construct a composite outcome by regressing $Z$ on $(1,Y)$ using ordinary least squares. However, the resulting treatment effect measure becomes design-dependent. Under unweighted inverse regression, $\Phi_{\textsc{zz}}^{\textsc{w}}$ and $\Phi_{\textsc{yy}}^{\textsc{w}}$ reduce to $\textup{var}(Z)$ and $\textup{cov}(Y)$, respectively, both of which depend on the treatment assignment probability $p$.

The fact that $\tau^{\textup{c}}\in[0,1)$  is related to the invariance of the composite outcome under nonsingular linear transformations. Moreover, $\tau^{\textup{c}}=0$ if and only if $\tau=0$, whereas $\tau^{\textup{c}}>0$ whenever $\tau\neq0$. In this sense, $\tau^{\textup{c}}$ is similar in spirit to recently proposed omnibus measures of heterogeneous
treatment effects \citep{chernozhukov2025fisher,yadlowsky2025evaluating}, which equal zero in the absence of treatment effect heterogeneity and become positive under meaningful heterogeneity.

Proposition \ref{CREnoX-equiv} also suggests that testing the null hypothesis $H_0: \tau= 0$ is equivalent to testing $\beta=0$.  To account for potential model misspecification, we must use the Wald test $\mathcal{W}^{\hat{\beta}} = \hat{\beta}^\T\hat{V}_{\smash{\beta}}^{-1}\hat{\beta}$, where
$\hat{V}_{\smash{\beta}}$ is the Huber--White robust covariance estimator of the asymptotic variance of $\hat{\beta}$ from the inverse weighted least squares regression that yields $\hat\beta$ \citep{huber1967,white1980}.  Under $H_0$,  $\mathcal{W}^{\hat{\beta}}$ converges in distribution to a chi-squared distribution with $L$ degrees of freedom. Importantly, inference based on $\mathcal{W}^{\hat{\beta}}$ remains valid even when the inverse regression model is misspecified.

\subsection{Asymptotic theory and statistical inference for the composite outcome}\label{subsec-CREnoX-asy}

Building on the equivalence relationships in Proposition \ref{CREnoX-equiv}, we investigate the asymptotic properties of  $\hat\tau^{\textup{c}}$. By the consistency of $\hat\beta$ and $\hat\tau$, we know that $\hat\tau^{\textup{c}}$ is consistent for $\tau^{\textup{c}}$. Theorem \ref{CREnoX-compATE-asy} then characterizes its asymptotic distribution.

\begin{theorem}\label{CREnoX-compATE-asy}
Assume complete randomization with $Z\ind\{Y(1),Y(0)\}$.

(i) If $\tau\neq0$, then $n^{1/2}(\hat\tau^{\textup{c}}-\tau^{\textup{c}})\rightarrow N(0, 4\beta^\T\textup{cov}\{R(Z,Y)\}\beta)$ in distribution as $n\rightarrow\infty$, where
\begin{equation*}
\begin{aligned}
R(Z,Y)&=\big(W\tilde{Y}\tilde{Y}^{\T}-\Phi_{\textsc{yy}}^{\textsc{w}}\big)\beta-(W-2)\Phi_{\textsc{yy}}^{\textsc{w}}\beta/2+2W\varepsilon\tilde{Y},
\end{aligned}
\end{equation*}
with $\tilde{Y}=Y-E(W)^{-1}E(WY)$ and $\varepsilon = Z-\beta_0-\beta^{\T}Y$ being the population residual from the weighted least squares regression of $Z$ on $(1, Y)$  with weights in \eqref{CRE-eq-weight}.

(ii) If $\tau=0$, then
$n\hat{\tau}^{\textup{c}}\rightarrow\sum_{\ell=1}^{L}\lambda_{\ell}\chi^2_{\ell}(1)$ in distribution as $n\rightarrow\infty$,  where $\chi^2_{\ell}(1)\ (\ell=1,\ldots, L)$ are independent chi-squared distributions with one degree of freedom and $\lambda_{\ell}\ (\ell=1,\ldots, L)$ are eigenvalues of
$$\Gamma=2(\Phi_{\textsc{yy}}^{\textsc{w}})^{-1/2}E\big(W^2\varepsilon^2\tilde{Y}\tilde{Y}^{\T}\big)(\Phi_{\textsc{yy}}^{\textsc{w}})^{-1/2}.$$
\end{theorem}

By Theorem \ref{CREnoX-compATE-asy}, the asymptotic distribution of $\hat{\tau}^{\textup{c}}$ depends on whether $\tau$ is zero. It is normal when $\tau\neq0$ and is a weighted sum of chi-squared distributions when $\tau=0$. Theorem \ref{CREnoX-compATE-asy} accounts for the uncertainty in both $\hat{\beta}$ and $\hat{\tau}$. We can use it to construct large-sample confidence intervals for $\tau^{\textup{c}}$. Specifically, when $\tau\neq0$, we construct a Wald-type confidence interval using the asymptotic normal distribution, with the variance $4\beta^\T\textup{cov}\{R(Z,Y)\}\beta$ estimated by its sample analogue. When $\tau=0$, we produce the interval based on the quantiles of the asymptotic weighted sum of chi-squared distributions, with the weights $\lambda_{\ell}\ (\ell=1,\ldots, L)$ estimated by the eigenvalues of the sample analogue of $\Gamma$. Because the true value of $\tau$ is typically unknown in practice, we also suggest procedures that yield valid confidence intervals for $\tau^{\textup{c}}$ without requiring knowledge of whether $\tau=0$. We provide the details in the Supplementary Material.

\section{Observational studies with multiple outcomes}\label{sec-OBSnoX}

\subsection{Observational studies, strong ignorability, propensity score, and average treatment effect}\label{sec-OBS-ATE}

We now extend the idea of inverse regression in completely randomized experiments to observational studies. Unless otherwise stated, we retain the notation introduced in Section \ref{sec-CREnoX}. Let $X=(X_1,\ldots, X_K)$ denote the vector of pretreatment covariates, and define the propensity score as $e(X)=\pr(Z=1\mid X)$ \citep{rosenbaum1983central}. Under strong ignorability $Z\ind\{Y(1), Y(0)\} \mid X$  and overlap $0<e(X)<1$, we can express the average treatment effect $\tau=E\{Y(1)-Y(0)\}$ as
\begin{equation}\label{OBS-eq-invdiff}
\tau=E\left\{\frac{ZY}{e(X)}-\frac{(1-Z)Y}{1-e(X)}\right\},
\end{equation}
which generalizes \eqref{CRE-eq-invdiff} by replacing the constant treatment probability $p$ with the propensity score $e(X)$. It also implies that $\tau$ equals the coefficient of $Z$ from the element-wise weighted least squares regression of $Y$ on $(1, Z)$ with weights
\begin{equation}\label{OBS-eq-weight}
W=\frac{Z}{e(X)}+\frac{1-Z}{1-e(X)},
\end{equation}
as in \eqref{CRE-eq-wls}. The weight in \eqref{OBS-eq-weight} reduces to \eqref{CRE-eq-weight} when $e(X)$ is a constant. We therefore continue to use $W$ for the inverse probability weights.
For each outcome $Y_{\ell}$, the sample weighted least squares estimator of the coefficient of $Z$  in the regression of $Y_{\ell}$ on $(1, Z)$ is identical to the Hajek estimator of $\tau_{\ell}$ \citep[Chapter 14]{ding2024first}.

Building on the weighted least squares representation of $\tau$, we construct a composite outcome by inversely regressing $Z$ on $(1, Y)$ using the weights in \eqref{OBS-eq-weight}. Let $(\beta_{0}, \beta) = \argmin_{b_0,b} E\{W(Z-b_0-b^{\T}Y)^2\}$ denote the population regression coefficients. The slope coefficient satisfies $\beta=(\beta_{1}, \ldots, \beta_{L})^{\T}=(\Phi_{\textsc{yy}}^{\textsc{w}})^{-1}\Phi_{\textsc{yz}}^{\textsc{w}}$, where $\Phi_{\textsc{yy}}^{\textsc{w}}$ and $\Phi_{\textsc{yz}}^{\textsc{w}}$ are defined in Section \ref{subsec-CREnoX-ATE}. We then define the composite outcome as
$$Y^{\textup{c}} =\beta^{\T}Y=\sum_{\ell = 1}^L \beta_{\ell} Y_{\ell},$$
with corresponding average treatment effect $\tau^{\textup{c}}=\beta^{\T}\tau$.

Let $x_i=(x_{i1},\ldots, x_{iK})^{\T}$ denote the observed covariates for unit $i$, and assume that the data $(z_i,x_i,y_i(0), y_i(1))\ (i=1,\ldots,n)$ are independent and identically distributed. Define the sample weights $\hat{w}_i=z_i/\hat{e}(x_i)+(1-z_i)/\{1-\hat{e}(x_i)\}$, where $\hat{e}(x_i)$ denotes the estimated propensity score for unit $i$. We estimate $\tau^{\textup{c}}$ by replacing the population quantities $\beta$ and $\tau$ with their sample analogues. Specifically, we estimate $\beta$ as the coefficient of $y_i$ from the weighted regression of $z_i$ on $(1, y_i)$ with weights $\hat{w}_i$.  The resulting estimator is $\hat{\beta}_{\textsc{os}}=(\phi_{yy}^{\hat{w}})^{-1}\phi_{yz}^{\hat{w}}$, where  $\phi_{yy}^{\hat{w}}=S^{\hat{w}}_{\smash{yy}}-S^{\hat{w}}_{\smash{y1}}(S^{\hat{w}}_{\smash{11}})^{-1}S^{\hat{w}}_{\smash{1y}}$ and $\phi_{yz}^{\hat{w}}=S^{\hat{w}}_{\smash{yz}}-S^{\hat{w}}_{\smash{y1}}(S^{\hat{w}}_{\smash{11}})^{-1}S^{\hat{w}}_{\smash{1z}}$. Similarly, we estimate $\tau$ as the coefficient of $z_i$ from the component-wise weighted regression of $y_i$ on $(1, z_i)$ with the same weights $\hat{w}_i$. The resulting estimator is
$\hat{\tau}_{\textsc{os}}=(\phi_{zz}^{\hat{w}})^{-1}\phi_{yz}^{\hat{w}}$, where $\phi_{zz}^{\hat{w}}=S^{\hat{w}}_{\smash{zz}}-S^{\hat{w}}_{\smash{z1}}(S^{\hat{w}}_{\smash{11}})^{-1}S^{\hat{w}}_{\smash{1z}}$. Finally, we estimate the average treatment effect of the composite outcome by
$$\hat{\tau}_{\textsc{os}}^{\textup{c}}=\hat{\beta}_{\textsc{os}}^{\T}\hat{\tau}_{\textsc{os}}.$$

\subsection{Equivalence between the inverse regression coefficients and average treatment effects}\label{sec-OBS-equiv}

We next establish the equivalence between $\hat\beta_{\textsc{os}}$ and $\hat\tau_{\textsc{os}}$ and that between $\beta_{\textsc{os}}$ and $\tau$ in observational studies.

\begin{proposition}\label{OBS-equiv}
(i) (Sample version) We have
$$\hat{\beta}_{\textup{\textsc{os}}}=\phi_{zz}^{\hat{w}}\big(\phi_{yy}^{\hat{w}}\big)^{-1}
\hat{\tau}_{\textup{\textsc{os}}}=\frac{\hat\Sigma_{\textup{\textsc{os}}}^{-1}\hat\tau_{\textup{\textsc{os}}}}{1+\hat\tau_{\textup{\textsc{os}}}^\T \hat\Sigma_{\textup{\textsc{os}}}^{-1}\hat\tau_{\textup{\textsc{os}}}},$$
where $\hat\Sigma_{\textup{\textsc{os}}}=(\phi_{zz}^{\hat{w}})^{-1}[\hat{\textup{cov}}_{\hat{w}}\{Y(1)\}+\hat{\textup{cov}}_{\hat{w}}\{Y(0)\}]$ with $\hat{\textup{cov}}_{\hat{w}}\{Y(1)\}=n^{-1}\sum_{i=1}^{n}\hat{w}_iz_i\{y_i-\bar{y}_{\hat{w}}(1)\}\{y_i-\bar{y}_{\hat{w}}(1)\}^{\T}$,
$\hat{\textup{cov}}_{\hat{w}}\{Y(0)\}=n^{-1}\sum_{i=1}^{n}\hat{w}_i(1-z_i)\{y_i-\bar{y}_{\hat{w}}(0)\}\{y_i-\bar{y}_{\hat{w}}(0)\}^{\T}$,
$\bar{y}_{\hat{w}}(1)=$ $\sum_{i=1}^{n}\hat{w}_iz_iy_i/(\sum_{i=1}^{n}\hat{w}_iz_i)$, and $\bar{y}_{\hat{w}}(0)=\sum_{i=1}^{n}\hat{w}_i(1-z_i)y_i/\{\sum_{i=1}^{n}\hat{w}_i(1-z_i)\}$.

(ii) (Population version) Under strong ignorability $Z\ind$ $\{Y(1),Y(0)\} \mid X$ and overlap $0<e(X)<1$, we have
\begin{equation*}
\begin{aligned}
\beta&=\Phi_{\textsc{zz}}^{\textsc{w}}\big(\Phi_{\textsc{yy}}^{\textsc{w}}\big)^{-1}\tau
=\frac{\Sigma^{-1}\tau}{1+\tau^\T\Sigma^{-1}\tau},
\end{aligned}
\end{equation*}
where $\Phi_{\textsc{zz}}^{\textsc{w}}$ and $\Sigma$ are defined in Proposition \ref{CREnoX-equiv}.
\end{proposition}

Proposition \ref{OBS-equiv} reduces to Proposition \ref{CREnoX-equiv} when $\hat{w}_i=z_i/p+(1-z_i)/(1-p)$ and $W=Z/p+(1-Z)/(1-p)$. In particular, the population summary measure in observational studies is identical to that in completely randomized experiments: $\tau^{\textup{c}}=\beta^{\T}\tau=\tau^\T \Sigma^{-1}\tau/(1+\tau^\T \Sigma^{-1}\tau)\in[0,1)$. Consequently, $\tau^{\textup{c}}$ remains
design-independent in observational studies. As in completely randomized experiments, the estimator of the summary measure satisfies $\hat{\tau}_{\textsc{os}}^{\textup{c}}=\hat\tau_{\textup{\textsc{os}}}^\T \hat\Sigma_{\textup{\textsc{os}}}^{-1}\hat\tau_{\textup{\textsc{os}}}/(1+\hat\tau_{\textup{\textsc{os}}}^\T \hat\Sigma_{\textup{\textsc{os}}}^{-1}\hat\tau_{\textup{\textsc{os}}})\in[0,1)$, because the positive definiteness of $\hat\Sigma_{\textup{\textsc{os}}}$ ensures that $\hat\tau_{\textup{\textsc{os}}}^\T \hat\Sigma_{\textup{\textsc{os}}}^{-1}\hat\tau_{\textup{\textsc{os}}}\geq0$.

Proposition \ref{OBS-equiv} also implies that testing $\tau=0$ is equivalent to testing $\beta=0$. We therefore define the Wald test for $H_0: \tau=0$ as
$\mathcal{W}^{\hat{\beta}_{\textsc{os}}} = \hat{\beta}_{\textsc{os}}^{\T}\hat{V}_{\smash{\beta_{\textsc{os}}}}^{-1}\hat{\beta}_{\textsc{os}}$, where $\hat{V}_{\smash{\beta_{\textsc{os}}}}$ is a consistent estimator of the asymptotic variance of $\hat{\beta}_{\textsc{os}}$.  Importantly, $\hat{V}_{\smash{\beta_{\textsc{os}}}}$ is not the Huber--White robust covariance estimator of the asymptotic variance of $\hat{\beta}_{\textsc{os}}$ from the inverse weighted regression that yields $\hat{\beta}_{\textsc{os}}$, because the latter ignores the uncertainty from estimating the propensity score. See the Supplementary Material for detailed asymptotic analysis and the next subsection for a related discussion. If the propensity score model is correctly specified, then under $H_0$ and valid asymptotic variance estimation, $\mathcal{W}^{\hat{\beta}_{\textsc{os}}}$ converges to a chi-squared distribution with $L$ degrees of freedom, even when the inverse regression model of $Z$ on $(1, Y)$ is misspecified.

\subsection{Asymptotic theory and statistical inference for the composite outcome} \label{subsec-OBS-asy}

To characterize the asymptotic properties of the estimator $\hat{\tau}_{\textsc{os}}^{\textup{c}}$, we posit a parametric model $e(X; \alpha)$ for the propensity score, indexed by the parameter $\alpha$, such as a logistic regression model. Let $\hat{e}(x_i)=e(x_i; \hat{\alpha})$ denote the estimated propensity score, where $\hat\alpha$ solves the estimating equation $\sum_{i=1}^{n}S(z_i,x_i; \alpha)=0$. Define the information matrix as $I(\alpha)=E\{-\nabla_{\alpha}S(Z, X; \alpha)\}$, where $\nabla_{\alpha}$ denotes the gradient with respect to $\alpha$.
If the propensity score model is correctly specified with $e(X)=e(X; \alpha_0)$ for some $\alpha_0$, then  $\hat{\beta}_{\textsc{os}}$ and $\hat{\tau}_{\textsc{os}}$ consistently estimate $\beta$ and $\tau$, respectively.
Consequently, $\hat{\tau}_{\textsc{os}}^{\textup{c}}$ is consistent for $\tau^{\textup{c}}$.

\begin{theorem}\label{OBS-compATE-asy}
Assume that strong ignorability $Z\ind\{Y(1), Y(0)\} \mid X$ and overlap $0<e(X)<1$ hold, and  the propensity score model $e(X; \alpha)$ is correctly specified.

(i) If $\tau\neq0$, then
$n^{1/2}(\hat{\tau}^{\textup{c}}_{\textup{\textsc{os}}}-\tau^{\textup{c}})\rightarrow N(0, 4\beta^\T\textup{cov}\{R_{\textup{\textsc{os}}}(Z,X,Y)\}\beta)$ in distribution as $n\rightarrow\infty$, where

\begin{equation*}
\begin{aligned}
R_{\textup{\textsc{os}}}(Z,X,Y)
&=\Big[\big(W\tilde{Y}\tilde{Y}^{\T}-\Phi_{\textsc{yy}}^{\textsc{w}}\big)\beta
+E\big\{\tilde{Y}\tilde{Y}^{\T}\beta\nabla_{\alpha}w(Z,X;\alpha_0)^{\T}\big\} I(\alpha_0)^{-1}S(Z,X;\alpha_0)\Big]\\
&\quad-\Big[(W-2)+E\big\{\nabla_{\alpha}w(Z,X;\alpha_0)^{\T}\big\}I(\alpha_0)^{-1}S(Z,X;\alpha_0)\Big]\Phi_{\textsc{yy}}^{\textsc{w}}\beta/2
+2\psi_{\textsc{y}{\scriptscriptstyle\varsigma}}(Z,X,Y),
\end{aligned}
\end{equation*}
and
\begin{equation*}
\begin{aligned}
\psi_{\textsc{y}{\scriptscriptstyle\varsigma}}(Z,X,Y)&=\varsigma W\tilde{Y}+E\big\{\varsigma\tilde{Y}\nabla_{\alpha}w(Z,X;\alpha_0)^{\T}\big\}I(\alpha_0)^{-1}S(Z,X;\alpha_0),\\
\nabla_{\alpha}w(Z,X;\alpha_0)&=\big[-Z/e(X; \alpha_0)^2+(1-Z)/\{1-e(X; \alpha_0)\}^2\big]\nabla_{\alpha}e(X; \alpha_0),
\end{aligned}
\end{equation*}
with $\tilde{Y}$ defined in Theorem \ref{CREnoX-compATE-asy}, $w(Z,X;\alpha)=Z/e(X; \alpha)+(1-Z)/\{1-e(X; \alpha)\}$,  and $\varsigma=Z-\beta_0-\beta^{\T}Y$ being the population residual from the weighted least squares regression of $Z$ on $(1, Y)$ with weights in \eqref{OBS-eq-weight}.

(ii) If $\tau=0$, then
$n\hat{\tau}_{\textup{\textsc{os}}}^{\textup{c}}\rightarrow\sum_{\ell=1}^{L}\lambda_{\textup{\textsc{os}}\ell}\chi^2_{\ell}(1)$ in distribution as $n\rightarrow\infty$, where $\lambda_{\textup{\textsc{os}}\ell}\ (\ell=1,\ldots, L)$ are eigenvalues of  $$\Gamma_{\textup{\textsc{os}}}=2(\Phi_{\textsc{yy}}^{\textsc{w}})^{-1/2} E\big\{\psi_{\textsc{y}{\scriptscriptstyle\varsigma}}(Z,X,Y)\psi_{\textsc{y}{\scriptscriptstyle\varsigma}}(Z,X,Y)^{\T}\big\}(\Phi_{\textsc{yy}}^{\textsc{w}})^{-1/2}.$$
\end{theorem}

As in completely randomized experiments, $\hat{\tau}_{\textsc{os}}^{\textup{c}}$ is asymptotically normal when $\tau\neq0$ and converges to a weighted sum of chi-squared distributions when $\tau=0$. Beyond the uncertainty in $\hat{\beta}_{\textsc{os}}$ and $\hat{\tau}_{\textsc{os}}$, Theorem \ref{OBS-compATE-asy} also accounts for the uncertainty from estimating the propensity score. Although the resulting asymptotic variance is more complex than that in Theorem \ref{CREnoX-compATE-asy}, it retains an analogous structure. In particular, compared with $R(Z,Y)$ in Theorem \ref{CREnoX-compATE-asy}, $R_{\textup{\textsc{os}}}(Z,X,Y)$ contains additional terms involving the gradient $\nabla_{\alpha}w(Z,X;\alpha_0)$, the score function $S(Z,X;\alpha_0)$, and the inverse information matrix $I(\alpha_0)^{-1}$, which capture the variability induced by estimating the propensity score. The asymptotic variance $4\beta^\T\textup{cov}\{R_{\textup{\textsc{os}}}(Z,X,Y)\}\beta$ reduces to $4\beta^\T\textup{cov}\{R(Z,Y)\}\beta$ in Theorem \ref{CREnoX-compATE-asy} when the propensity score is constant. We construct confidence intervals for $\tau^{\textup{c}}_{\textsc{os}}$ using sample-based estimators of the asymptotic variance and eigenvalues, with additional details provided in the Supplementary Material. We assume a parametric propensity score model for $e(X)$ to permit a tractable theoretical analysis of $\hat{\tau}_{\textsc{os}}^{\textup{c}}$. The proposed inverse regression framework could also incorporate more flexible, nonparametric propensity score estimators, such as the sieve estimator of \cite{hirano2003efficient}. Establishing the corresponding asymptotic theory would require a more technical analysis, which we leave for future research.

In this paper, we use the plug-in weighted least squares estimator of $\tau^{\textup{c}}$, because it is simple, arises naturally from the proposed inverse regression framework, and leads to elegant theoretical results. As an alternative, we could construct a double machine learning estimator \citep{chernozhukov2018double} by expressing $\tau$ and $\Sigma$ in terms of the first and second moments of the potential outcomes. Let $m_z=E\{Y(z)\}$ and $q_z=E\{Y(z)Y(z)^{\T}\}$ for $z=1,0$. Then $\tau=m_1-m_0$ and $\Sigma=2\{(q_1-m_1m_1^{\T})+(q_0-m_0m_0^{\T})\}$. We can estimate $m_z$ and $q_z$ by double machine learning, with the nuisance functions $e(X)$, $E\{Y\mid Z=z, X\}$, and $E\{YY^{\T}\mid Z=z, X\}$ estimated by flexible machine learning methods. Substituting the resulting estimates of $m_z$ and $q_z$ yields double machine learning estimators of $\tau$ and $\Sigma$, and hence of $\tau^{\textup{c}}$. We omit the technical details.

\section{Simulation studies}\label{sec-simu}

\subsection{Comparison of tests for multiple continuous outcomes}\label{sec-power-cont}

We conduct a simulation study to evaluate the performance of the test based on the proposed treatment effect measure $\hat{\tau}^{\textup{c}}$, denoted by $\mathcal{W}^{\hat{\tau}^{\textup{c}}}=n\hat{\tau}^{\textup{c}}$, for testing $H_0: \tau= 0$ in completely randomized experiments. We compare it with the Wald test based on  $\hat{\beta}$, $\mathcal{W}^{\hat{\beta}} = \hat{\beta}^\T\hat{V}_{\smash{\beta}}^{-1}\hat{\beta}$, and the standard Hotelling's $T^2$ test based on $\hat{\tau}$, denoted by  $\mathcal{W}^{\hat{\tau}}$. Let $\bar{y}_w=S^{w}_{\smash{y1}}(S^{w}_{\smash{11}})^{-1}$ and $\bar{z}_w=S^{w}_{\smash{z1}}(S^{w}_{\smash{11}})^{-1}$, so that  $\phi_{yy}^{w}=n^{-1}\sum_{i=1}^{n}w_i(y_i-\bar{y}_w)(y_i-\bar{y}_w)^{\T}$.
These three test statistics have similar forms:
\begin{equation}\label{simu-eq-teststats}
\begin{aligned}
&\mathcal{W}^{\hat{\tau}^{\textup{c}}}=n\phi_{zz}^{w}\hat{\tau}^{\T}\bigg\{n^{-1}\sum_{i=1}^{n}w_i(y_i-\bar{y}_w)(y_i-\bar{y}_w)^{\T}\bigg\}^{-1}
\hat{\tau}\\
&\hspace{0.8cm}=n\phi_{zz}^{w}\hat{\tau}^{\T}\big[\hat{\textup{cov}}\{Y(1)\}+\hat{\textup{cov}}\{Y(0)\}
+\phi_{zz}^{w}\hat{\tau}\hat{\tau}^{\T}\big]^{-1}\hat{\tau},\\
&\mathcal{W}^{\hat{\beta}}
=n(\phi_{zz}^{w})^2\hat{\tau}^{\T}\bigg\{n^{-1}\sum_{i=1}^{n}w_i^2\hat\varepsilon_i^2(y_i-\bar{y}_w)(y_i-\bar{y}_w)^\T\bigg\}^{-1}\hat{\tau},\\
&\mathcal{W}^{\hat{\tau}}=n\hat{\tau}^{\T}\big[\bar{z}^{-1}\hat{\textup{cov}}\{Y(1)\}+(1-\bar{z})^{-1}\hat{\textup{cov}}\{Y(0)\}\big]^{-1}\hat{\tau},
\end{aligned}
\end{equation}
where $\hat{\varepsilon}_i = z_i-\hat{\beta}_0-\hat{\beta}^{\T}y_i$ is the residual from the weighted least squares regression of $z_i$ on $(1, y_i)$ with weights $w_i$, and $\hat{\beta}_0=\bar{z}_w-\hat{\beta}^{\T}\bar{y}_w$.
All three test statistics are quadratic forms in $\hat{\tau}$, but they use different weighting matrices.
In the Supplementary
Material, we show that $\mathcal{W}^{\hat{\beta}}$ and $\mathcal{W}^{\hat{\tau}}$ are asymptotically equivalent under $H_0$ and the local alternatives $H_{1n}: \tau=hn^{-1/2}$, where $h$ is a fixed nonzero vector. By contrast,  $\mathcal{W}^{\hat{\tau}^{\textup{c}}}$ is generally not asymptotically equivalent to either $\mathcal{W}^{\hat{\tau}}$ or $\mathcal{W}^{\hat{\beta}}$, except under $H_0$ and $H_{1n}$ when $p=1/2$ or $\textup{cov}\{Y(1)\}/p=\textup{cov}\{Y(0)\}/(1-p)$. The relative performance of the three tests depends on $\tau$, $p$, $\textup{cov}\{Y(1)\}$, and $\textup{cov}\{Y(0)\}$.

\begin{table}
\caption{Empirical type I errors of $\mathcal{W}^{\hat{\tau}^{\textup{c}}}$, $\mathcal{W}^{\hat{\beta}}$, and $\mathcal{W}^{\hat{\tau}}$ at the nominal significance level of 0.05.}\label{tab-typeIerror}
\begin{center}
\begin{tabular}{ccccccc}
\hline
Scenario&1&2&3&4&5&6\\
$\mathcal{W}^{\hat{\tau}^{\textup{c}}}$&0.048 &0.052 &0.054 &0.056 &0.060 &0.057\\
$\mathcal{W}^{\hat{\beta}}$                          &0.057 &0.064 &0.056 &0.062 &0.070 &0.061\\
$\mathcal{W}^{\hat{\tau}}$             &0.069 &0.063 &0.055 &0.061 &0.064 &0.069\\
\hline
\end{tabular}
\end{center}
\end{table}

To examine these theoretical results, we independently generate the potential outcome $y_i(1)=(y_{i1}(1),\ldots, y_{iL}(1))^{\T}$ from
an $L$-dimensional normal distribution with mean vector $\tau=(\tau_1,\ldots,\tau_L)^{\T}$ and covariance matrix $\Omega_1$, and
$y_i(0)=(y_{i1}(0),\ldots, y_{iL}(0))^{\T}$ from an  $L$-dimensional normal distribution with mean zero and covariance matrix $\Omega_0$. The matrix $\Omega_1$ has constant diagonal entries $\sigma_1^2$ and constant off-diagonal entries $\rho_1$, and $\Omega_0$ has constant diagonal entries $\sigma_0^2$ and constant off-diagonal entries $\rho_0$. We independently draw each treatment indicator $z_i$ from a Bernoulli distribution with success probability $p$. We fix $L=10$ and vary the number $K$ of nonzero elements of $\tau$ from 0 to 10. We evaluate the empirical type I error rate at $K=0$, and the empirical power when $K\geq1$.  For $K\geq1$, we set $\tau_{\ell}=5n^{-1/2}$ for $\ell=1,\ldots,K$ and $\tau_{\ell}=0$ for $\ell=K+1,\ldots,L$. We consider six scenarios for the treatment probability $p$ and covariance matrices $\Omega_1$ and $\Omega_0$:

Scenario 1: $p=0.2$, $\sigma^2_1=3$, $\rho_1=-0.06$, $\sigma^2_0=1$, and $\rho_0=0.3$;

Scenario 2: $p=0.2$, $\sigma^2_1=1$, $\rho_1=0.3$,  $\sigma^2_0=3$, and $\rho_0=-0.06$;

Scenario 3: $p=0.5$, $\sigma^2_1=3$, $\rho_1=-0.06$, $\sigma^2_0=1$, and $\rho_0=0.3$;

Scenario 4: $p=0.5$, $\sigma^2_1=1$, $\rho_1=0.3$, $\sigma^2_0=3$, and $\rho_0=-0.06$;

Scenario 5: $p=0.8$, $\sigma^2_1=3$, $\rho_1=-0.06$, $\sigma^2_0=1$, and $\rho_0=0.3$;

Scenario 6: $p=0.8$, $\sigma^2_1=1$, $\rho_1=0.3$, $\sigma^2_0=3$,  and $\rho_0=-0.06$.

For each scenario, we simulate 1000 experiments, each with $n=1000$ units. We calculate the empirical type I error and power of each test using its asymptotic null distribution at the nominal significance level of $0.05$.

\begin{figure}
\centering
\includegraphics[width=0.8\linewidth]{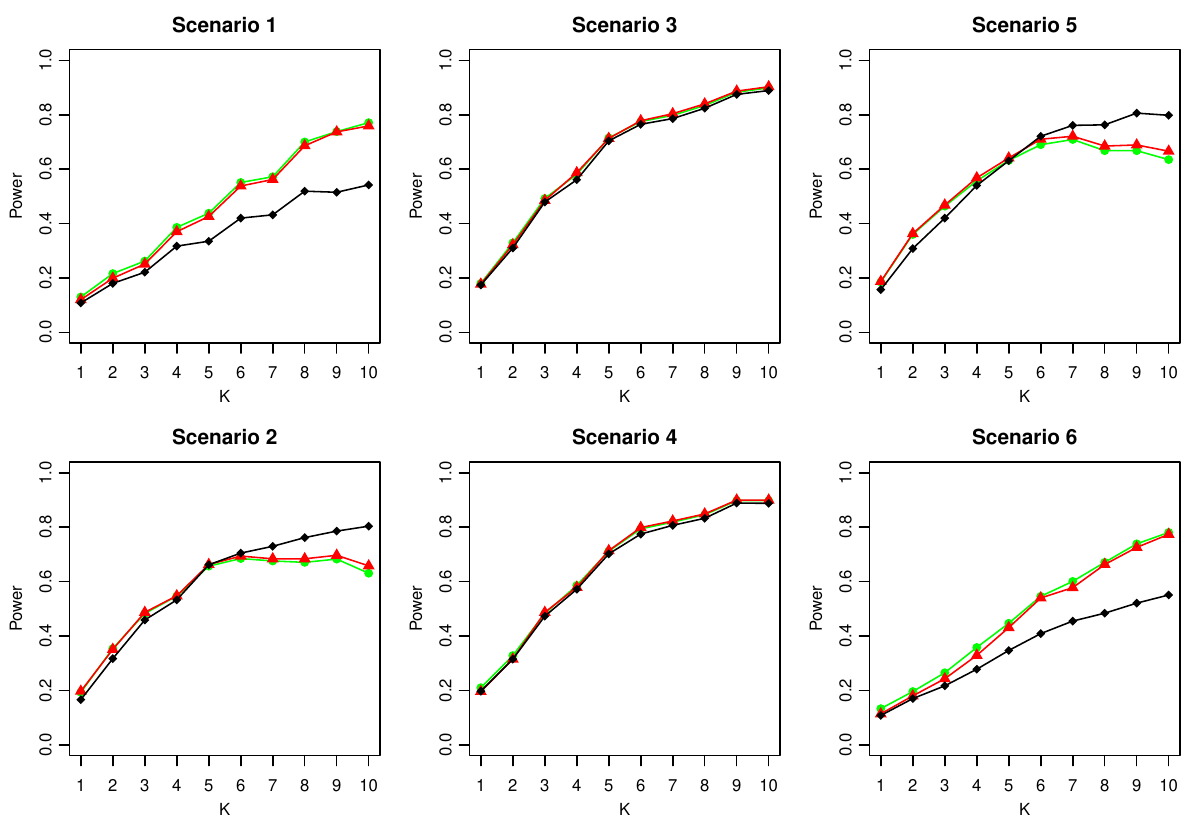}
\caption{Empirical powers of $\mathcal{W}^{\hat{\tau}^{\textup{c}}}$ (black diamonds), $\mathcal{W}^{\hat{\beta}}$ (red triangles), and $\mathcal{W}^{\hat{\tau}}$ (green circles) as the number $K$ of nonzero treatment effects varies from 1 to 10.}
\label{fig-power}
\end{figure}

Table \ref{tab-typeIerror} reports the empirical type I error rates. The tests $\mathcal{W}^{\hat{\beta}}$ and $\mathcal{W}^{\hat{\tau}}$ show slightly inflated type I error rates when $p=0.2$ or $p=0.8$, whereas $\mathcal{W}^{\hat{\tau}^{\textup{c}}}$ controls the type I error more stably. This stability may result from estimating the asymptotic null distribution of $\mathcal{W}^{\hat{\tau}^{\textup{c}}}$  from the data.
Figure \ref{fig-power} presents the empirical powers under the six scenarios. The powers of all three tests generally increases with $K$. The tests $\mathcal{W}^{\hat{\beta}}$ and $\mathcal{W}^{\hat{\tau}}$ have nearly identical power in every scenario, supporting their asymptotic equivalence under the local alternatives $H_{1n}$. By contrast, the power of $\mathcal{W}^{\hat{\tau}^{\textup{c}}}$ generally differs from those of the other two tests. The test $\mathcal{W}^{\hat{\tau}^{\textup{c}}}$ is less powerful than $\mathcal{W}^{\hat{\beta}}$ and $\mathcal{W}^{\hat{\tau}}$ in Scenarios 1 and 6. In Scenarios 2 and 5, $\mathcal{W}^{\hat{\tau}^{\textup{c}}}$ has similar or slightly lower power than $\mathcal{W}^{\hat{\beta}}$ and $\mathcal{W}^{\hat{\tau}}$ for small $K$, but becomes more powerful as $K$ increases. In Scenarios 3 and 4, where $p=0.5$, all three tests have nearly identical powers. These results agree with our theoretical comparisons of the three tests.

\subsection{Comparison of tests for multiple rare outcomes}\label{sec-power-rare}

We also conduct a simulation study to compare the three tests for multiple rare binary outcomes in a practically relevant regime where $L/n$ converges to a positive constant. The asymptotic theory developed in Theorem \ref{CREnoX-compATE-asy} does not apply to this regime, and we leave the corresponding asymptotic analysis of the three tests for future research. We generate the binary potential outcomes by first simulating latent continuous variables and then dichotomizing them. Specifically, we independently generate $\eta_{i}(1)=(\eta_{i1}(1), \ldots, \eta_{iL}(1))^T$ and $\eta_{i}(0)=(\eta_{i1}(0), \ldots, \eta_{iL}(0))^T$ from $L$-dimensional normal distributions with mean zero and covariance matrices $\Omega_1$ and $\Omega_0$, respectively. Let $\sigma_{z\ell\ell}^2$ denote the $\ell$th diagonal entry of $\Omega_z$ for $z=0,1$. We then define
\begin{equation*}
\begin{aligned}
y_{i\ell}(1) = 1\big(\eta_{i\ell}(1)/\sigma_{1\ell\ell}>\Phi^{-1}(1-\tau_{1\ell})\big) \quad &(i=1,\ldots,n; \ell=1,\ldots,L),\\
y_{i\ell}(0) = 1\big(\eta_{i\ell}(0)/\sigma_{0\ell\ell}>\Phi^{-1}(1-\tau_{0\ell})\big) \quad &(i=1,\ldots,n; \ell=1,\ldots,L),
\end{aligned}
\end{equation*}
where $1(\cdot)$ denotes the indicator function, $\Phi^{-1}(\cdot)$ denotes the quantile function of the standard normal distribution, $\tau_{1\ell}=E\{Y_{\ell}(1)\}$, and $\tau_{0\ell}=E\{Y_{\ell}(0)\}$.
We independently draw $z_i$ from a Bernoulli distribution with success probability $p$.
For each $Y_{\ell}(0)$, we set $\tau_{0\ell}$  to a value drawn from the uniform distribution on $(0.1, 0.2)$. We fix $L=50$ and vary the number $K$ of nonzero elements of $\tau$ from 0 to 50. For $K\geq 1$, we set $\tau_{\ell}=8(nL)^{-1/2}$ for $\ell=1,\ldots,K$ and $\tau_{\ell}=0$ for $\ell=K+1,\ldots,L$. We then set $\tau_{1\ell}=\tau_{0\ell}+\tau_{\ell}$. We use the same covariance structures for $\Omega_1$ and $\Omega_0$ and  consider the same six scenarios for $p$, $\Omega_1$, and $\Omega_0$ as in Section \ref{sec-power-cont}. For each scenario, we simulate 1000 experiments with $n=1000$ units. We implement each test using a bootstrap procedure with $B=2000$ replications and evaluate its the empirical type I error and power at the nominal significance level of $0.05$.

\begin{figure}
\centering
\includegraphics[width=0.8\linewidth]{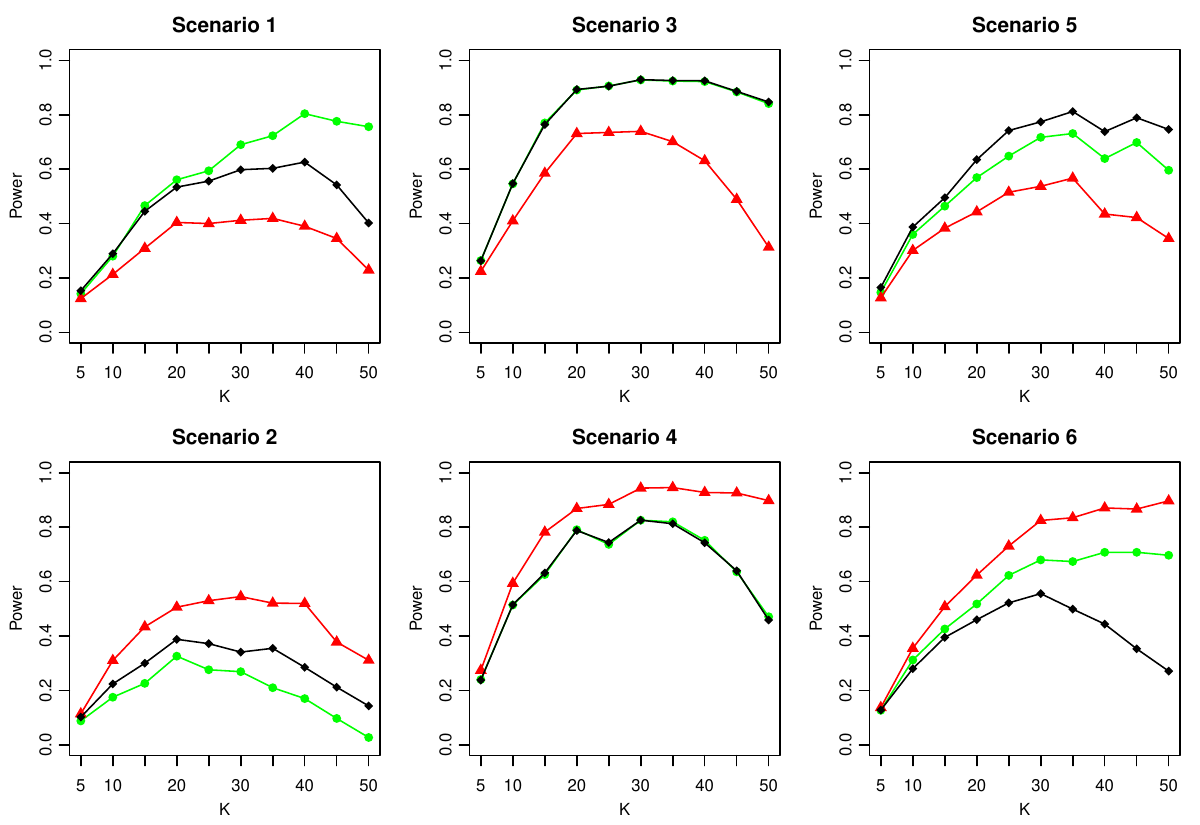}
\caption{Empirical powers of $\mathcal{W}^{\hat{\tau}^{\textup{c}}}$ (black diamonds), $\mathcal{W}^{\hat{\beta}}$ (red triangles), and $\mathcal{W}^{\hat{\tau}}$ (green circles) for rare outcomes as the number $K$ of nonzero treatment effects varies from 5 to 50.}
\label{fig-power-binaryrare}
\end{figure}

All three tests control the type I error rate adequately under the bootstrap procedure, so we omit these results and focus on power.  Figure \ref{fig-power-binaryrare} presents the empirical powers under the six scenarios. In most scenarios, the powers of the three tests initially increase with $K$ and then level off or decrease, although the value of $K$ that maximizes power varies across tests.
In contrast to the results for common outcomes in Figure \ref{fig-power},  the three tests can exhibit substantial power differences in the rare outcome setting. Here,  $\textup{cov}\{Y(1)\}$ and $\textup{cov}\{Y(0)\}$ have the same asymptotic order as $\tau$, or equivalently $\beta$, under the alternative hypotheses. Consequently,  $\mathcal{W}^{\hat{\beta}}$ and $\mathcal{W}^{\hat{\tau}}$ are no longer asymptotically equivalent.

Nevertheless, in Scenarios 3 and 4, where $p=0.5$, $\mathcal{W}^{\hat{\tau}^{\textup{c}}}$ and $\mathcal{W}^{\hat{\tau}}$ have nearly identical powers. Although the entries of $\textup{cov}\{Y(1)\}$ and $\textup{cov}\{Y(0)\}$ are small, the additional term $\tau\tau^{\T}$ in the weighting matrix of $\mathcal{W}^{\hat{\tau}^{\textup{c}}}$ in \eqref{simu-eq-teststats} remains asymptotically negligible relative to $2\textup{cov}\{Y(1)\}+2\textup{cov}\{Y(0)\}$. No test uniformly dominates the other two. The test $\mathcal{W}^{\hat{\tau}}$ has the highest power in Scenarios 1 and 3, whereas $\mathcal{W}^{\hat{\beta}}$ has the highest power in Scenarios 2, 4, and 6. The test $\mathcal{W}^{\hat{\tau}^{\textup{c}}}$ has the highest power in Scenarios 3 and 5. These results show that the relative power of the three tests depends strongly on the treatment probability and the covariance structure of the potential outcomes.

\section{Real-data examples}\label{sec-RealExample}

\subsection{A completely randomized experiment}\label{sec-RealExample-CRE}

We apply the proposed method to data from an 18-month randomized clinical trial evaluating the effect of a family-based behavioral intervention on dietary intake among youth with type 1 diabetes \citep{nansel2015improving}. The trial randomized 136 youth-parent dyads in a 1:1 ratio to the intervention and control groups. Children's dietary intake was assessed at six study visits: baseline and every three months following the intervention. Dietary quality was measured using 14 outcomes: Healthy Eating Index 2005, whole plant food density,  total fruit, whole fruit, total vegetables, dark green/orange vegetables and legumes, total grains, whole grains, dairy, meat and beans, oils, saturated fat, sodium, and calories from solid fats, alcohol, and added sugars. The first two are overall diet quality indicators, and the remaining twelve correspond to specific food categories.

\begin{figure}
\centering
\includegraphics[width=0.9\linewidth]{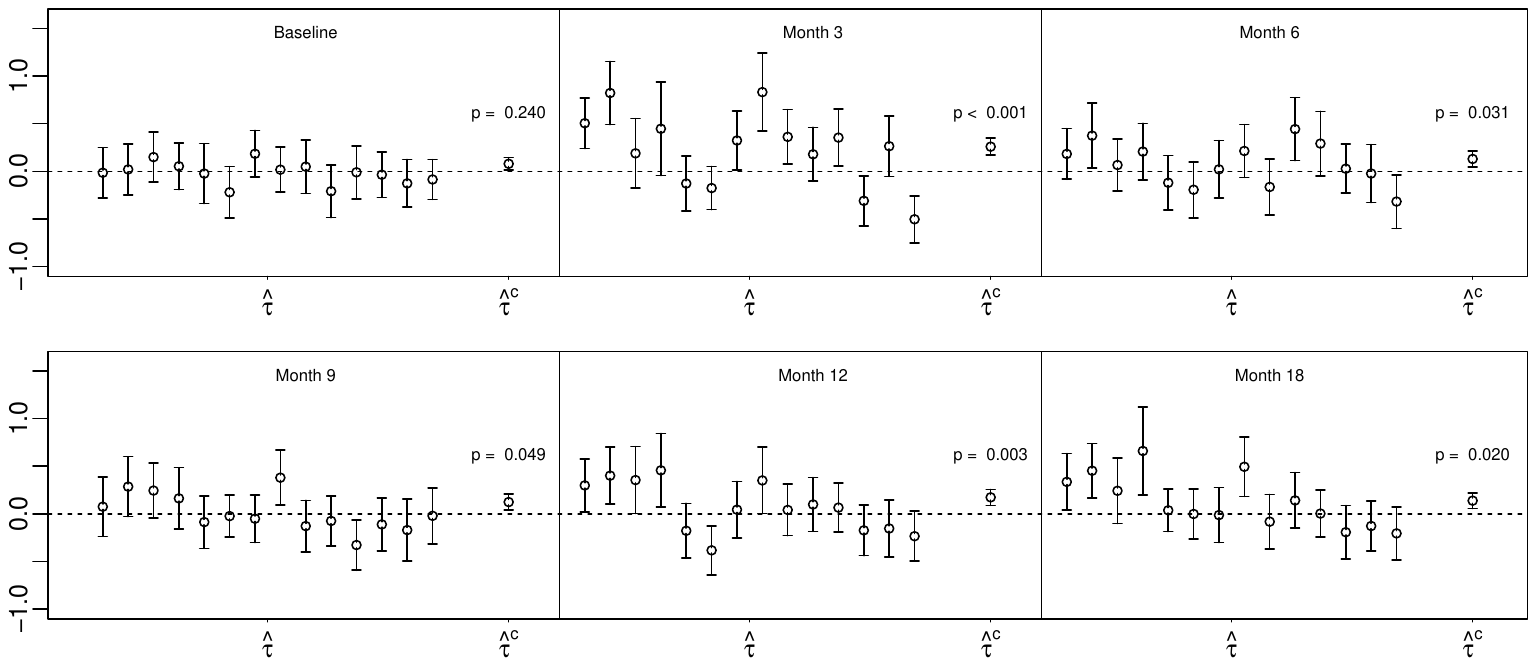}
\caption{Point estimates and 95\% confidence intervals for intervention effects on single dietary outcomes ($\hat\tau$) and the composite outcome ($\hat{\tau}^{\textup{c}}$) at six study visits from the type 1 diabetes trial analysed in Section \ref{sec-RealExample-CRE}. The $p$-values are from the tests based on $\hat{\tau}^{\textup{c}}$. }
\label{fig1-dietary}
\end{figure}

As an illustration,  we construct a composite outcome at each visit by combining 14 outcomes using the method in Section \ref{sec-CREnoX}. Figure \ref{fig1-dietary} presents the estimated intervention effects and associated 95\% confidence intervals for both single outcomes and the composite outcome across all visits.
Except at baseline, the composite outcome has $p$-values below 0.05, indicating significant overall effects of the intervention on the outcomes. The effect on the composite outcome is statistically significant whenever at least one single outcome is significant.

\subsection{An observational study} \label{sec-RealExample-OBS}

Studies have shown that smoking alters the concentrations of various vitamins \citep{Schectman1989,Bashar2004}. We
use the method in Section \ref{sec-OBSnoX} to study how smoking influences serum vitamin levels and related laboratory test results. Our analysis uses data from the 2005--2006 U.S. National Health and Nutrition Examination Survey, comparing daily smokers to never smokers. Daily smokers are defined as individuals who smoked every day for the
last 30 days and averaged at least 10 cigarettes each day, while never smokers smoked fewer than 100 cigarettes in their lifetime. We adjust for five categorical observed confounders: (i) gender (female or male), (ii) age  ($<40$, 40--50, or $\geq50$ years), (iii) education level ($<$ high school, high school, or $>$ high school), (iv) BMI ($<30$, 30--35, or $\geq35$), and (v) federal poverty level (income $<2\times$ poverty or $\geq2\times$ poverty).  Propensity scores are estimated using the logistic regression.

\begin{figure}
\centering
\includegraphics[width=0.45\linewidth]{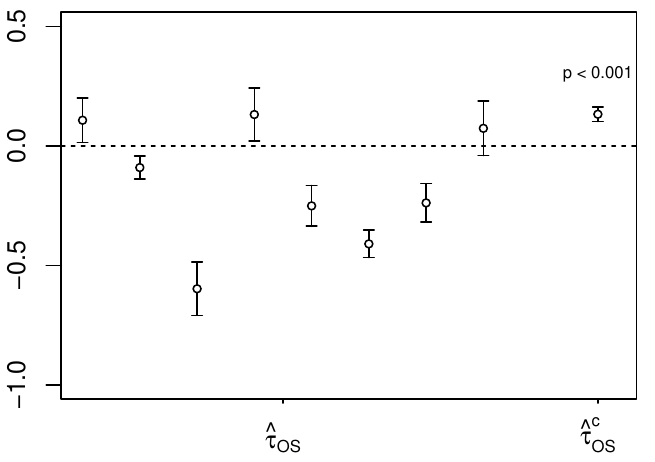}
\caption{Point estimates and 95\% confidence intervals for treatment effects on single outcomes ($\hat\tau_{\textsc{os}}$) and the composite outcome ($\hat\tau_{\textsc{os}}^{\textup{c}}$) for the 2005--2006 National Health and Nutrition Examination Survey data analysed in Section \ref{sec-RealExample-OBS}. The $p$-value is from the test based on $\hat\tau_{\textsc{os}}^{\textup{c}}$.}
\label{fig2-nhanes}
\end{figure}

We examine eight laboratory outcomes: serum levels of vitamin A, B12, C, D, and E, as well as beta-carotene, folate, and C-reactive protein. The final sample includes 2521 individuals aged 12 to 85, with 524 daily smokers and 1997 never smokers. To evaluate the overall effect of smoking, we construct a composite outcome from these eight outcomes. Figure \ref{fig2-nhanes}
presents point estimates and 95\% confidence intervals for the composite outcome as well as the single outcomes. Since the single outcomes vary widely in scale, we standardize each using the standard deviation in the never smoker group to facilitate visual comparison in the figure. This standardization does not affect the composite outcome based on inverse regression, as explained in Section \ref{subsec-CREnoX-ATE}. The composite outcome yields a $p$-value
less than $0.001$, indicating a highly significant overall effect of smoking on serum vitamin levels. Except for C-reactive protein, all single outcomes differ significantly between daily smokers and never smokers, consistent with previous findings \citep{Schectman1989,Bashar2004}.

\section{Extensions} \label{sec-exts}

\subsection{Covariate adjustment}\label{subsec-ext-covadj}

An important advantage of the inverse regression approach is its flexibility in incorporating additional covariates within a regression framework. We extend the proposed method to both completely randomized experiments and observational studies with covariates. For notational simplicity, we continue to use $X$ for the covariates and $x_i$ for the observed covariates of unit $i$.

For completely randomized experiments, we inversely regress $Z$ on $(1, X, Y)$ using the weights in \eqref{CRE-eq-weight} to construct a composite outcome.  Let $\beta_{\textsc{a}}$ denote the coefficient of $Y$, where the subscript ``A'' signifies ``covariate adjustment''. We define the composite outcome  $Y_{\textsc{a}}^{\textup{c}} =\beta_{\textsc{a}}^{\T}Y$, with corresponding average treatment effect $\tau_{\textsc{a}}^{\textup{c}}=\beta_{\textsc{a}}^{\T}\tau$. To estimate $\tau_{\textsc{a}}^{\textup{c}}$, we first estimate $\beta_{\textsc{a}}$ by the coefficient of $y_i$ from the weighted least squares regression of $z_i$ on $(1, x_i, y_i)$ with weights $w_i=z_i/p+(1-z_i)/(1-p)$ and denote the resulting estimator by $\hat{\beta}_{\textsc{a}}$. We then estimate $\tau$ by the coefficient of $z_i$ from the component-wise weighted least squares regression of $y_i$ on $(1, x_i, z_i)$ using the same weights, and denote the estimator by $\hat{\tau}_{\textsc{a}}$. The covariate-adjusted estimator of $\tau_{\textsc{a}}^{\textup{c}}$ is $\hat\tau_{\textsc{a}}^{\textup{c}}=\hat\beta_{\textsc{a}}^{\T}\hat\tau_{\textsc{a}}$.


For observational studies, we directly include $X$ in the weighted inverse regression to construct the composite outcome,
as in completely randomized experiments. Let
$\beta_{\textsc{a}}$ denote the coefficient of $Y$ from the weighted least squares regression of $Z$ on $(1, X, Y)$ with weights $W$ in \eqref{OBS-eq-weight}. We define the composite outcome  $Y_{\textsc{a}}^{\textup{c}} =\beta_{\textsc{a}}^{\T}Y$ and its average treatment effect $\tau_{\textsc{a}}^{\textup{c}}=\beta_{\textsc{a}}^{\T}\tau$. We estimate $\tau_{\textsc{a}}^{\textup{c}}$ by $\hat\tau_{\textsc{os,a}}^{\textup{c}}=\hat{\beta}_{\textsc{os,a}}^{\T}\hat{\tau}_{\textsc{os,a}}$, where $\hat{\beta}_{\textsc{os,a}}$ is the coefficient of $y_i$ from the weighted regression of $z_i$ on $(1, x_i, y_i)$ with weights $\hat{w}_i=z_i/\hat{e}(x_i)+(1-z_i)/\{1-\hat{e}(x_i)\}$ and $\hat{\tau}_{\textsc{os,a}}$ is the coefficient of $z_i$ from the component-wise weighted regression of $y_i$ on $(1, z_i, x_i)$ using the same weights.

For both designs with covariate adjustment, we establish equivalence results and derive asymptotic properties analogous to those in Sections \ref{sec-CREnoX}--\ref{sec-OBSnoX}, with details in the Supplementary Material. The covariate-adjusted coefficient vector $\beta_{\textsc{a}}$ depends on both the moments of the potential outcomes and the covariates included in the inverse regression model. Consequently, different covariate sets may produce different inverse regression coefficient vectors and composite outcomes.  As in the setting without covariate adjustment, each covariate-adjusted inverse regression coefficient vector represents a treatment-aligned direction that maximizes the corresponding standardized mean difference. For a fixed design, covariate adjustment yields a maximum standardized mean difference at least as large as that obtained without covariate adjustment. More generally, expanding the covariate set does not decrease the maximum standardized mean difference.

\subsection{Stratified randomized experiments with multiple outcomes} \label{subsec-ext-SRE}

The inverse regression method extends naturally to stratified randomized experiments.
Stratified randomization is widely used to balance baseline covariates and improve estimation efficiency \citep{miratrix:2013, imbens2015causal, liu2020regression}. It partitions experimental units into strata based on a discrete covariate and conducts independent completely randomized experiments within each stratum. Consider an experiment with $n$ units divided into $S$ strata. Let $C\in\{1,\ldots, S\}$ denote the stratification variable, and let stratum $s$ contains $n_s$ units. For each stratum $s$, define the stratum-specific average treatment effect, treatment probability, and stratum probability as
$\tau_{\smash{[s]}}=E\{Y(1)-Y(0)\mid C=s\}$,   $p_{\smash{[s]}}=\pr(Z=1\mid C=s)$, and  $\pi_{s}=\pr(C=s)$, respectively. The overall average treatment effect satisfies $\tau=\sum_{s=1}^{S}\pi_s\tau_{\smash{[s]}}$.

There are two primary strategies for estimating treatment effects in stratified randomized experiments: the stratification strategy \citep{liu2020regression, ding2024first}  and the regression strategy \citep{angrist2009mostly}. The stratification strategy first estimates $\tau_{\smash{[s]}}$ within each stratum and then aggregates the stratum-specific estimates using the stratum proportions $n_s/n$. The methods and results in Section \ref{sec-OBSnoX} apply directly to the stratification strategy. Stratified randomization ensures $Z\ind\{Y(1), Y(0)\} \mid C$, so we can represent the stratified experiment as an observational study with  $C$  as a pretreatment covariate and  propensity score $e(C)=\sum_{s=1}^Sp_{[s]}I(C=s)$. Following Section \ref{sec-OBS-ATE}, we construct a composite outcome by inversely regressing $Z$ on $(1, Y)$ with weights $W=Z/e(C)+(1-Z)/\{1-e(C)\}$, and define the corresponding overall treatment effect measure. We can establish an equivalence result and asymptotic theory analogous to those in Proposition \ref{OBS-equiv} and Theorem \ref{OBS-compATE-asy}, respectively. Because $e(C)$ is known by design and constant within each stratum, the asymptotic variance does not include uncertainty from propensity score estimation. To avoid repetition, we omit the technical details and implement the procedure in our R package. We next consider the regression strategy.

The regression strategy is an alternative, popular method in social sciences. It pools observations across strata and adjusts for stratification by including  treatment and stratum indicators
in a linear regression. Assume that $(z_i, c_i, y_{i}(0), y_i(1))\ (i=1,\ldots, n)$  are independent copies of $(Z, C, Y(0),Y(1))$. Let $G=(G_{1},\ldots,G_{S})^{\T}$ denote the vector of stratum indicators, where $G_s=I(C=s)$, and let $g_i=(I(c_i=1),\ldots,I(c_i=s))^{\T}$ denote the observed stratum indicators for unit $i$. The regression strategy estimates the treatment effect using the coefficient of $z_i$ from the component-wise ordinary least squares regression of $y_i$ on $(g_i, z_i)$. In general, the coefficient does not consistently estimate $\tau$. Instead, it consistently estimates the variance-weighted stratum-specific treatment effects \citep{angrist1998estimating}:
\begin{equation}\label{indATE-regressStra}
\begin{aligned}
\tau_{\textsc{sr}}=(\tau_{\textsc{sr}1},\ldots, \tau_{\textsc{sr}L})^{\T}=\frac{\sum_{s=1}^{S}\pi_{s}p_{[s]}(1-p_{[s]})\tau_{[s]}}{\sum_{s=1}^{S}\pi_{s}p_{[s]}(1-p_{[s]})}.
\end{aligned}
\end{equation}
Thus, $\tau_{\textsc{sr}}$ is a weighted average of $\tau_{[s]}$, with weights proportional to $\pi_{s}p_{[s]}(1-p_{[s]})$,
the product of the stratum proportion and the variance of the treatment within that stratum.
The estimand $\tau_{\textsc{sr}}$  coincides with $\tau$ when either the treatment probabilities or the stratum-specific treatment effects are constant across strata, i.e., $p_{[s]}=p$ or $\tau_{[s]} = \tau$ for all $s$.

We can also characterize $\tau_{\textsc{sr}}$ using weighted least squares regression. Specifically, $\tau_{\textsc{sr}}$ equals the coefficient of $Z$ from the element-wise weighted least squares regression of $Y$ on $(G, Z)$ with overlap weights \citep{li2018balancing}
\begin{equation}\label{SRE-eq-weight}
W_{\textsc{sr}}=Z\{1-e(C)\}+(1-Z)e(C),
\end{equation}
which assign each unit the probability of receiving the opposite treatment. Motivated by the weighted regression representation, we construct a composite outcome by inversely regressing $Z$ on $(G, Y)$ using the same weights. Let $\beta_{\textsc{sr}}$ denote the coefficient of $Y$, and define $Y_{\textsc{sr}}^{\textup{c}} =\beta_{\textsc{sr}}^{\T}Y$ and $\tau_{\textsc{sr}}^{\textup{c}}=\beta_{\textsc{sr}}^{\T}\tau_{\textsc{sr}}$. We estimate $\tau^{\textup{c}}_{\textsc{sr}}$ using the sample analogues of $\beta_{\textsc{sr}}$ and $\tau_{\textsc{sr}}$, denoted by $\hat{\beta}_{\textsc{sr}}$ and $\hat{\tau}_{\textsc{sr}}$. For unit $i$, let $w_{\textsc{sr}i}=z_i\{1-e(c_i)\}+(1-z_i)e(c_i)$. We obtain $\hat{\beta}_{\textsc{sr}}$ as the coefficient of $y_i$ from the weighted least squares regression of $z_i$ on $(g_i,y_i)$, and obtain $\hat{\tau}_{\textsc{sr}}$ as the coefficient of $z_i$ from the component-wise weighted regression of $y_i$ on $(g_i,z_i)$, using  $w_{\textsc{sr}i}$in both regressions. The resulting estimator of $\tau_{\textsc{sr}}^{\textup{c}}$  is $\hat\tau^{\textup{c}}_{\smash{\textsc{sr}}}=\hat{\beta}_{\textsc{sr}}^{\T}\hat{\tau}_{\textsc{sr}}.$

To adjust for covariates $X$, we regress $Z$ on $(G, X, Y)$ with weights $W_{\textsc{sr}}$ in \eqref{SRE-eq-weight}. Let $\beta_{\textsc{sr,a}}$ denote the coefficient of $Y$. We define  $Y_{\textsc{sr,a}}^{\textup{c}} =\beta_{\textsc{sr,a}}^{\T}Y$ and $\tau_{\textsc{sr,a}}^{\textup{c}}=\beta_{\textsc{sr,a}}^{\T}\tau_{\textsc{sr}}$. A direct weighted least squares regression of $y_i$ on $(g_i, z_i, x_i)$ does not consistently estimate $\tau_{\textsc{sr}}$. Indeed,  $\tau_{\textsc{sr}}$ equals the coefficient of $Z$ from the unadjusted weighted regression of $Y$ on $(G,Z)$ with weights $W_{\textsc{sr}}$. Consequently, the product of the covariate-adjusted estimators of $\beta_{\textsc{sr,a}}$ and $\tau_{\textsc{sr}}$ does not consistently estimate $\tau_{\textsc{sr,a}}^{\textup{c}}$.

Under $Z\ind\{Y(1),Y(0), X\}\mid C$, the coefficient of $Z$ in the element-wise weighted least squares regression of $X$ on $(G, Z)$ equals zero. This property motivates a three-step covariate-adjusted procedure that treats $X$ as outcomes and uses the estimated treatment effect of a composite covariate to adjust for covariates. First, regress $z_i$ on $(g_i, x_i, y_i)$ with weights $w_{\textsc{sr}i}$ and obtain the coefficients  $\hat{\beta}_{\textsc{sr,a}}$ for $y_i$ and $\hat{\beta}_{\textsc{srx,a}}$ for $x_i$. Define $y^{\textup{c}}_{\textsc{sr,a}i}=\hat{\beta}_{\textsc{sr,a}}^{\T}y_i$ and $x^{\textup{c}}_{\textsc{sr,a}i}=\hat{\beta}_{\textsc{srx,a}}^{\T}x_i$.  Second, regress $y^{\textup{c}}_{\textsc{sr,a}i}$ on $(g_i, z_i)$ using the weights $w_{\textsc{sr}i}$, and denote the coefficient of $z_i$ by $\hat{\tau}^{\textup{c}}_{\textsc{sry,a}}$. Third, regress $x^{\textup{c}}_{\textsc{sr,a}i}$ on $(g_i, z_i)$ using the same weights, and denote the coefficient of $z_i$ by $\hat{\tau}^{\textup{c}}_{\textsc{srx,a}}$.
We estimate $\tau_{\textsc{sr,a}}^{\textup{c}}$ by
$\hat{\tau}^{\textup{c}}_{\textsc{sr,a}}= \hat{\tau}^{\textup{c}}_{\textsc{sry,a}}-\hat{r}_{\textup{opt}}\hat{\tau}^{\textup{c}}_{\textsc{srx,a}}$,  where $\hat{r}_{\textup{opt}}=\hat{\textup{cov}}(\hat{\tau}^{\textup{c}}_{\textsc{sry,a}},  \hat{\tau}^{\textup{c}}_{\textsc{srx,a}})/\hat{\textup{var}}(\hat{\tau}^{\textup{c}}_{\textsc{srx,a}})$. The quantity $\hat{r}_{\textup{opt}}$  estimates the coefficient that minimizes the variance among the estimators of the form $\hat{\tau}^{\textup{c}}_{\textsc{sry,a}}-r\hat{\tau}^{\textup{c}}_{\textsc{srx,a}}$, where $\hat{\textup{cov}}$ and $\hat{\textup{var}}$ denote the sample covariance and variance, respectively.

For both the unadjusted and covariate-adjusted procedures, we establish equivalence results and asymptotic theory analogous to those for completely randomized experiments and observational studies. We provide the details in the Supplementary Material.

\subsection{Inverse logistic regression}\label{subsec-ext-logistic}

Since $Z$ is binary, most people may feel logistic regression of $Z$ on $Y$ is more natural.  \cite{Zhang2014} applied an inverse logistic regression framework to jointly analyse multiple clinical outcomes, formulating it as a semiparametric density ratio model for the outcome distributions in the two treatment groups. They used the inverse logistic regression coefficients to simultaneously test the equality of the conditional outcome distributions given treatment.  Beyond linear regression, we also explore the use of inverse logistic regression to combine multiple outcomes.

For completely randomized experiments, we construct a composite outcome by fitting the inverse weighted logistic regression of $Z$ on $Y$:
$\logit\{\pr(Z=1\mid Y)\}=\gamma_0+\gamma^{\T}Y$, using the weights $W=Z/p+(1-Z)/(1-p)$. Let $\hat\gamma$ denote the weighted maximum likelihood estimator of $\gamma$ with weights $w_i=z_i/p+(1-z_i)/(1-p)$. Let $\gamma^*$  denote the probability limit of $\hat\gamma$, or equivalently,  the unique maximizer of the population weighted log-likelihood. We define the composite outcome as $Y^{\textup{logit,c}} =(\gamma^*)^{\T}Y$ and the corresponding average treatment effect as $\tau^{\textup{logit,c}}=(\gamma^*)^{\T}\tau$. We estimate $\tau^{\textup{logit,c}}$ by $\hat{\tau}^{\textup{logit,c}}=\hat\gamma^{\T}\hat\tau$, where $\hat\tau$ is defined in Section \ref{subsec-CREnoX-ATE}. For observational studies, we construct a composite outcome analogously by fitting the inverse weighted logistic regression of $Z$ on $Y$ with weights $W=Z/e(X)+(1-Z)/\{1-e(X)\}$. The Supplementary Material provides further details.

For both completely randomized experiments and observational studies, we establish an asymptotic equivalence between the inverse logistic regression coefficient estimator $\hat\gamma$ and the treatment effect estimator $\hat\tau$ when $\tau=0$. The equivalence supports  the use of the inverse logistic regression coefficients to jointly test for treatment effects across multiple outcomes, consistent with the approach of \cite{Zhang2014}.
When $\tau\neq0$,  however, the asymptotic equivalence no longer holds, making the interpretation of the inverse logistic regression coefficients challenging. Therefore, inverse logistic regression is primarily useful for testing the null hypothesis of no causal effects.

\section*{Acknowledgement}
We thank the associate editor and two reviewers for insightful comments that improved this paper.
Zhang was supported by the National Key R\&D Program of China (2022YFA1004800). Li was supported by the National Natural Science Foundation of China (12325110, 12288201). Ding was supported by the U.S. National Science Foundation (1945136, 2514234). Li is also affiliated with the School of Mathematical Sciences, University of Chinese Academy of Sciences, China.

\section*{Supplementary material} \label{SM}

The Supplementary Material includes proofs, extensions, and details omitted in the main text.

\bibliographystyle{apalike}
 \bibliography{paper-ref}

\newpage

\begin{center}
\LARGE{Supplementary Material}
\end{center}

\appendix

\setcounter{section}{0}
\global\long\def\thesection{S\arabic{section}}%
\newcommand*{\QEDB}{\null\nobreak\hfill\ensuremath{\square}}%

\pagenumbering{arabic} 
\renewcommand*{\thepage}{S\arabic{page}}
\renewcommand{\thetheorem}{S\arabic{theorem}}
\renewcommand{\theproposition}{S\arabic{proposition}}
\renewcommand{\thelemma}{S\arabic{lemma}}
\renewcommand{\thefigure}{S\arabic{figure}}
\setcounter{equation}{0}
\renewcommand{\theequation}{S\arabic{equation}}

\medskip
\noindent Section \ref{sec::lemmas} presents some useful lemmas.

\noindent Section \ref{sec::tests-addresult} presents additional results that complement the main paper.

\noindent Section \ref{sec::CRE-OBS-withX} provides extensions to covariate adjustment.

\noindent Section \ref{sec::varweight-SRE} provides an extension to a variance-weighted treatment effect in stratified randomized experiments.

\noindent Section \ref{sec::logistic} extends the inverse regression framework to logistic regression.

\noindent Section \ref{sec::proof-CRE-noX} gives proofs of results in Section 2 of the main paper.

\noindent Section \ref{sec::proof-OBS-noX} gives proofs of results in Section 3 of the main paper.

\noindent Section \ref{sec::secSupp-proof} gives proofs of results in Sections \ref{sec::CRE-OBS-withX}--\ref{sec::logistic}.

\medskip
\noindent {\bf Notations:} Let $\stackrel{d}\rightarrow$ denote convergence in distribution and $\stackrel{p}\rightarrow$ denote convergence in probability. For a  random vector $a_n$ and a random variable $b_n$, we write $a_n=o_p(b_n)$ if  $a_n/b_n\rightarrow0$ in probability as $n\rightarrow\infty$ and $a_n=O_p(b_n)$ if for any $\kappa>0$, there exists a constant $C_\kappa>0$ such that $\sup_n\pr(\|a_n\|>C_\kappa |b_n|)<\kappa$. Let $\textup{cov}(\cdot)$ denote the covariance of a random vector. Let $I_d$ denote the $d\times d$ identity matrix, and $0_d, 1_d\in\mathbb{R}^d$ denote the vectors with all 0's and 1's.  We abbreviate weighted least squares as WLS, Frisch--Waugh--Lovell as FWL, Law of Large Numbers as LLN, and Central Limit Theorem as CLT.

\section{Some useful lemmas}\label{sec::lemmas}

\begin{lemma}[Rank-one update formula]\label{Woodburyformula}
Suppose that $A\in\mathbb{R}^{L\times L}$ is an invertible square matrix and $\mu,\nu\in\mathbb{R}^{L}$ are column vectors. Then $A+\mu\nu^{\T}$ is invertible if and only if $1+\nu^{\T}A^{-1}\mu\neq0$. In this case,
\begin{equation*}
(A+\mu\nu^\T)^{-1}=A^{-1}-\frac{A^{-1}\mu\nu^\T A^{-1}}{1+\nu^\T A^{-1}\mu}.
\end{equation*}
\end{lemma}

\begin{lemma}\label{sampleEig-Consist}
Let $\Gamma$ and $\hat\Gamma$ be two symmetric matrices with the eigenvalues $\lambda_1\geq\cdots\geq\lambda_L$ and
$\hat\lambda_1\geq\cdots\geq\hat\lambda_L$, respectively. Assume that  $\|\hat\Gamma-\Gamma\|_{\infty}=o_p(1)$.  Then for any $1\leq\ell\leq L$, we have
$$\hat\lambda_{\ell}-\lambda_{\ell}=o_p(1).$$
\end{lemma}
\begin{proof}
Applying Weyl's inequality yields
\begin{equation*}
\lambda_{\min}(\hat\Gamma-\Gamma)\leq\hat\lambda_{\ell}-\lambda_{\ell}
\leq\lambda_{\max}(\hat\Gamma-\Gamma),
\end{equation*}
where $\lambda_{\min}(A)$ and $\lambda_{\max}(A)$ are the smallest and largest eigenvalues of $A$, respectively.
Then we have
$$|\hat\lambda_{\ell}-\lambda_{\ell}|<\sup\limits_{\|u\|=1}|u^\T(\hat\Gamma-\Gamma)u|=o_p(1).$$
\end{proof}

\begin{lemma}[\citeauthor{Rudin:1976} \citeyear{Rudin:1976}, Theorem 7.17]\label{lemma-Rudin1976}
Suppose that $\{f_N(x)\}_{N=1}^{\infty}$ is a sequence of functions, differentiable on $[a,b]$ and $\{f_N(x)\}_{N=1}^{\infty}$ converges for some point $x_0$ on $[a,b]$. If the sequence of derivatives, $\{f'_N(x)\}_{N=1}^{\infty}$, converges uniformly on $[a,b]$, then $\{f_N(x)\}_{N=1}^{\infty}$  converges uniformly on $[a,b]$ to a function $f(x)$, and
\begin{equation*}
\lim\limits_{N\rightarrow\infty}f'_N(x)=f'(x), \quad x\in[a,b].
\end{equation*}
\end{lemma}

\begin{lemma}[\citeauthor{Newey1994McFadden} \citeyear{Newey1994McFadden}, Theorem 2.7]\label{lemma-NeweyMcFadden1994}
If there is a function $Q_0(\theta)$ such that (a) $Q_N(\theta)=Q_0(\theta)+o_p(1)$ for all $\theta\in\Theta$;
(b) $Q_0(\theta)$ is uniquely maximized at $\theta_0$; and (c) $\theta_0$ is an element of the interior of a convex set $\Theta$ and $Q_N(\theta)$ is concave, then the maximizer of $Q_N(\theta)$, denoted by $\hat{\theta}_N$, exists with probability approaching one and $\hat{\theta}_N=\theta_0+o_p(1)$.
\end{lemma}

\section{Additional results}\label{sec::tests-addresult}

\subsection{Confidence interval for the treatment effect $\tau^{\textup{c}}$ in completely randomized experiments}
\label{subsec::CRE-noX-CI}

We use Theorem 1 to construct confidence intervals for the treatment effect $\tau^{\textup{c}}$ on the composite outcome. When $\tau\neq0$,  a Wald-type $(1-\alpha)$-level confidence interval for $\tau^{\textup{c}}$ is
$$\big[\hat\tau^{\textup{c}}-n^{-1/2}\Phi^{-1}(1-\alpha/2)\hat{V}^{1/2},\  \hat\tau^{\textup{c}}+n^{-1/2}\Phi^{-1}(1-\alpha/2)\hat{V}^{1/2}\big],$$
where $\Phi^{-1}(1-\alpha/2)$ is the $1-\alpha/2$ quantile of the standard normal distribution, $\hat{V}=4\hat{\beta}^\T(n^{-1}\sum_{i=1}^{n}r_ir_i^{\T})\hat{\beta}$ with $r_i=(w_i\tilde{y}_i\tilde{y}_i^{\T}-\phi_{yy}^{w})\hat{\beta}-(w_i-2)\phi_{yy}^{w}\hat{\beta}/2+2w_i\hat{\varepsilon}_i\tilde{y}_i$, $\tilde{y}_i=y_i-\sum_{i=1}^{n}w_i$ $y_i/(\sum_{i=1}^{n}w_i)$,
and $\hat{\varepsilon}_i=z_i-\hat\beta_0-\hat\beta^{\T}y_i$ being the residual from the WLS fit of $z_i$ on $(1, y_i)$ with weights $w_i$.
Here, $(\hat\beta_0, \hat\beta^{\T})^{\T}$ denotes the WLS coefficient estimator.

When $\tau=0$, the limiting distribution of $n\hat\tau^{\textup{c}}$ involves a weighted sum of independent chi-squared random variables. Let
$\hat\Gamma=2(\phi_{yy}^{w})^{-1/2}
(n^{-1}\sum_{i=1}^{n}w_i^2\hat\varepsilon_i^2\tilde{y}_i
\tilde{y}_i^{\T})(\phi_{yy}^{w})^{-1/2}$ denote the sample analogue of $\Gamma$.
Let $\hat\lambda_1, \ldots, \hat\lambda_L$ denote the eigenvalues of $\hat\Gamma$, and let $\hat{\Upsilon}$ denote the distribution function of $\sum_{\ell=1}^{L}\hat\lambda_{\ell}\chi^2_{\ell}(1)$. A $(1-\alpha)$-level confidence interval for $\tau^{\textup{c}}$ is $$\big[\hat\tau^{\textup{c}}-n^{-1}\hat{\Upsilon}^{-1}(1-\alpha/2),\  \hat\tau^{\textup{c}}-n^{-1}\hat{\Upsilon}^{-1}(\alpha/2)\big],$$
where $\hat{\Upsilon}^{-1}(\alpha/2)$ and $\hat{\Upsilon}^{-1}(1-\alpha/2)$ are the $\alpha/2$ and $1-\alpha/2$ quantiles of $\hat{\Upsilon}$, respectively. This confidence interval has an asymptotic coverage probability $1-\alpha$, because Lemma \ref{sampleEig-Consist} ensures that the estimated eigenvalues $\hat\lambda_{\ell}$ are consistent for $\lambda_{\ell}$.

In practice, the value of $\tau$ is typically unknown, making it challenging to select the correct confidence interval. One possible approach is to construct separate confidence intervals for $\tau=0$ and $\tau\neq0$, and then take their union. An alternative is a two-step procedure that first tests $\tau=0$ using the Wald statistic $\mathcal{W} = \hat{\beta}^\T\hat{V}_{\smash{\beta}}^{-1}\hat{\beta}$ at a significance level $\alpha_1$ (often set to $\alpha/2$), and then selects the appropriate asymptotic distribution to construct a $(1-\alpha+\alpha_1)$-level confidence interval. If the observed Wald test statistic falls in the rejection region, the normal distribution is used; otherwise, the confidence interval is based on the weighted sum of chi-squared distributions.

\subsection{Asymptotic distribution of the Wald test in observational studies}
\label{subsec::OBS-noX-wald}

To derive the null distribution of the Wald statistic $\mathcal{W}^{\hat{\beta}_{\textsc{os}}}$, we first establish the asymptotic distribution of $\hat{\beta}_{\textsc{os}}$. Recall that $\varsigma=Z-\beta_0-\beta^{\T}Y$ is the population residual from the WLS regression of $Z$ on $(1, Y)$ with weights $W=Z/e(X)+(1-Z)/\{1-e(X)\}$.  Define the residual
$\varsigma_i=z_i-\beta_{0}-\beta^{\T}y_i$, $i=1,\ldots,n$.

\begin{lemma}\label{AsyDis-Beta-OBS}
Under standard regularity conditions, assume $0<e(X)<1$ and that the propensity score model $e(X; \alpha)$ is correctly specified. Then
$$n^{1/2}(\hat\beta_{\textup{\textsc{os}}}-\beta)\rightarrow N\big(0,  (\Phi_{\textsc{yy}}^{\textsc{w}})^{-1} E\{\psi_{\textsc{y}{\scriptscriptstyle\varsigma}}(Z,X,Y)\psi_{\textsc{y}{\scriptscriptstyle\varsigma}}(Z,X,Y)^{\T}\}
(\Phi_{\textsc{yy}}^{\textsc{w}})^{-1}\big)$$
in distribution as $n\rightarrow\infty$, where $\psi_{\textsc{y}{\scriptscriptstyle\varsigma}}(Z,X,Y)$ is defined in Theorem 2.
\end{lemma}

\begin{theorem}\label{asyDis-wald-OBS}
Suppose the conditions in Lemma \ref{AsyDis-Beta-OBS} hold. Under the null hypothesis $H_0: \tau=0$, we have
$\mathcal{W}^{\hat{\beta}_{\textup{\textsc{os}}}} = \hat{\beta}_{\textup{\textsc{os}}}^{\T}\hat{V}_{\smash{\beta_{\textup{\textsc{os}}}}}^{-1}\hat{\beta}_{\textup{\textsc{os}}}\rightarrow \chi^2(L)$ in distribution as $n\rightarrow\infty$, where
$$\hat{V}_{\smash{\beta_{\textup{\textsc{os}}}}}=n^{-1}(\phi_{yy}^{\hat{w}})^{-1}\bigg\{n^{-1}\sum_{i=1}^{n} \hat{\psi}_{\textsc{y}{\scriptscriptstyle\varsigma}}(z_i,x_i,y_i)\hat{\psi}_{\textsc{y}{\scriptscriptstyle\varsigma}}(z_i,x_i,y_i)^{\T}\bigg\}(\phi_{yy}^{\hat{w}})^{-1},
$$
and $\hat{\psi}_{\textsc{y}{\scriptscriptstyle\varsigma}}(z_i,x_i,y_i)=\hat\varsigma_iw(z_i,x_i;\hat\alpha)\tilde{y}_{\textup{\textsc{os}}i}
+n^{-1}\sum_{i=1}^{n}\{\hat{\varsigma}_i\tilde{y}_{\textup{\textsc{os}}i}\nabla_{\alpha}w(z_i,x_i;\hat\alpha)^{\T}\}\hat{I}(\hat\alpha)^{-1}S(z_i,x_i;\hat\alpha)$ with $\tilde{y}_{\textup{\textsc{os}}i}=y_i-\sum_{i=1}^{n}\hat{w}_iy_i/(\sum_{i=1}^{n}\hat{w}_i)$, $\hat{I}(\hat\alpha)=-n^{-1}\sum_{i=1}^{n}\nabla_{\alpha}S(z_i, x_i; \hat\alpha)$,  $\hat{\varsigma}_i =z_i-\hat{\beta}_{\textup{\textsc{os}}0}-\hat{\beta}_{\textup{\textsc{os}}}^{\T}y_i$ being the residual from the WLS regression of $z_i$ on $(1, y_i)$,  and
\begin{equation*}
\begin{aligned}
&\nabla_{\alpha}w(z_i,x_i;\hat\alpha)=\big[-z_ie(x_i; \hat\alpha)^{-2}+(1-z_i)\{1-e(x_i; \hat\alpha)\}^{-2}\big]\nabla_{\alpha}e(x_i; \hat\alpha).
\end{aligned}
\end{equation*}
\end{theorem}
\begin{proof}
The result follows directly from Lemma \ref{AsyDis-Beta-OBS} and the consistency of $\hat{V}_{\smash{\beta_{\textup{\textsc{os}}}}}$.
\end{proof}

\subsection{Confidence interval for the treatment effect $\tau^{\textup{c}}_{\textup{\textsc{os}}}$ in observational studies}
\label{subsec::OBS-noX-CI}

We construct confidence intervals for $\tau^{\textup{c}}$ using the asymptotic distributions established in Theorem 2. We estimate the asymptotic variance and eigenvalues in the asymptotic distributions by their sample analogues. When $\tau\neq0$, a Wald-type $(1-\alpha)$-level confidence interval for $\tau^{\textup{c}}$ is
$$\big[\hat\tau_{\textsc{os}}^{\textup{c}}- n^{-1/2}\Phi^{-1}(1-\alpha/2)\hat{V}_{\textsc{os}}^{-1/2},\ \hat\tau_{\textsc{os}}^{\textup{c}}+ n^{-1/2}\Phi^{-1}(1-\alpha/2)\hat{V}_{\textsc{os}}^{-1/2}\big],$$
where $\hat{V}_{\textsc{os}}=4\hat{\beta}_{\textsc{os}}^\T(n^{-1}\sum_{i=1}^{n}r_{\textsc{os}i}{r_{\textsc{os}i}}^{\T})\hat{\beta}_{\textsc{os}}$ and
\begin{equation*}
\begin{aligned}
r_{\textsc{os}i}&=\big(\hat{w}_i\tilde{y}_{\textsc{os}i}\tilde{y}_{\textsc{os}i}^{\T}-\phi_{yy}^{\hat{w}}\big)\hat\beta_{\textsc{os}}
+\bigg\{n^{-1}\sum_{i=1}^{n}\tilde{y}_{\textsc{os}i}\tilde{y}_{\textsc{os}i}^{\T}\hat\beta_{\textsc{os}}\nabla_{\alpha}w(z_i,x_i;\hat\alpha)^{\T}\bigg\}
\hat{I}(\hat\alpha)^{-1}S(z_i,x_i;\hat\alpha)\\
&\quad-\bigg[(\hat{w}_i-2)+\bigg\{n^{-1}\sum_{i=1}^{n} \nabla_{\alpha}w(z_i,x_i;\hat\alpha)^{\T}\bigg\}\hat{I}(\hat\alpha)^{-1}S(z_i,x_i;\hat\alpha)\bigg]\phi_{yy}^{\hat{w}}\hat\beta_{\textsc{os}}/2
+2\hat{\psi}_{\textsc{y}{\scriptscriptstyle\varsigma}}(z_i,x_i,y_i),\\
\end{aligned}
\end{equation*}
with $\tilde{y}_{\textup{\textsc{os}}i}$,   $\nabla_{\alpha}w(z_i,x_i;\hat\alpha)$, $\hat{I}(\hat\alpha)$, and $\hat{\psi}_{\textsc{y}{\scriptscriptstyle\varsigma}}(z_i,x_i,y_i)$ defined in Theorem \ref{asyDis-wald-OBS}.
When $\tau=0$, we construct a $(1-\alpha)$-level confidence interval as
$$\big[\hat\tau_{\textsc{os}}^{\textup{c}}-n^{-1}\hat{\Upsilon}_{\textsc{os}}^{-1}(1-\alpha/2),\  \hat\tau_{\textsc{os}}^{\textup{c}}-n^{-1}\hat{\Upsilon}_{\textsc{os}}^{-1}(\alpha/2)\big],$$
where $\hat{\Upsilon}_{\textsc{os}}$ denotes the distribution function of $\sum_{\ell=1}^{L}\hat\lambda_{\textsc{os}\ell}\chi^2_{\ell}(1)$  and $\hat\lambda_{\textsc{os}\ell}\ (\ell=1,\ldots, L)$ are the eigenvalues of $\hat\Gamma_{\textsc{os}}=2(\phi_{yy}^{\hat{w}})^{-1/2}\{n^{-1}\sum_{i=1}^{n} \hat{\psi}_{\textsc{y}{\scriptscriptstyle\varsigma}}(z_i,x_i,y_i)\hat{\psi}_{\textsc{y}{\scriptscriptstyle\varsigma}}(z_i,x_i,y_i)^{\T}\}
(\phi_{yy}^{\hat{w}})^{-1/2}$.
By Lemma S2, the eigenvalue estimators $\hat\lambda_{\textsc{os}\ell}$ are consistent for $\lambda_{\textsc{os}\ell}$, ensuring the asymptotic validity of the confidence interval. When $\tau$ is unknown, we can obtain a valid confidence interval for $\tau^{\textup{c}}$ by applying the procedure described in Section \ref{subsec::CRE-noX-CI} for completely randomized experiments.

\subsection{Asymptotic comparison of $\mathcal{W}^{\hat{\tau}^{\textup{c}}}$, $\mathcal{W}^{\hat{\beta}}$, and $\mathcal{W}^{\hat{\tau}}$ }

Recall the formulas for $\mathcal{W}^{\hat{\tau}^{\textup{c}}}$, $\mathcal{W}^{\hat{\beta}}$, and $\mathcal{W}^{\hat{\tau}}$ from Equation (6) in the main paper. From the proof of Proposition 1, $\phi_{yy}^{w}=\hat{\textup{cov}}\{Y(1)\}+\hat{\textup{cov}}\{Y(0)\}+\phi_{zz}^{w}\hat{\tau}\hat{\tau}^{\T}$. Thus, we can write  $\mathcal{W}^{\hat{\tau}^{\textup{c}}}$ as
\begin{equation*}\label{simu-eq-compostest2}
\begin{aligned}
\mathcal{W}^{\hat{\tau}^{\textup{c}}}=n\phi_{zz}^{w}\hat{\tau}^{\T}\big[\hat{\textup{cov}}\{Y(1)\}+\hat{\textup{cov}}\{Y(0)\}
+\phi_{zz}^{w}\hat{\tau}\hat{\tau}^{\T}\big]^{-1}
\hat{\tau}.
\end{aligned}
\end{equation*}
Under the null hypothesis $H_0: \tau= 0$, $\mathcal{W}^{\hat{\tau}^{\textup{c}}}\stackrel{d}\rightarrow \sum_{\ell=1}^{L}\lambda_{\ell}\chi^2_{\ell}(1)$, where $\lambda_{\ell}\ (\ell=1,\ldots, L)$ are the eigenvalues of
$\Gamma=2(\Phi_{\textsc{yy}}^{\textsc{w}})^{-1/2}E(W^2\varepsilon^2\tilde{Y}\tilde{Y}^{\T})(\Phi_{\textsc{yy}}^{\textsc{w}})^{-1/2}$ and $\tilde{Y}=Y-E(W)^{-1}E(WY)$. By contrast, $\mathcal{W}^{\hat{\beta}}\stackrel{d}\rightarrow \chi^2_{\ell}(L)$ and $\mathcal{W}^{\hat{\tau}}\stackrel{d}\rightarrow \chi^2_{\ell}(L)$.
By the LLN,
\begin{equation*}
\begin{aligned}
&\phi_{zz}^{w}\stackrel{p}\rightarrow \Phi_{\textsc{zz}}^{\textsc{w}}=E(WZ^2)-E(WZ)E(W)^{-1}E(WZ)=1/2,\\
&\phi_{yy}^{w}\stackrel{p}\rightarrow
\Phi_{\textsc{yy}}^{\textsc{w}}=E(WYY^{\T})-E(WY)E(W)^{-1}E(WY^{\T})=
\textup{cov}\{Y(1)\}+\textup{cov}\{Y(0)\}+\tau\tau^{\T}/2.
\end{aligned}
\end{equation*}
Similarly,
$$\bar{z}^{-1}\hat{\textup{cov}}\{Y(1)\}+(1-\bar{z})^{-1}\hat{\textup{cov}}\{Y(0)\}=\textup{cov}\{Y(1)\}/p+\textup{cov}\{Y(0)\}/(1-p)+o_p(1).$$
The consistency of the Huber--White robust covariance estimator implies that
\begin{equation*}
\begin{aligned}
n^{-1}\sum_{i=1}^{n}w_i^2\hat\varepsilon_i^2(y_i-\bar{y}_w)(y_i-\bar{y}_w)^\T
= E\big(W^2\varepsilon^2\tilde{Y}\tilde{Y}^{\T}\big)+o_p(1),\\
\end{aligned}
\end{equation*}
where $\varepsilon=\{Z-E(W)^{-1}E(WZ)\}-\beta^{\T}\{Y-E(W)^{-1}E(WY)\}$. Expanding this expectation gives
\begin{equation}\label{invcoef-middlematrix}
\begin{aligned}
E\big(W^2\varepsilon^2\tilde{Y}\tilde{Y}^{\T}\big)
&=\textup{cov}\{Y(1)\}/(4p)+\textup{cov}\{Y(0)\}/\{4(1-p)\}\\
&\quad-2E\big[W^2\{Z-E(W)^{-1}E(WZ)\}(\beta^{\T}\tilde{Y})\tilde{Y}\tilde{Y}^{\T}\big]
+E\big\{W^2(\beta^{\T}\tilde{Y})^2\tilde{Y}\tilde{Y}^{\T}\big\}.
\end{aligned}
\end{equation}

We next use  these probability limits to compare $\mathcal{W}^{\hat{\tau}^{\textup{c}}}$, $\mathcal{W}^{\hat{\beta}}$, and $\mathcal{W}^{\hat{\tau}}$ pairwise.

\vspace{0.2cm}
(i) \textit{Comparison between $\mathcal{W}^{\hat{\beta}}$} and $\mathcal{W}^{\hat{\tau}}$. Under $H_0$, $\beta=0$, and therefore
\begin{equation}\label{invcoef-middlematrix-firstorder}
E(W^2\varepsilon^2\tilde{Y}\tilde{Y}^{\T})=\textup{cov}\{Y(1)\}/(4p)+\textup{cov}\{Y(0)\}/\{4(1-p)\}+o_p(1).
\end{equation}
Since $\phi_{zz}^{w}\stackrel{p}\rightarrow1/2$, it follows that
\begin{equation*}
\begin{aligned}
(\phi_{zz}^{w})^{-2}\bigg\{n^{-1}\sum_{i=1}^{n}w_i^2\hat\varepsilon_i^2(y_i-\bar{y}_w)(y_i-\bar{y}_w)^\T\bigg\}
=\textup{cov}\{Y(1)\}/p+\textup{cov}\{Y(0)\}/(1-p)+o_p(1).
\end{aligned}
\end{equation*}
Thus,  $\mathcal{W}^{\hat{\beta}}$ and $\mathcal{W}^{\hat{\tau}}$ are asymptotically equivalent under $H_0$, that is,
$\mathcal{W}^{\hat{\tau}}=\mathcal{W}^{\hat{\beta}}+o_p(1)$. Under the local alternatives $H_{1n}: \tau=n^{-1/2}h$, we have $\beta=O_p(n^{-1/2})$. The last two terms in \eqref{invcoef-middlematrix-firstorder} are then asymptotically negligible, so Equation \eqref{invcoef-middlematrix-firstorder} continues to hold. Consequently, $\mathcal{W}^{\hat{\beta}}$ and $\mathcal{W}^{\hat{\tau}}$ remain asymptotically equivalent under $H_{1n}$. Under a fixed alternative with $\tau\neq0$, the last two terms in \eqref{invcoef-middlematrix} are $O_p(1)$. Hence,
$$(\phi_{zz}^{w})^{-2}\bigg\{n^{-1}\sum_{i=1}^{n}w_i^2\hat\varepsilon_i^2(y_i-\bar{y}_w)(y_i-\bar{y}_w)^\T\bigg\}=\textup{cov}\{Y(1)\}/p+\textup{cov}\{Y(0)\}/(1-p)+O_p(1).$$
Therefore, $\mathcal{W}^{\hat{\beta}}$ and $\mathcal{W}^{\hat{\tau}}$ are generally not asymptotically equivalent under fixed alternatives.

\vspace{0.2cm}
(ii) \textit{Comparison between $\mathcal{W}^{\hat{\tau}^{\textup{c}}}$ and $\mathcal{W}^{\hat{\tau}}$}. Applying the LLN and Slutsky's theorem yields
$$(\phi_{zz}^{w})^{-1}\bigg\{n^{-1}\sum_{i=1}^{n}w_i(y_i-\bar{y}_w)(y_i-\bar{y}_w)^{\T}\bigg\}\stackrel{p}\rightarrow
(\Phi_{\textsc{zz}}^{\textsc{w}})^{-1}\Phi_{\textsc{yy}}^{\textsc{w}}=
2\textup{cov}\{Y(1)\}+2\textup{cov}\{Y(0)\}+\tau\tau^{\T}.$$
Under $H_0$, $(\Phi_{\textsc{zz}}^{\textsc{w}})^{-1}\Phi_{\textsc{yy}}^{\textsc{w}}=2\textup{cov}\{Y(1)\}+2\textup{cov}\{Y(0)\}$, whereas the weighting matrix of $\mathcal{W}^{\hat{\tau}}$ converges to $\textup{cov}\{Y(1)\}/p+\textup{cov}\{Y(0)\}/(1-p)$. The two asymptotic limits are equal only when $p=1/2$ or $\textup{cov}\{Y(1)\}/p=\textup{cov}\{Y(0)\}/(1-p)$. The same comparison applies under $H_{1n}$, because $\tau\tau^{\T}=O(n^{-1})$ and therefore vanishes asymptotically. Thus, under $H_0$ and $H_{1n}$, $\mathcal{W}^{\hat{\tau}^{\textup{c}}}$ and $\mathcal{W}^{\hat{\tau}}$ are generally not asymptotically equivalent, except when $p=1/2$ or $\textup{cov}\{Y(1)\}/p=\textup{cov}\{Y(0)\}/(1-p)$.
Under fixed alternatives $\tau\neq0$, the additional term $\tau\tau^{\T}$ remains in the asymptotic limit of the weighting matrix of $\mathcal{W}^{\hat{\tau}^{\textup{c}}}$. Consequently, the two statistics are generally not asymptotically equivalent under fixed alternatives, even when the two conditions above hold.

\vspace{0.2cm}
(ii) \textit{Comparison between $\mathcal{W}^{\hat{\tau}^{\textup{c}}}$ and $\mathcal{W}^{\hat{\beta}}$}. Because $\mathcal{W}^{\hat{\beta}}$ and $\mathcal{W}^{\hat{\tau}}$ are asymptotically equivalent under $H_0$ and $H_{1n}$, the comparison between $\mathcal{W}^{\hat{\tau}^{\textup{c}}}$ and $\mathcal{W}^{\hat{\tau}}$ also applies to  $\mathcal{W}^{\hat{\tau}^{\textup{c}}}$ and $\mathcal{W}^{\hat{\beta}}$.

\vspace{0.2cm}
In summary, $\mathcal{W}^{\hat{\tau}}$ and  $\mathcal{W}^{\hat{\beta}}$ are asymptotically equivalent under $H_0$ and the local alternatives $H_{1n}$, but generally not under fixed alternatives. The test $\mathcal{W}^{\hat{\tau}^{\textup{c}}}$ is generally not asymptotically equivalent to either  $\mathcal{W}^{\hat{\beta}}$ or $\mathcal{W}^{\hat{\tau}}$. Under $H_0$ and $H_{1n}$, asymptotic equivalence between $\mathcal{W}^{\hat{\tau}^{\textup{c}}}$ and either $\mathcal{W}^{\hat{\beta}}$ or $\mathcal{W}^{\hat{\tau}}$ holds only when  $p=1/2$ or $\textup{cov}\{Y(1)\}/p=\textup{cov}\{Y(0)\}/(1-p)$; these conditions do not ensure equivalence under fixed alternatives.

\section{Extensions to covariate adjustment}
\label{sec::CRE-OBS-withX}

\subsection{Completely randomized experiments with covariates in inverse regression}
\label{subsec::CRE-withX}

Next, we extend the inverse regression method in Section 2  to completely randomized experiments with covariates $X=(X_1,\ldots, X_K)$. Let $\mathring{X}=(1,X^{\T})^{\T}$. By the FWL theorem for WLS \citep[Theorem 19.2]{ding2025linear}, $\beta_{\textup{\textsc{a}}}$ equals the WLS coefficient of $Z$ on the residualized outcomes $\check{Y}$ with weights $W=Z/p+(1-Z)/(1-p)$, where $\beta_{\textsc{a}}=E(W\check{Y}\check{Y}^{\T})^{-1}E(W\check{Y}Z)$ and $\check{Y}=Y-E(WY\mathring{X}^{\T})E(W \mathring{X}\mathring{X}^{\T})^{-1}\mathring{X}=Y-E(WY\mid X)/E(W\mid X)$. Here,  $\check{Y}$ is the residual from the element-wise WLS of $Y$ on $\mathring{X}$ with weights $W$. Similarly, let $\check{Z}$ denote the residual from the WLS of $Z$ on $\mathring{X}$ with weights $W$. Thus,
$\check{Z}=Z-E(WZ\mathring{X}^{\T})E(W \mathring{X}\mathring{X}^{\T})^{-1}\mathring{X}$.

Recall that the treatment effect estimator for the composite outcome is $\hat\tau_{\textsc{a}}^{\textup{c}}=\hat\beta_{\textsc{a}}^{\T}\hat\tau_{\textsc{a}}$, where $\hat\beta_{\textsc{a}}$ is the coefficient of $y_i$ from the WLS fit of $z_i$ on $(1, x_i, y_i)$ and $\hat\tau_{\textsc{a}}$ is the coefficient of $z_i$ from the component-wise WLS fit of $y_i$ on $(1, x_i, z_i)$ with weights $w_i=z_i/p+(1-z_i)/(1-p)$. Alternatively, we can compute $\hat{\tau}_{\textsc{a}}^{\textup{c}}$ using a two-step WLS procedure. First, we regress $z_i$ on $(1, x_i, y_i)$ to obtain the coefficient  $\hat{\beta}_{\textsc{a}}$ and compute $y_{\textsc{a}i}^{\textup{c}}=\hat{\beta}_{\textsc{a}}^{\T}y_i$. Second, we regress $y_{\textsc{a}i}^{\textup{c}}$ on $(1, z_i, x_i)$ to obtain the coefficient of $z_i$, which equals $\hat{\tau}_{\textsc{a}}^{\textup{c}}$.  Let $\mathring{x}=(1_n,x)$ and $\Lambda=\diag\{w_1,\ldots, w_n\}$. Applying the FWL theorem, we obtain $\hat{\beta}_{\textup{\textsc{a}}}=(\check{y}^{\T}\Lambda\check{y})^{-1}
(\check{y}^{\T}\Lambda z)$, where $\check{y}=y-\mathring{x}(\mathring{x}^{\T}\Lambda\mathring{x})^{-1}(\mathring{x}^{\T}\Lambda y)$ denotes the residual matrix from the column-wise WLS fit $y$ on $\mathring{x}$. Similarly, we obtain $\hat{\tau}_{\textup{\textsc{a}}}=(\check{z}^{\T}\Lambda\check{z})^{-1}
(y^{\T}\Lambda\check{z})$, where $\check{z}=z-\mathring{x}(\mathring{x}^{\T}\Lambda\mathring{x})^{-1}(\mathring{x}^{\T}\Lambda z)$.

The following proposition establishes an equivalence relationship between the inverse regression coefficient estimator $\hat{\beta}_{\textsc{a}}$  and the treatment effect estimator $\hat{\tau}_{\textsc{a}}$, as well as that between their corresponding estimands, $\beta_{\textsc{a}}$ and $\tau$.

\begin{proposition}\label{Equiv-CR-withX}
(i) (Sample version) We have
$$\hat{\beta}_{\textup{\textsc{a}}}=(\check{z}^{\T}\Lambda\check{z})(\check{y}^{\T}\Lambda\check{y})^{-1}\hat{\tau}_{\textup{\textsc{a}}}
=\frac{\hat{\Sigma}_{\textsc{a}}^{-1}\hat{\tau}_{\textsc{a}}}{1+\hat{\tau}_{\textsc{a}}^{\T}\Sigma_{\textsc{a}}^{-1}\hat{\tau}_{\textsc{a}}},$$
where
\begin{equation*}
\begin{aligned}
\hat{\Sigma}_{\textup{\textsc{a}}}&=(\check{z}^{\T}\Lambda \check{z})^{-1}\big([\hat{\textup{cov}}\{Y(1)\}+\hat{\textup{cov}}\{Y(0)\}]
-[\hat{\textup{cov}}\{Y(1), X\}+\hat{\textup{cov}}\{Y(0), X\}]
\{\hat{\textup{cov}}_{1}(X)+\hat{\textup{cov}}_{0}(X)\}^{-1}\\
&\hspace{7.2cm}\times
[\hat{\textup{cov}}\{Y(1), X\}+\hat{\textup{cov}}\{Y(0), X\}]^{\T}\big)
\end{aligned}
\end{equation*}
with
$\hat{\textup{cov}}\{Y(1)\}$, $\hat{\textup{cov}}\{Y(0)\}$, $\bar{y}(1)$, and $\bar{y}(0)$ defined in Proposition 1, $\hat{\textup{cov}}\{Y(1),X\}=(np)^{-1}$ $\sum_{i=1}^{n}z_i\{y_i-\bar{y}(1)\}(x_i-\bar{x}_{1})^{\T}$,
$\hat{\textup{cov}}\{Y(0), X\}=\{n(1-p)\}^{-1}\sum_{i=1}^{n}(1-z_i)\{y_i-\bar{y}(0)\}(x_i-\bar{x}_{0})^{\T}$, $\hat{\textup{cov}}_{1}(X)=(np)^{-1}\sum_{i=1}^{n}z_i(x_i-\bar{x}_{1})(x_i-\bar{x}_{1})^{\T}$,
$\hat{\textup{cov}}_{0}(X)=\{n(1-p)\}^{-1}\sum_{i=1}^{n}(1-z_i)(x_i-\bar{x}_{0})(x_i$ $-\bar{x}_{0})^{\T}$, $\bar{x}_1=\sum_{i=1}^{n}z_ix_i/$ $(\sum_{i=1}^{n}z_i)$, and $\bar{x}_0=\sum_{i=1}^{n}(1-z_i)x_i/\{\sum_{i=1}^{n}(1-z_i)\}$.

(ii) (Population version) Under complete randomization with $Z\ind\{Y(1),Y(0), X\}$, we have
$$\beta_{\textup{\textsc{a}}}=E(W\check{Z}^2)E(W\check{Y}\check{Y}^{\T})^{-1}\tau=\frac{\Sigma_{\textsc{a}}^{-1}\tau}{1+\tau^{\T}\Sigma_{\textsc{a}}^{-1}\tau},
$$
where $E(W\check{Z}^2)=1/2$ and
$\Sigma_{\textup{\textsc{a}}}=2[\textup{cov}\{Y(1)\}+\textup{cov}\{Y(0)\}]-\textup{cov}\{Y(1)+Y(0), X\}\textup{cov}(X)^{-1}$ $
\textup{cov}\{X, Y(1)+Y(0)\}$.
\end{proposition}

Proposition \ref{Equiv-CR-withX} shows that testing the null hypothesis $H_0: \tau= 0$ is equivalent to testing $\beta_{\textsc{a}}=0$. This equivalence motivates a Wald test for $H_0$ based on $\hat{\beta}_{\textsc{a}}$. Specifically, we define
$\mathcal{W}^{\hat{\beta}_{\textsc{a}}} = \hat\beta_{\textsc{a}}^{\T}\hat{V}_{\smash{\beta_{\textsc{a}}}}^{-1}\hat\beta_{\textsc{a}}$, where $\hat{V}_{\smash{\beta_{\textsc{a}}}}$ is the Huber--White robust estimator of the asymptotic variance of $\hat\beta_{\textsc{a}}$ from the inverse weighted regression that yields $\hat\beta_{\textsc{a}}$.
Under $H_0$,  $\mathcal{W}^{\hat{\beta}_{\textsc{a}}}$ converges in distribution to a chi-squared distribution with $L$ degrees of freedom. Moreover, inference based on $\mathcal{W}^{\hat{\beta}_{\textsc{a}}}$ remains valid even if the linear model for $Z$ given  $(X,Y)$ is misspecified.

We next study the asymptotic properties of $\hat{\tau}_{\textsc{a}}^{\textup{c}}$.  By the FWL theorem, the coefficient of $Z$ from the element-wise WLS regression of $Y$ on $(1, X, Z)$ with weights $W$ equals $E(W\check{Z}^2)^{-1}E(W\check{Z}Y)$. Under $Z\ind\{Y(1),Y(0), X\}$, $E(W\check{Z}^2)^{-1}E(W\check{Z}Y)=\tau$. Hence, $\hat{\tau}_{\textsc{a}}$ consistently estimate $\tau$. As a WLS estimator, $\hat{\beta}_{\textsc{a}}$ consistently estimate $\beta_{\textsc{a}}$.
 Therefore, $\hat{\tau}_{\textsc{a}}^{\textup{c}}=\hat{\beta}_{\textsc{a}}^{\T}\hat{\tau}_{\textsc{a}}$ consistently estimates $\tau_{\textsc{a}}^{\textup{c}}$. The following theorem establishes the asymptotic distribution of $\hat{\tau}_{\textsc{a}}^{\textup{c}}$.

\begin{theorem}\label{AsyDis-BetaTau-CR-withX}
Assume complete randomization with $Z\ind\{Y(1),Y(0), X\}$.

(i) If $\tau\neq0$, we have
$n^{1/2}(\hat{\tau}_{\textup{\textsc{a}}}^{\textup{c}}-\tau_{\textup{\textsc{a}}}^{\textup{c}})\rightarrow N(0, 4\beta_{\textup{\textsc{a}}}^\T\textup{cov}\{R_{\textup{\textsc{a}}}(Z,X,Y)\}\beta_{\textup{\textsc{a}}})$  in distribution as $n\rightarrow\infty$,
where
\begin{equation*}
\begin{aligned}
R_{\textup{\textsc{a}}}(Z,X,Y)=\big\{W\check{Y}\check{Y}^{\T}-E(W\check{Y}\check{Y}^{\T})\big\}\beta_{\textup{\textsc{a}}}
-(W-2)E(W\check{Y}\check{Y}^{\T})\beta_{\textup{\textsc{a}}}/2
+2W\varepsilon_{\textup{\textsc{a}}}\check{Y},
\end{aligned}
\end{equation*}
with $\varepsilon_{\textsc{a}}$ being the population residual from the WLS regression of $Z$ on $(1, X, Y)$.

(ii) If $\tau=0$, we have
$n\hat{\tau}_{\textup{\textsc{a}}}^{\textup{c}}\rightarrow\sum_{\ell=1}^{L}\lambda_{\textup{\textsc{a}}\ell}\chi^2_{\ell}(1)$  in distribution as $n\rightarrow\infty$,
where $\lambda_{\textup{\textsc{a}}\ell}$ $(\ell=1,\ldots, L)$ are the eigenvalues of
$$\Gamma_{\textup{\textsc{a}}}=2E(W\check{Y}\check{Y}^{\T})^{-1/2} E(W^2\varepsilon_{\textup{\textsc{a}}}^2\check{Y}\check{Y}^{\T})
E(W\check{Y}\check{Y}^{\T})^{-1/2}.$$
\end{theorem}

As in Theorem 1, the asymptotic distribution of the average treatment effect estimator $\hat{\tau}_{\textsc{a}}^{\textup{c}}=\hat\beta_{\textsc{a}}^{\T}\hat{\tau}_{\textsc{a}}$ depends on the value of $\tau$. Specifically, when $\tau\neq0$, $\hat{\tau}_{\textsc{a}}^{\textup{c}}$ is asymptotically normal, and when  $\tau=0$, it converges to a weighted sum of chi-squared distributions. Furthermore, if $X$ is empty, Theorem \ref{AsyDis-BetaTau-CR-withX} reduces to Theorem 1. We can construct confidence intervals for $\tau_{\textsc{a}}^{\textup{c}}$ by replacing the asymptotic variance and eigenvalues with their sample analogues. Because $\tau$ is generally unknown in practice, we recommend using the procedure described in Section \ref{subsec::CRE-noX-CI} to construct a confidence interval for $\tau_{\textsc{a}}^{\textup{c}}$.

\subsection{Observational studies with covariates in inverse regression}
\label{subsec::OBS-withX}

We extend the method in Section 3 to include covariates $X$ in inverse weighted regression. By the FWL theorem, $\beta_{\textsc{a}}$ equals the  WLS coefficient from the regression of $Z$ on  the residualized outcomes $\check{Y}$ with weights $W=Z/e(X)+(1-Z)/\{1-e(X)\}$. Specifically,  $\beta_{\textsc{a}}=E(W\check{Y}\check{Y}^{\T})^{-1}E(W\check{Y}Z)$, where $\check{Y}=Y-E(WY\mathring{X}^{\T})E(W \mathring{X}\mathring{X}^{\T})^{-1}\mathring{X}$.

Recall that the treatment effect estimator for the composite outcome is $\hat\tau_{\textsc{os,a}}^{\textup{c}}=\hat{\beta}_{\textsc{os,a}}^{\T}\hat{\tau}_{\textsc{os,a}}$, where $\hat\beta_{\textsc{os, a}}$ is the coefficient of $y_i$ from the WLS regression of $z_i$ on $(1, x_i, y_i)$ and $\hat\tau_{\textsc{os,a}}$ is the coefficient of $z_i$ from the component-wise WLS regression of $y_i$ on $(1, x_i, z_i)$. Both regressions use weights $\hat{w}_i=z_i/\hat{e}(x_i)+(1-z_i)/\{1-\hat{e}(x_i)\}$.
Analogous to Section \ref{subsec::CRE-withX}, we can compute $\hat\tau_{\textsc{os,a}}^{\textup{c}}$ using a two-step WLS procedure. First, we regress $z_i$ on $(1, x_i, y_i)$ with weights $\hat{w}_i$ to obtain $\hat{\beta}_{\textsc{os,a}}$, and  construct the composite outcome $y_{\textsc{os,a}i}^{\textup{c}}=\hat{\beta}_{\textsc{os,a}}^{\T}y_i$. Second, we regress $y_{\textsc{os,a}i}^{\textup{c}}$ on $(1, z_i, x_i)$ using the same weights. The coefficient of $z_i$ from this regression equals $\hat\tau_{\textsc{os,a}}^{\textup{c}}$.  Let $\hat{\Lambda}_{\textsc{os}}=\diag\{\hat{w}_1,\ldots, \hat{w}_n\}$. By the FWL theorem, $\hat{\beta}_{\textup{\textsc{os,a}}}=(\check{y}_{\textsc{os}}^{\T}\hat{\Lambda}_{\textsc{os}}\check{y}_{\textsc{os}})^{-1}
(\check{y}_{\textsc{os}}^{\T}\hat{\Lambda}_{\textsc{os}}z)$, where $\check{y}_{\textsc{os}}=y-\mathring{x}(\mathring{x}^{\T}\hat{\Lambda}_{\textsc{os}}\mathring{x})^{-1}(\mathring{x}^{\T}\hat{\Lambda}_{\textsc{os}}y)$ is the residual matrix from the column-wise WLS regression of $y$ on $\mathring{x}$ with weights $\hat{w}_i$. Similarly, we obtain $\hat{\tau}_{\textup{\textsc{os,a}}}=(\check{z}_{\textsc{os}}^{\T}\hat{\Lambda}_{\textsc{os}}\check{z}_{\textsc{os}})^{-1}
(y^{\T}\hat{\Lambda}_{\textsc{os}}\check{z}_{\textsc{os}})$, where $\check{z}_{\textsc{os}}=z-\mathring{x}(\mathring{x}^{\T}\hat{\Lambda}_{\textsc{os}}\mathring{x})^{-1}(\mathring{x}^{\T}\hat{\Lambda}_{\textsc{os}}z)$ is the residual vector from the  WLS regression of $z$ on $\mathring{x}$ with weights $\hat{w}_i$.

The following proposition establishes the equivalence between $\hat{\beta}_{\textsc{os,a}}$ and $\hat{\tau}_{\textsc{os,a}}$, as well as that between $\beta_{\textsc{a}}$ and $\tau$ at the population level.

\begin{proposition}\label{Equiv-OBS-withX}
(i) (Sample version) We have
$$\hat{\beta}_{\textup{\textsc{os,a}}}=(\check{z}_{\textup{\textsc{os}}}^{\T}\hat{\Lambda}_{\textup{\textsc{os}}}\check{z}_{\textup{\textsc{os}}})
(\check{y}_{\textup{\textsc{os}}}^{\T}\hat{\Lambda}_{\textup{\textsc{os}}}\check{y}_{\textup{\textsc{os}}})^{-1}
\hat{\tau}_{\textup{\textsc{os,a}}}
=\frac{\hat{\Sigma}_{\textup{\textsc{os,a}}}^{-1}\hat{\tau}_{\textup{\textsc{os,a}}}}{1+\hat{\tau}_{\textup{\textsc{os,a}}}^{\T}
\Sigma_{\textup{\textsc{os,a}}}^{-1}\hat{\tau}_{\textup{\textsc{os,a}}}},$$
where
\begin{equation*}
\begin{aligned}
\hat{\Sigma}_{\textup{\textsc{os,a}}}&=(\check{z}_{\textup{\textsc{os}}}^{\T}\hat{\Lambda}_{\textup{\textsc{os}}}\check{z}_{\textup{\textsc{os}}})^{-1}
\big([\hat{\textup{cov}}_{\hat{w}}\{Y(1)\}+\hat{\textup{cov}}_{\hat{w}}\{Y(0)\}]
\\
&\hspace{3cm}-[\hat{\textup{cov}}_{\hat{w}}\{Y(1), X\}+\hat{\textup{cov}}_{\hat{w}}\{Y(0), X\}]
\{\hat{\textup{cov}}_{\hat{w}1}(X)+\hat{\textup{cov}}_{\hat{w}0}(X)\}^{-1}\\
&\hspace{3.4cm}\times
[\hat{\textup{cov}}_{\hat{w}}\{Y(1), X\}+\hat{\textup{cov}}_{\hat{w}}\{Y(0), X\}]^{\T}\big),
\end{aligned}
\end{equation*}
with $\hat{\textup{cov}}_{\hat{w}}\{Y(1)\}$ and $\hat{\textup{cov}}_{\hat{w}}\{Y(0)\}$ defined in Proposition 2,
$\hat{\textup{cov}}_{\hat{w}1}(X)=n^{-1}\sum_{i=1}^{n}\hat{w}_iz_i(x_i-\bar{x}_{\hat{w}1})(x_i-\bar{x}_{\hat{w}1})^{\T}$, $\hat{\textup{cov}}_{\hat{w}0}(X)=n^{-1}\sum_{i=1}^{n}\hat{w}_i(1-z_i)(x_i-\bar{x}_{\hat{w}0})(x_i-\bar{x}_{\hat{w}0})^{\T}$, $\hat{\textup{cov}}_{\hat{w}}\{Y(1), X\}=n^{-1}$ $\sum_{i=1}^{n}\hat{w}_iz_i\{y_i-\bar{y}_{\hat{w}}(1)\}(x_i-\bar{x}_{\hat{w}1})^{\T}$, $\hat{\textup{cov}}_{\hat{w}}\{Y(0), X\}=n^{-1}\sum_{i=1}^{n}\hat{w}_i(1-z_i)\{y_i-\bar{y}_{\hat{w}}(0)\}(x_i-\bar{x}_{\hat{w}0})^{\T}$,
$\bar{x}_{\hat{w}1}=\sum_{i=1}^{n}\hat{w}_iz_ix_i/(\sum_{i=1}^{n}\hat{w}_iz_i)$, and $\bar{x}_{\hat{w}0}=\sum_{i=1}^{n}\hat{w}_i(1-z_i)x_i/\{\sum_{i=1}^{n}\hat{w}_i(1-z_i)\}$.

(ii) (Population version) Under strong ignorability $Z\ind$ $\{Y(1), Y(0)\} \mid X$ and overlap $0<e(X)<1$, we have
\begin{equation*}
\begin{aligned}
\beta_{\textsc{a}}&=E(W\check{Z}^2)
E(W\check{Y}\check{Y}^{\T})^{-1}\tau=\frac{\Sigma_{\textsc{a}}^{-1}\tau}{1+\tau^{\T}\Sigma_{\textsc{a}}^{-1}\tau},
\end{aligned}
\end{equation*}
where $E(W\check{Z}^2)=1/2$ and
$\Sigma_{\textsc{a}}=2[\textup{cov}\{Y(1)\}+\textup{cov}\{Y(0)\}]-\textup{cov}\{Y(1)+Y(0), X\}\textup{cov}(X)^{-1}$ $
\textup{cov}\{X, Y(1)+Y(0)\}$.
\end{proposition}

\begin{proof}
The proof follows the same arguments as those used in  Proposition \ref{Equiv-CR-withX} and is therefore omitted.
\end{proof}

Proposition \ref{Equiv-OBS-withX} shows that testing the null hypothesis $H_0: \tau= 0$ is equivalent to testing $\beta_{\textsc{a}}=0$. This equivalence motivates a Wald test based on $\hat{\beta}_{\textsc{os,a}}$ : $\mathcal{W}^{\hat{\beta}_{\textsc{os,a}}} = \hat\beta_{\textsc{os,a}}^{\T}\hat{V}_{\smash{\beta_{\textsc{os,a}}}}^{-1}\hat\beta_{\textsc{os,a}}$, where $\hat{V}_{\smash{\beta_{\textsc{os,a}}}}$ is a consistent estimator of the asymptotic variance of $\hat\beta_{\textsc{os,a}}$. We provide an explicit expression for $\mathcal{W}^{\hat{\beta}_{\textsc{os,a}}}$ below.
To establish the  asymptotic distribution of $\hat\beta_{\textsc{os,a}}$,  we assume a parametric propensity score model $e(X; \alpha)$,
as in Lemma \ref{AsyDis-Beta-OBS}. Let $\hat{e}(x_i)=e(x_i; \hat{\alpha})$ be the estimated propensity score, where $\hat\alpha$ solves the estimating equation $\sum_{i=1}^{n}S(z_i,x_i; \alpha)=0$. We also define the information matrix as $I(\alpha)=E\{-\nabla_{\alpha}S(Z, X; \alpha)\}$. The following lemma gives the asymptotic distribution of $\hat\beta_{\textsc{os,a}}$.

\begin{lemma}\label{AsyDis-Beta-OBS-withX}
Under standard regularity conditions, assume $0<e(X)<1$ and that the propensity score model $e(X; \alpha)$ is correctly specified. Then we have
$$n^{1/2}(\hat\beta_{\textsc{os,a}}-\beta_{\textsc{a}})\rightarrow N\big(0,  E(W\check{Y}\check{Y}^{\T})^{-1} E\{\psi_{\textsc{y}{\scriptscriptstyle\varsigma_{\textup{\textsc{a}}}}}(Z,X,Y)\psi_{\textsc{y}{\scriptscriptstyle\varsigma_{\textup{\textsc{a}}}}}(Z,X,Y)^{\T}\}
E(W\check{Y}\check{Y}^{\T})^{-1}\big)$$
in distribution as $n\rightarrow\infty$, where
$\psi_{\textsc{y}{\scriptscriptstyle\varsigma_{\textup{\textsc{a}}}}}(Z,X,Y)=\varsigma_{\textup{\textsc{a}}} W\check{Y}
+E\{\varsigma_{\textup{\textsc{a}}}\check{Y}\nabla_{\alpha}w(Z,X;\alpha_0)^{\T}\}I(\alpha_0)^{-1}S(Z,$ $X;\alpha_0)$
with $\nabla_{\alpha}w(Z,X;\alpha_0)$ defined in Theorem 2 and $\varsigma_{\textup{\textsc{a}}}$ being the population residual from the WLS regression of $Z$ on $(1, X, Y)$.
\end{lemma}

\begin{proof}
The proof parallels that of Lemma \ref{AsyDis-Beta-OBS} and we therefore omit the details.
\end{proof}

\begin{theorem}\label{asyDis-wald-OBS-withX}
Suppose the conditions in Lemma \ref{AsyDis-Beta-OBS-withX} hold. Under the null hypothesis $H_0: \tau=0$,  we have
$\mathcal{W}^{\hat{\beta}_{\textsc{os,a}}} = \hat{\beta}_{\textup{\textsc{os,a}}}^{\T}\hat{V}_{\smash{\beta_{\textup{\textsc{os,a}}}}}^{-1}\hat{\beta}_{\textup{\textsc{os,a}}}\rightarrow \chi^2(L)$ in distribution as $n\rightarrow\infty$, where
\begin{equation*}
\begin{aligned} \hat{V}_{\smash{\beta_{\textup{\textsc{os,a}}}}}&=n^{-1}(n^{-1}\check{y}_{\textup{\textsc{os}}}^{\T}\hat{\Lambda}_{\textup{\textsc{os}}}\check{y}_{\textup{\textsc{os}}})^{-1}\bigg\{n^{-1}\sum_{i=1}^{n} \hat{\psi}_{\textsc{y}{\scriptscriptstyle\varsigma_{\textup{\textsc{a}}}}}(z_i,x_i,y_i)\hat{\psi}_{\textsc{y}{\scriptscriptstyle\varsigma_{\textup{\textsc{a}}}}}(z_i,x_i,y_i)^{\T}\bigg\}
(n^{-1}\check{y}_{\textup{\textsc{os}}}^{\T}\hat{\Lambda}_{\textup{\textsc{os}}}\check{y}_{\textup{\textsc{os}}})^{-1},
\end{aligned}
\end{equation*}
and
$\hat{\psi}_{\textsc{y}{\scriptscriptstyle\varsigma_{\textup{\textsc{a}}}}}(z_i,x_i,y_i)=\hat\varsigma_{\textup{\textup{\textsc{a}}}i}w(z_i,x_i;\alpha_0)\check{y}_{\textup{\textsc{os}}i}
+n^{-1}\sum_{i=1}^{n}\{\hat{\varsigma}_{\textup{\textsc{a}}i}\check{y}_{\textup{\textsc{os}}i}\nabla_{\alpha}w(z_i,x_i;\hat\alpha)^{\T}\}\hat{I}(\hat\alpha)^{-1}S(z_i,x_i;\hat\alpha)
$
with $\check{y}_{\textup{\textsc{os}}i}=y_i-(y^{\T}\hat{\Lambda}_{\textup{\textsc{os}}}\mathring{x})(\mathring{x}^{\T}\hat{\Lambda}_{\textup{\textsc{os}}}\mathring{x})^{-1}\mathring{x}_i$, $\hat{I}(\hat\alpha)=n^{-1}\sum_{i=1}^{n}\{-\nabla_{\alpha}S(z_i, x_i; \hat\alpha)\}$, $\hat\varsigma_{\textup{\textup{\textsc{a}}}i}$ being the residual from the WLS regression of $z_i$ on $(1, x_i, y_i)$, and $\nabla_{\alpha}w(z_i,x_i;\hat\alpha)$ defined in Theorem \ref{asyDis-wald-OBS}.
 \end{theorem}

\begin{proof}
The result follows directly from Lemma \ref{AsyDis-Beta-OBS-withX} and the consistency of $\hat{V}_{\smash{\beta_{\textup{\textsc{os,a}}}}}$.
\end{proof}

Proposition \ref{Equiv-OBS-withX} implies that $\tau=E(W\check{Z}^2)^{-1}E(WZ\check{Y})=E(W\check{Z}^2)^{-1}E(W\check{Z}Y)$, which equals the coefficient of $Z$ from the element-wise WLS regression of $Y$ on $(1, X, Z)$ with weight $W=Z/e(X)+(1-Z)/\{1-e(X)\}$. If the propensity score model is correctly specified with $e(X)=e(X; \alpha_0)$ for some $\alpha_0$, the estimator $\hat{\tau}_{\textsc{os,a}}$ consistently estimate $\tau$. Similarly, the WLS estimator $\hat{\beta}_{\textsc{os,a}}$ consistently estimate $\beta_{\textsc{a}}$ under correct specification of the propensity score model.  Consequently, $\hat{\tau}_{\textsc{os,a}}^{\textup{c}}$ is consistent for $\tau_{\textsc{a}}^{\textup{c}}$. We next derive the asymptotic distribution of $\hat{\tau}_{\textsc{os,a}}^{\textup{c}}$ under correct specification of the propensity score model.


\begin{theorem}\label{OBS-compATE-asy-withX}
Assume that strong ignorability $Z\ind\{Y(1), Y(0)\} \mid X$ and overlap $0<e(X)<1$ hold, and  the propensity score model $e(X; \alpha)$ is correctly specified.

(i) If $\tau\neq0$, then
$n^{1/2}(\hat{\tau}^{\textup{c}}_{\textup{\textsc{os,a}}}-\tau^{\textup{c}}_{\textup{\textsc{a}}})\rightarrow N(0, 4\beta_{\textsc{a}}^\T\textup{cov}\{R_{\textup{\textsc{os,a}}}(Z,X,Y)\}\beta_{\textsc{a}})$ in distribution as $n\rightarrow\infty$,
where
\begin{equation*}
\begin{aligned}
&R_{\textup{\textsc{os,a}}}(Z,X,Y)\\
&=\Big(\big\{W\check{Y}\check{Y}^{\T}-E(W\check{Y}\check{Y}^{\T})\big\}\beta_{\textsc{a}}
+E\big\{\check{Y}\check{Y}^{\T}\beta_{\textsc{a}}\nabla_{\alpha}w(Z,X;\alpha_0)^{\T}\big\} I(\alpha_0)^{-1}S(Z,X;\alpha_0)\Big)\\
&\quad-\big[(W-2)+E\{\nabla_{\alpha}w(Z,X;\alpha_0)^{\T}\}I(\alpha_0)^{-1}S(Z,X;\alpha_0)\big]E(W\check{Y}\check{Y}^{\T})\beta_{\textsc{a}}/2
+2\psi_{\textsc{y}{\scriptscriptstyle\varsigma_{\textsc{a}}}}(Z,X,Y),
\end{aligned}
\end{equation*}
with
$\nabla_{\alpha}w(Z,X;\alpha_0)$ and $\psi_{\textsc{y}{\scriptscriptstyle\varsigma_{\textsc{a}}}}$ defined in Lemma \ref{AsyDis-Beta-OBS-withX}.

(ii) If $\tau=0$, then
$n\hat{\tau}_{\textup{\textsc{os,a}}}^{\textup{c}}\rightarrow\sum_{\ell=1}^{L}\lambda_{\textup{\textsc{os,a}}\ell}\chi^2_{\ell}(1)$ in distribution as $n\rightarrow\infty$, where $\lambda_{\textup{\textsc{os,a}}\ell}$ $(\ell=1,\ldots, L)$ are eigenvalues of  $$\Gamma_{\textup{\textsc{os,a}}}=2E(W\check{Y}\check{Y}^{\T})^{-1/2} E\{\psi_{\textsc{y}{\scriptscriptstyle\varsigma_{\textsc{a}}}}(Z,X,Y)\psi_{\textsc{y}{\scriptscriptstyle\varsigma_{\textsc{a}}}}(Z,X,Y)^{\T}\}
E(W\check{Y}\check{Y}^{\T})^{-1/2}.$$
\end{theorem}

\begin{proof}
The proof parallels that of Theorem 2 and we therefore omit the details.
\end{proof}

As in Theorem 2, $\hat{\tau}_{\textup{\textsc{os,a}}}^{\textup{c}}$ is asymptotically normal when $\tau\neq0$ and converges to a weighted sum of chi-squared distributions when $\tau=0$. Beyond the uncertainty in $\hat{\beta}_{\textsc{os,a}}$ and $\hat{\tau}_{\textsc{os,a}}$, Theorem \ref{OBS-compATE-asy-withX}  accounts for the uncertainty from estimating the propensity score.  When $\mathring{X}=1$,  it reduces to Theorem 3.  We can construct confidence intervals for $\tau^{\textup{c}}_{\textsc{a}}$ by replacing the asymptotic variance and eigenvalues with their sample analogues, with technical details omitted.

\section{Variance-weighted treatment effect in stratified randomized experiments}\label{sec::varweight-SRE}

\subsection{Variance-weighted estimator} \label{subsec-SREnoX-varweightATE}


Recall that the variance-weighted stratum-specific treatment effect $\tau_{\textsc{sr}}$ equals the coefficient of $Z$ from the element-wise WLS regression of $Y$ on $(G, Z)$ with weights
$W_{\textsc{sr}}=Z\{1-e(C)\}+(1-Z)e(C)$,
where $C\in\{1,\ldots, S\}$ and $e(C)=\sum_{s=1}^Sp_{[s]}I(C=s)$. Let $(\beta_{\textsc{srg}}, \beta_{\textsc{sr}}) = \argmin_{b_{\textsc{g}},b} E\{W_{\textsc{sr}}(Z-b_{\textsc{g}}^{\T}G-b^\T Y)^2\}$ denote the population coefficients from the WLS regression of $Z$ on $(G, Y)$ with weights $W_{\textsc{sr}}$. By the FWL theorem, we can express $\beta_{\textsc{sr}}$ as the WLS coefficient of $Z$ on  the residualized outcomes $\tilde{Y}_{\textsc{sr}}$ with the same weight $W_{\textsc{sr}}$, where $\beta_{\textsc{sr}}=E(W_{\textsc{sr}}\tilde{Y}_{\textsc{sr}}\tilde{Y}_{\textsc{sr}}^{\T})^{-1}E(W_{\textsc{sr}}\tilde{Y}_{\textsc{sr}}Z)$ and $\tilde{Y}_{\textsc{sr}}=Y-E(W_{\textsc{sr}} YG^{\T})E(W_{\textsc{sr}} GG^{\T})^{-1}G=Y-E(W_{\textsc{sr}} Y\mid C)/E(W_{\textsc{sr}}\mid C)$. Here, $\tilde{Y}_{\textsc{sr}}$ is the residual from the element-wise WLS regression of $Y$ on $G$ with weight $W_{\textsc{sr}}$, which centers $Y$ by its stratum-specific weighted means.

Recall that $\hat\tau^{\textup{c}}_{\smash{\textsc{sr}}}=\hat{\beta}_{\textsc{sr}}^{\T}\hat{\tau}_{\textsc{sr}},$ where $\hat{\beta}_{\textsc{sr}}$ is the coefficient of $y_i$ from the WLS regression of $z_i$ on $(g_i,y_i)$ with weights $w_{\textsc{sr}i}=z_i\{1-e(c_i)\}+(1-z_i)e(c_i)$, and $\hat{\tau}_{\textsc{sr}}$ is the coefficient of $z_i$ from the component-wise weighted regression of $y_i$ on $(g_i,z_i)$ using the same weights. Applying the FWL theorem, we obtain $\hat{\beta}_{\textsc{sr}}=(\tilde{y}_{\textsc{sr}}^{\T}\Lambda_{\textsc{sr}}\tilde{y}_{\textsc{sr}})^{-1}
\tilde{y}_{\textsc{sr}}^{\T}\Lambda_{\textsc{sr}}z$ and $\hat{\tau}_{\textsc{sr}}=(\tilde{z}_{\textsc{sr}}^{\T}\Lambda_{\textsc{sr}}\tilde{z}_{\textsc{sr}})^{-1}
y^{\T}\Lambda_{\textsc{sr}}\tilde{z}_{\textsc{sr}}$, where
$\tilde{y}_{\textsc{sr}}=y-g(g^{\T}\Lambda_{\textsc{sr}}g)^{-1}(g^{\T}\Lambda_{\textsc{sr}}y)$ and $\tilde{z}_{\textsc{sr}}=z-g(g^{\T}\Lambda_{\textsc{sr}}g)^{-1}(g^{\T}\Lambda_{\textsc{sr}}z)$ are the WLS residuals from regressing $y$  and $z$ on $g$ with weights $w_{\textsc{sr}i}$, respectively.

\subsection{Equivalence between the inverse regression coefficients and average treatment effects}\label{subsec-SREnoX-equiv}

Analogous to Propositions 1 and 2, we establish the equivalence relationship between $\hat{\beta}_{\textsc{sr}}$ and $\hat{\tau}_{\textsc{sr}}$, as well as the corresponding population-level relationship between $\beta_{\textsc{sr}}$ and $\tau_{\textsc{sr}}$ under stratified randomization.  The sample-level relationship parallels the population-level one.

\begin{proposition}\label{SREnoX-equiv}
(i) (Sample version) We have
$$\hat{\beta}_{\textup{\textsc{sr}}}=(\tilde{z}_{\textup{\textsc{sr}}}^{\T}\Lambda_{\textup{\textsc{sr}}}\tilde{z}_{\textup{\textsc{sr}}})
(\tilde{y}_{\textup{\textsc{sr}}}^{\T}\Lambda_{\textup{\textsc{sr}}}\tilde{y}_{\textup{\textsc{sr}}})^{-1}\hat{\tau}_{\textup{\textsc{sr}}}
=\frac{\hat\Sigma_{\textup{\textsc{sr}}}^{-1}\hat\tau_{\textup{\textsc{sr}}}}{1+\hat\tau_{\textup{\textsc{sr}}}^\T \hat\Sigma_{\textup{\textsc{sr}}}^{-1}\hat\tau_{\textup{\textsc{sr}}}},$$
where
\begin{equation*}
\begin{aligned}
\hat{\Sigma}_{\textup{\textsc{sr}}}&=(\tilde{z}_{\textup{\textsc{sr}}}^{\T}\Lambda_{\textup{\textsc{sr}}}\tilde{z}_{\textup{\textsc{sr}}})^{-1}\sum_{s=1}^S\Big[(1-p_{[s]})n_{1s}\hat{\textup{cov}}\{Y(1)\mid C=s\}
+p_{[s]}n_{0s}\hat{\textup{cov}}\{Y(0)\mid C=s\}\\
&\hspace{3.2cm}+\frac{n_{1s}n_{0s}p_{[s]}(1-p_{[s]})}{n_{1s}(1-p_{[s]})+n_{0s}p_{[s]}}(\hat{\tau}_{[s]}-\hat\tau_{\textup{\textsc{sr}}})(\hat{\tau}_{[s]}-\hat\tau_{\textup{\textsc{sr}}})^{\T}\Big],
\end{aligned}
\end{equation*}
with $\hat{\textup{cov}}\{Y(1)\mid C=s\}=n_{1s}^{-1}\sum_{i=1}^nz_iI(c_i=s)\{y_i-\bar{y}_{[s]}(1)\}\{y_i-\bar{y}_{[s]}(1)\}^{\T}$, $\hat{\textup{cov}}\{Y(0)\mid C=s\}=n_{0s}^{-1}\sum_{i=1}^n(1-z_i)I(c_i=s)\{y_i-\bar{y}_{[s]}(0)\}\{y_i-\bar{y}_{[s]}(0)\}^{\T}$, $\bar{y}_{[s]}(1)=n_{1s}^{-1}\sum_{i=1}^ny_iz_iI(c_i=s)$, $\bar{y}_{[s]}(0)=n_{0s}^{-1}\sum_{i=1}^ny_i(1-z_i)I(c_i=s)$, $\hat{\tau}_{[s]}$ $=\bar{y}_{[s]}(1)-\bar{y}_{[s]}(0)$, $n_{1s}=\sum_{i=1}^nz_iI(c_i=s)$, and $n_{0s}=\sum_{i=1}^n(1-z_i)I(c_i=s)$.

(ii) (Population version) Under stratified randomization with $Z\ind\{Y(1),Y(0)\}\mid C$, we have
$$\beta_{\textup{\textsc{sr}}}=E(W_{\textup{\textsc{sr}}}\tilde{Z}_{\textup{\textsc{sr}}}^2)E(W_{\textup{\textsc{sr}}}\tilde{Y}_{\textup{\textsc{sr}}}
\tilde{Y}_{\textup{\textsc{sr}}}^{\T})^{-1}\tau_{\textup{\textsc{sr}}}
=\frac{\Sigma_{\textup{\textsc{sr}}}^{-1}\tau_{\textup{\textsc{sr}}}}{1+\tau_{\textup{\textsc{sr}}}^\T \Sigma_{\textup{\textsc{sr}}}^{-1}\tau_{\textup{\textsc{sr}}}},$$
where $\tilde{Z}_{\textsc{sr}}=Z-E(W_{\textsc{sr}}ZG^{\T})E(W_{\textsc{sr}} GG^{\T})^{-1}G=Z-E(W_{\textsc{sr}} Z|C)/E(W_{\textsc{sr}}|C)$,
$E(W_ {\textup{\textsc{sr}}}\tilde{Z}_{\textup{\textsc{sr}}}^2)=\sum_{s=1}^{S}$ $\pi_sp_{[s]}(1-p_{[s]})/2$, and
\begin{equation*}
\begin{aligned}
\Sigma_{\textup{\textsc{sr}}}&=E(W_{\textup{\textsc{sr}}}\tilde{Z}_{\textup{\textsc{sr}}}^2)^{-1}\sum_{s=1}^{S}\pi_sp_{[s]}(1-p_{[s]})\big[\textup{cov}\{Y(1)\mid C=s\}+\textup{cov}\{Y(0)\mid C=s\}\\
&\hspace{5.8cm}+(\tau_{[s]}-\tau_{\textup{\textsc{sr}}})(\tau_{[s]}-\tau_{\textup{\textsc{sr}}})^{\T}/2\big].
\end{aligned}
\end{equation*}
\end{proposition}

Proposition \ref{SREnoX-equiv} implies that $\hat\tau^{\textup{c}}_{\smash{\textsc{sr}}}=\hat\tau_{\textup{\textsc{sr}}}^\T \hat\Sigma_{\textup{\textsc{sr}}}^{-1}\hat\tau_{\textup{\textsc{sr}}}/(1+\hat\tau_{\textup{\textsc{sr}}}^\T \hat\Sigma_{\textup{\textsc{sr}}}^{-1}\hat\tau_{\textup{\textsc{sr}}})\in[0,1)$ and $\tau^{\textup{c}}_{\smash{\textsc{sr}}}=\tau_{\textup{\textsc{sr}}}^\T \Sigma_{\textup{\textsc{sr}}}^{-1}\tau_{\textup{\textsc{sr}}}/(1+\tau_{\textup{\textsc{sr}}}^\T \Sigma_{\textup{\textsc{sr}}}^{-1}\tau_{\textup{\textsc{sr}}})\in[0,1)$, since
$\hat\Sigma_{\textup{\textsc{sr}}}$ and $\Sigma_{\textup{\textsc{sr}}}$ are positive semi-definite. Unlike the overall treatment effects $\tau^{\textup{c}}$ and $\tau^{\textup{c}}_{\smash{\textsc{os}}}$ introduced for completely randomized experiments and observational studies, $\tau^{\textup{c}}_{\smash{\textsc{sr}}}$ depends on the stratum-specific treatment probabilities and therefore is not design-independent. The dependence arises because the definition of the associated  estimand $\tau_{\textsc{sr}}$ involves the treatment variances $p_{[s]}(1-p_{[s]})$. Proposition \ref{SREnoX-equiv} also implies that testing $\tau_{\textsc{sr}}=0$ is equivalent to testing $\beta_{\textsc{sr}}=0$. For the null hypothesis $H_{0\textsc{sr}}:\tau_{\textsc{sr}}=0$, we propose the Wald test: $\mathcal{W}^{\hat{\beta}_{\textsc{sr}}} = \hat{\beta}_{\textsc{sr}}^{\T}\hat{V}_{\smash{\beta_{\textsc{sr}}}}^{-1}\hat{\beta}_{\textsc{sr}}$, where $\hat{V}_{\smash{\beta_{\textsc{sr}}}}$ is the Huber--White robust covariance estimator of the asymptotic variance of $\hat{\beta}_{\textsc{sr}}$ from the inverse weighted regression that yields $\hat{\beta}_{\textsc{sr}}$. Under $H_{0\textsc{sr}}$, $\mathcal{W}^{\hat{\beta}_{\textsc{sr}}}$ asymptotically follows a chi-squared distribution with $L$ degrees of freedom, regardless of whether the inverse regression model of $Z$ on $(G, Y)$ is correctly specified.

\subsection{Asymptotic theory and statistical inference for the composite outcome}\label{subsec-SREnoX-asy}

Because the WLS coefficient estimators $\hat{\beta}_{\textsc{sr}}$ and $\hat{\tau}_{\textsc{sr}}$ are consistent for $\beta_{\textsc{sr}}$ and $\tau_{\textsc{sr}}$, respectively,  $\hat\tau^{\textup{c}}_{\textsc{sr}}$ is consistent for $\tau^{\textup{c}}_{\textsc{sr}}$. Theorem \ref{SREnoX-compATE-asy} below presents the asymptotic distribution of $\hat\tau^{\textup{c}}_{\textsc{sr}}$.

\begin{theorem}\label{SREnoX-compATE-asy}
Assume stratified randomization with $Z\ind\{Y(1),Y(0)\}\mid C$.

(i) If $\tau_{\textup{\textsc{sr}}}\neq0$, then
$n^{1/2}(\hat{\tau}^{\textup{c}}_{\textup{\textsc{sr}}}-\tau^{\textup{c}}_{\textup{\textsc{sr}}})\rightarrow N(0, E(W_{\textup{\textsc{sr}}}\tilde{Z}_{\textup{\textsc{sr}}}^2)^{-2}\beta_{\textup{\textsc{sr}}}^{\T}E[\textup{cov}\{R_{\textup{\textsc{sr}}}(Z,C,Y)\mid C\}]\beta_{\textup{\textsc{sr}}})$ in distribution as $n\rightarrow\infty$, where
\begin{equation*}
\begin{aligned}
R_{\textup{\textsc{sr}}}(Z,C,Y)&=\big\{W_{\textup{\textsc{sr}}}\tilde{Y}_{\textup{\textsc{sr}}}\tilde{Y}_{\textup{\textsc{sr}}}^{\T}
-E(W_{\textup{\textsc{sr}}}\tilde{Y}_{\textup{\textsc{sr}}}\tilde{Y}_{\textup{\textsc{sr}}}^{\T}\mid C)\big\}\beta_{\textup{\textsc{sr}}}\\
&\quad-\big\{W_{\textup{\textsc{sr}}}\tilde{Z}_{\textup{\textsc{sr}}}^2-E(W_{\textup{\textsc{sr}}}\tilde{Z}_{\textup{\textsc{sr}}}^2\mid C)\big\}E(W_{\textup{\textsc{sr}}}\tilde{Y}_{\textup{\textsc{sr}}}\tilde{Y}_{\textup{\textsc{sr}}}^{\T})\beta_{\textup{\textsc{sr}}}/E(W_{\textup{\textsc{sr}}}\tilde{Z}_{\textup{\textsc{sr}}}^2)
+2W_{\textup{\textsc{sr}}}\zeta\tilde{Y}_{\textup{\textsc{sr}}}.
\end{aligned}
\end{equation*}
and $\zeta = Z-\beta_{\textup{\textsc{srg}}}^{\T}G-\beta_{\textup{\textsc{sr}}}^{\T}Y$ is the population residual from the WLS regression of $Z$ on $(G, Y)$.

(ii) If $\tau_{\textup{\textsc{sr}}}=0$, then
$n\hat{\tau}^{\textup{c}}_{\textup{\textsc{sr}}}\rightarrow \sum_{\ell=1}^{L}\lambda_{\textup{\textsc{sr}}\ell}\chi^2_{\ell}(1)$ in distribution as $n\rightarrow\infty$,
where $\lambda_{\textup{\textsc{sr}}\ell}\ (\ell=1,\ldots, L)$ are eigenvalues of
$$\Gamma_{\textsc{sr}}=E(W_{\textup{\textsc{sr}}}\tilde{Z}_{\textup{\textsc{sr}}}^2)^{-1}E(W_{\textup{\textsc{sr}}}\tilde{Y}_{\textup{\textsc{sr}}}\tilde{Y}_{\textup{\textsc{sr}}}^{\T})^{-1/2}
\bigg\{\sum_{s=1}^{S}\pi_sE(W_{\textup{\textsc{sr}}}^2\zeta^2\tilde{Y}_{\textup{\textsc{sr}}}\tilde{Y}_{\textup{\textsc{sr}}}^{\T}\mid C=s)\bigg\}
E(W_{\textup{\textsc{sr}}}\tilde{Y}_{\textup{\textsc{sr}}}\tilde{Y}_{\textup{\textsc{sr}}}^{\T})^{-1/2}.$$
\end{theorem}

The asymptotic distribution of $\hat\tau^{\textup{c}}_{\textsc{sr}}$ parallels those of $\hat\tau^{\textup{c}}$ in completely randomized experiments and $\hat\tau^{\textup{c}}_{\textsc{os}}$ in observational studies. It is asymptotically normal when $\tau_{\textsc{sr}} \neq 0$ and a weighted sum of chi-squared distributions when $\tau_{\textsc{sr}}=0$. Theorem \ref{SREnoX-compATE-asy} accounts for the estimation uncertainty in both $\hat{\beta}_{\textsc{sr}}$ and $\hat{\tau}_{\textsc{sr}}$.

\subsection{Confidence interval for the treatment effect $\tau^{\textup{c}}_{\textup{\textsc{sr}}}$}
\label{subsec::SRE-noX-CI}

We use the asymptotic distributions in Theorem \ref{SREnoX-compATE-asy} to construct confidence intervals for $\tau^{\textup{c}}_{\textsc{sr}}$, with the asymptotic variance and eigenvalues estimated by their sample analogues. Define $\hat{E}(W_{\textsc{sr}}\tilde{Z}_{\textsc{sr}}^2\mid C=s)=n_s^{-1}\sum_{i=1}^{n}w_{\textsc{sr}i}(z_i-\bar{z}_{w,[s]})^2 I(c_i=s)$ and
$\hat{E}(W_{\textsc{sr}}\tilde{Y}_{\textsc{sr}}\tilde{Y}_{\textsc{sr}}^{\T}\mid C=s)=n_s^{-1}\sum_{i=1}^{n}w_{\textsc{sr}i}(y_i-\bar{y}_{w,[s]})(y_i-\bar{y}_{w,[s]})^{\T}I(c_i=s)$, where $\bar{z}_{w,[s]}=\sum_{i=1}^{n}w_{\textsc{sr}i}z_iI(c_i=s)/\{\sum_{i=1}^{n}w_{\textsc{sr}i}I(c_i=s)\}$ and $\bar{y}_{w,[s]}=\sum_{i=1}^{n}w_{\textsc{sr}i}y_iI(c_i=s)/\{\sum_{i=1}^{n}w_{\textsc{sr}i}I(c_i=s)\}$.
Thus, $\tilde{z}_{\textsc{sr}}^{\T}\Lambda_{\textsc{sr}}\tilde{z}_{\textsc{sr}}=\sum_{s=1}^{S}n_s\hat{E}(W_{\textsc{sr}}\tilde{Z}_{\textsc{sr}}^2\mid C=s)$ and $\tilde{y}_{\textsc{sr}}^{\T}\Lambda_{\textsc{sr}}\tilde{y}_{\textsc{sr}}=\sum_{s=1}^{S}n_s\hat{E}(W_{\textsc{sr}}\tilde{Y}_{\textsc{sr}}\tilde{Y}_{\textsc{sr}}^{\T}\mid C=s)$.
Let $\hat\zeta_i=z_i-\hat\beta_{\textsc{srg}}^{\T}g_i-\hat\beta_{\textsc{sr}}^{\T}y_i$ denote the residual from the WLS fit of $z_i$ on $(g_i, y_i)$ with the weights $w_{\textsc{sr}i}$.
When $\tau_{\textsc{sr}}\neq0$, a Wald-type $(1-\alpha)$-level confidence interval is
$$\big[\hat\tau_{\textsc{sr}}^{\textup{c}}- n^{-1/2}\Phi^{-1}(1-\alpha/2)\hat{V}_{\textsc{sr}}^{-1/2},\ \hat\tau_{\textsc{sr}}^{\textup{c}}+ n^{-1/2}\Phi^{-1}(1-\alpha/2)\hat{V}_{\textsc{sr}}^{-1/2}\big],$$
where
$\hat{V}_{\textsc{sr}}=\{n^{-1}\tilde{z}_{\textsc{sr}}^{\T}\Lambda_{\textsc{sr}}\tilde{z}_{\textsc{sr}}\}^{-2}\hat{\beta}_{\textsc{sr}}^{\T}
\{n^{-1}\sum_{s=1}^{S}\sum_{i=1}^{n}r_{\smash{[s]i}}r_{\smash{[s]i}}^{\T}I(c_i=s)\}\hat{\beta}_{\textsc{sr}},$
and
\begin{equation*}
\begin{aligned}
r_{\smash{[s]i}}&=\big\{w_{\textsc{sr}i}(y_i-\bar{y}_{w,[s]})(y_i-\bar{y}_{w,[s]})^{\T}-\hat{E}(W_{\textsc{sr}}\tilde{Y}_{\textsc{sr}}\tilde{Y}_{\textsc{sr}}^{\T}\mid C=s)\big\}\hat{\beta}_{\textsc{sr}}\\
&\quad-\big\{w_{\textsc{sr}i}(z_i-\bar{z}_{w,[s]})^2-\hat{E}(W_{\textsc{sr}}\tilde{Z}_{\textsc{sr}}^2\mid C=s)\big\}\frac{\tilde{y}_{\textsc{sr}}^{\T}\Lambda_{\textsc{sr}}\tilde{y}_{\textsc{sr}}}{\tilde{z}_{\textsc{sr}}^{\T}\Lambda_{\textsc{sr}}\tilde{z}_{\textsc{sr}}}\hat{\beta}_{\textsc{sr}}\\
&\quad+2w_{\textsc{sr}i}\hat\zeta_i(y_i-\bar{y}_{w,[s]}).
\end{aligned}
\end{equation*}
On the other hand, when $\tau_{\textsc{sr}}=0$,  a $(1-\alpha)$-level confidence interval for $\tau^{\textup{c}}_{\textsc{sr}}$ is
$$\big[\hat\tau_{\textsc{sr}}^{\textup{c}}-n^{-1}\hat{\Upsilon}_{\textsc{sr}}^{-1}(1-\alpha/2),\  \hat\tau_{\textsc{sr}}^{\textup{c}}-n^{-1}\hat{\Upsilon}_{\textsc{sr}}^{-1}(\alpha/2)\big],$$
where $\hat{\Upsilon}_{\textsc{sr}}$ is the distribution function of $\sum_{\ell=1}^{L}\hat\lambda_{\textsc{sr}\ell}\chi^2_{\ell}(1)$,  and $\hat\lambda_{\textsc{sr}\ell}\ (\ell=1,\ldots, L)$ are the eigenvalues of \begin{equation*}
\begin{aligned}
\hat{\Gamma}_{\textsc{sr}}
&=\big\{n^{-1}\tilde{z}_{\textsc{sr}}^{\T}\Lambda_{\textsc{sr}}\tilde{z}_{\textsc{sr}}\big\}^{-1}\big\{n^{-1}\tilde{y}_{\textsc{sr}}^{\T}\Lambda_{\textsc{sr}}\tilde{y}_{\textsc{sr}}\big\}^{-1/2}\\
&\quad\times
\bigg\{n^{-1}\sum_{s=1}^{S} \sum_{i=1}^{n}w_{\textsc{sr}i}^2\hat\zeta_i^2(y_i-\bar{y}_{w,[s]})(y_i-\bar{y}_{w,[s]})^{\T}I(c_i=s)\bigg\}
\big\{n^{-1}\tilde{y}_{\textsc{sr}}^{\T}\Lambda_{\textsc{sr}}\tilde{y}_{\textsc{sr}}\big\}^{-1/2}.
\end{aligned}
\end{equation*}
Lemma S2 ensures that the estimated eigenvalues $\hat\lambda_{\textsc{sr}\ell}$ are consistent for $\lambda_{\textsc{sr}\ell}$, the eigenvalues of $\Gamma_{\textsc{sr}}$. Hence, the resulting confidence interval achieves the correct asymptotic coverage probability.
When $\tau_{\textsc{sr}}$ is unknown, we can construct a valid confidence interval for $\tau_{\textsc{sr}}^{\textup{c}}$ by using the methods in Section \ref{subsec::CRE-noX-CI} for completely randomized experiments.

\subsection{Stratified randomized experiments with covariates in inverse regression}\label{subsec::SRE-withX}

We extend the method in Section \ref{subsec-SREnoX-varweightATE}  to stratified randomized experiments that include covariates $X$ beyond the stratification variable $C$. Recall that $\tau_{\textsc{sr}}$ is the coefficient of $Z$ from the element-wise WLS regression of $Y$ on $(G, Z)$ with weights $W_{\textsc{sr}}=Z\{1-e(C)\}+(1-Z)e(C)$. Similarly, let $\tau_{\textsc{srx}}$ denote the coefficient of $Z$ from the element-wise WLS regression of $X$ on $(G, Z)$ with the same weights. By the FWL theorem, $\tau_{\textsc{srx}}=E(W_{\textsc{sr}}\tilde{Z}_{\textsc{sr}}\tilde{Z}_{\textsc{sr}}^{\T})^{-1}E(W_{\textsc{sr}}\tilde{Z}_{\textsc{sr}}X)$. Under $Z\ind\{Y(1),Y(0), X\}\mid C$, we have $E(W_{\textsc{sr}}\tilde{Z}_{\textsc{sr}}X)=0$, implying $\tau_{\textsc{srx}}=0$.

Let $\hat{\tau}_{\textsc{sr}}$ and $\hat{\tau}_{\textsc{srx}}$ denote the estimators of $\tau_{\textsc{sr}}$ and $\tau_{\textsc{srx}}$, obtained as the coefficients of $z_i$ from the component-wise WLS regression of $y_i$ and $x_i$ on $(g_i,z_i)$ with weights $w_{\textsc{sr}i}=z_i\{1-e(c_i)\}+(1-z_i)e(c_i)$. Recall that $\Lambda_{\textsc{sr}}=\diag\{w_{\textsc{sr}1},\ldots, w_{\textsc{sr}n}\}$.
By the FWL theorem, we obtain $\hat{\tau}_{\textsc{sr}}=(\tilde{z}_{\textsc{sr}}^{\T}\Lambda_{\textsc{sr}}\tilde{z}_{\textsc{sr}})^{-1}
y^{\T}\Lambda_{\textsc{sr}}\tilde{z}_{\textsc{sr}}$ and $\hat{\tau}_{\textsc{srx}}=(\tilde{z}_{\textsc{sr}}^{\T}\Lambda_{\textsc{sr}}\tilde{z}_{\textsc{sr}})^{-1}
x^{\T}\Lambda_{\textsc{sr}}\tilde{z}_{\textsc{sr}}$, where $\tilde{z}_{\textsc{sr}}=z-g(g^{\T}\Lambda_{\textsc{sr}}g)^{-1}(g^{\T}\Lambda_{\textsc{sr}}z)$.
Recall that $\hat{\beta}_{\textsc{srx,a}}$ and $\hat{\beta}_{\textsc{sr,a}}$ are the coefficients of $x_i$ and $y_i$ in the WLS fit of $z_i$ on $(g_i, x_i, y_i)$ with weights $w_{\textsc{sr}i}$. We define the treatment effect estimators for the composite outcome and covariate as $\hat{\tau}_{\textsc{sry,a}}^{\textup{c}}=\hat{\beta}_{\textsc{sr,a}}^{\T}\hat\tau_{\textsc{sr}}$ and $\hat{\tau}_{\textsc{srx,a}}^{\textup{c}}=\hat{\beta}_{\textsc{srx,a}}^{\T}\hat\tau_{\textsc{srx}}$. Building on $\hat{\tau}_{\textsc{sry,a}}^{\textup{c}}$ and $\hat{\tau}_{\textsc{srx,a}}^{\textup{c}}$, we construct a class of covariate-adjusted estimators for $\tau_{\textsc{sr,a}}^{\textup{c}}$: $$\hat{\tau}_{\textsc{sr,a}}^{\textup{c}}(r)=\hat{\tau}_{\textsc{sry,a}}^{\textup{c}}-r\hat{\tau}_{\textsc{srx,a}}^{\textup{c}},\quad r\in\mathcal{R}.$$
The optimal coefficient $r_{\textup{opt}}$ that minimizes the variance of $\hat{\tau}_{\textsc{sr,a}}^{\textup{c}}(r)$ is $$r_{\textup{opt}}=\frac{\textup{cov}(\hat{\tau}_{\textsc{sry,a}}^{\textup{c}},  \hat{\tau}_{\textsc{srx,a}}^{\textup{c}})}{\textup{var}(\hat{\tau}_{\textsc{srx,a}}^{\textup{c}})}
=\frac{\textup{var}(\hat{\tau}_{\textsc{sry,a}}^{\textup{c}}+\hat{\tau}_{\textsc{srx,a}}^{\textup{c}})
-\textup{var}(\hat{\tau}_{\textsc{sry,a}}^{\textup{c}}-\hat{\tau}_{\textsc{srx,a}}^{\textup{c}})}{4\textup{var}(\hat{\tau}_{\textsc{srx,a}}^{\textup{c}})}.$$
We estimate $r_{\textup{opt}}$ by its sample analogue:
$$\hat{r}_{\textup{opt}}=\frac{\hat{\textup{cov}}(\hat{\tau}_{\textsc{sry,a}}^{\textup{c}},  \hat{\tau}_{\textsc{srx,a}}^{\textup{c}})}{\hat{\textup{var}}(\hat{\tau}_{\textsc{srx,a}}^{\textup{c}})}=\frac{\hat{\textup{var}}
(\hat{\tau}_{\textsc{sry,a}}^{\textup{c}}+\hat{\tau}_{\textsc{srx,a}}^{\textup{c}})-\hat{\textup{var}}(\hat{\tau}_{\textsc{sry,a}}^{\textup{c}}
-\hat{\tau}_{\textsc{srx,a}}^{\textup{c}})}{4\hat{\textup{var}}(\hat{\tau}_{\textsc{srx,a}}^{\textup{c}})},$$
where the variance estimates $\hat{\textup{var}}
(\hat{\tau}_{\textsc{sry,a}}^{\textup{c}}+\hat{\tau}_{\textsc{srx,a}}^{\textup{c}})$, $\hat{\textup{var}}(\hat{\tau}_{\textsc{sry,a}}^{\textup{c}}
-\hat{\tau}_{\textsc{srx,a}}^{\textup{c}})$, and $\hat{\textup{var}}(\hat{\tau}_{\textsc{srx,a}}^{\textup{c}})$ can be obtained from Theorem \ref{AsyDis-BetaTau-SR-withX} below. Hence, the final estimator of $\tau_{\textsc{sr,a}}^{\textup{c}}$ is
$\hat{\tau}^{\textup{c}}_{\textsc{sr,a}}= \hat{\tau}^{\textup{c}}_{\textsc{sry,a}}-\hat{r}_{\textup{opt}}\hat{\tau}^{\textup{c}}_{\textsc{srx,a}}$.

For notational simplicity, we define $U=(X^{\T}, Y^{\T})^{\T}$, $u=(x, y)$, $\beta_{\textsc{sru,a}}= (\beta_{\textsc{srx,a}}^{\T}, \beta_{\textsc{sr,a}}^{\T})^{\T}$, $\tau_{\textsc{sru}}= (\tau_{\textsc{srx}}^{\T}, \tau_{\textsc{sr}}^{\T})^{\T}$,
$\hat\beta_{\textsc{sru,a}}= (\hat\beta_{\textsc{srx,a}}^{\T}, \hat\beta_{\textsc{sr,a}}^{\T})^{\T}$, and $\hat\tau_{\textsc{sru}}= (\hat\tau_{\textsc{srx}}^{\T}, \hat\tau_{\textsc{sr}}^{\T})^{\T}$. Let $\tilde{U}_{\textsc{sr}}=(\tilde{X}_{\textsc{sr}}^{\T}, \tilde{Y}_{\textsc{sr}}^{\T})^{\T}$ and $\tilde{u}_{\textsc{sr}}=(\tilde{x}_{\textsc{sr}}, \tilde{y}_{\textsc{sr}})$, where $\tilde{X}_{\textsc{sr}}=X-E(W_{\textsc{sr}} XG^{\T})E(W_{\textsc{sr}} GG^{\T})^{-1}G$ is the residual from the element-wise WLS regression of $X$ on $G$ with weight $W_{\textsc{sr}}$ and $\tilde{x}_{\textsc{sr}}=x-g(g^{\T}\Lambda_{\textsc{sr}}g)^{-1}(g^{\T}\Lambda_{\textsc{sr}}x)$ is the residual from the WLS regression of $x$  on $g$ with weights $w_{\textsc{sr}i}$. Hence, $\tilde{U}_{\textsc{sr}}=U-E(W_{\textsc{sr}} UG^{\T})E(W_{\textsc{sr}} GG^{\T})^{-1}G=U-E(W_{\textsc{sr}} U\mid C)/E(W_{\textsc{sr}}\mid C)$.
By the FWL theorem, $\beta_{\textup{\textsc{sru,a}}}=E(W_{\textsc{sr}}\tilde{U}_{\textsc{sr}}\tilde{U}_{\textsc{sr}}^{\T})^{-1}E(W_{\textsc{sr}}\tilde{U}_{\textsc{sr}}Z)$ and $\hat\beta_{\textup{\textsc{sru,a}}}=(\tilde{u}_{\textsc{sr}}^{\T}\Lambda_{\textsc{sr}}\tilde{u}_{\textsc{sr}})^{-1}
\tilde{u}_{\textsc{sr}}^{\T}\Lambda_{\textsc{sr}}z$.

The next proposition establishes the equivalence between the inverse WLS regression coefficient $\hat\beta_{\textsc{sru,a}}$ and the forward regression coefficient $\hat{\tau}_{\textup{\textsc{sru}}}$, as well as that between $\beta_{\textup{\textsc{sru,a}}}$ and  $\tau_{\textup{\textsc{sru}}}$.

\begin{proposition}\label{Equiv-SR-withX}
(i) (Sample version) We have
$$\hat\beta_{\textup{\textsc{sru,a}}}=(\tilde{z}_{\textup{\textsc{sr}}}^{\T}\Lambda_{\textup{\textsc{sr}}}\tilde{z}_{\textup{\textsc{sr}}})
(\tilde{u}_{\textup{\textsc{sr}}}^{\T}\Lambda_{\textup{\textsc{sr}}}\tilde{u}_{\textup{\textsc{sr}}})^{-1}\hat{\tau}_{\textup{\textsc{sru}}}.$$

(ii) (Population version) Under stratified randomization with $Z\ind\{Y(1),Y(0), X\}\mid C$, we have
$$
\beta_{\textup{\textsc{sru,a}}}=E(W_{\textup{\textsc{sr}}}\tilde{Z}_{\textup{\textsc{sr}}}^2)E(W_{\textup{\textsc{sr}}}\tilde{U}_{\textup{\textsc{sr}}}
\tilde{U}_{\textup{\textsc{sr}}}^{\T})^{-1}\tau_{\textup{\textsc{sru}}},$$
where $E(W_ {\textup{\textsc{sr}}}\tilde{Z}_{\textup{\textsc{sr}}}^2)=\sum_{s=1}^{S}\pi_sp_{[s]}(1-p_{[s]})/2$.
\end{proposition}

\begin{proof}
The proof follows the same arguments as those used in Proposition \ref{SREnoX-equiv} after replacing $Y$ with $U$ and is therefore omitted.
\end{proof}

Since $\tau_{\textsc{srx}}=0$, Proposition  \ref{Equiv-SR-withX} implies that testing $\tau_{\textsc{sr}}=0$ is equivalent to testing $\tau_{\textsc{sru}}=0$, which in turn is equivalent to testing $\beta_{\textsc{sru,a}}=0$. Based on $\hat{\beta}_{\textsc{sru,a}}$, we construct a Wald test for the null hypothesis $H_{0\textsc{sr}}:\tau_{\textsc{sr}}=0$. The test statistic is $\mathcal{W}^{\hat{\beta}_{\textsc{sru,a}}} = \hat{\beta}_{\textsc{sru,a}}^{\T}\hat{V}_{\smash{\beta_{\textsc{sru,a}}}}^{-1}\hat{\beta}_{\textsc{sru,a}}$, where $\hat{V}_{\smash{\beta_{\textsc{sru,a}}}}$ is the Huber--White robust covariance estimator for the asymptotic variance of $\hat{\beta}_{\textsc{sru,a}}$ from the inverse WLS regression that yields $\hat{\beta}_{\textsc{sru,a}}$.
Under $H_{0\textsc{sr}}$, $\mathcal{W}^{\hat{\beta}_{\textsc{sru,a}}}$ asymptotically follows a chi-squared distribution with $L$ degrees of freedom, regardless of whether the linear model of $Z$ on $(G, X, Y)$ is correctly specified.

Next, we study the asymptotic properties of the estimator $\hat{\tau}_{\textsc{sr,a}}^{\textup{c}}(r)=\hat{\tau}^{\textup{c}}_{\textsc{sry,a}}-r\hat{\tau}^{\textup{c}}_{\textsc{srx,a}}$ for a fixed constant $r$. The consistency of $\hat{\beta}_{\textsc{srx,a}}$, $\hat{\beta}_{\textsc{sr,a}}$,
$\hat{\tau}_{\textsc{srx}}$,
and $\hat{\tau}_{\textsc{sr}}$ ensures that  $\hat{\tau}_{\textsc{sr,a}}^{\textup{c}}(r)$ is consistent for $\tau_{\textsc{sr,a}}^{\textup{c}}$. The following theorem states its  asymptotic distribution.

\begin{theorem}\label{AsyDis-BetaTau-SR-withX}
Assume stratified randomization with $Z\ind\{Y(1),Y(0), X\}\mid C$.

(i) If $\tau_{\textup{\textsc{sr}}}\neq0$, we have
$n^{1/2}\{\hat{\tau}_{\textup{\textsc{sr,a}}}^{\textup{c}}(r)-\tau_{\textup{\textsc{sr,a}}}^{\textup{c}}\}\rightarrow  N(0, E(W_{\textup{\textsc{sr}}}\tilde{Z}_{\textup{\textsc{sr}}}^2)^{-2}\beta_{\textup{\textsc{sru,a}}}^{\T}E[\textup{cov}\{R_{\textup{\textsc{sr,a}}}(Z,C,$ $X,Y)\mid C\}]\beta_{\textup{\textsc{sru,a}}})$ in distribution as $n\rightarrow\infty$, where
\begin{equation*}
\begin{aligned}
R_{\textup{\textsc{sr,a}}}(Z,C,X,Y)&=D\big\{W_{\textup{\textsc{sr}}}\tilde{U}_{\textup{\textsc{sr}}}\tilde{U}_{\textup{\textsc{sr}}}^{\T}
-E(W_{\textup{\textsc{sr}}}\tilde{U}_{\textup{\textsc{sr}}}\tilde{U}_{\textup{\textsc{sr}}}^{\T}\mid C)\big\}\beta_{\textup{\textsc{sru,a}}}\\
&\quad-D\big\{W_{\textup{\textsc{sr}}}\tilde{Z}_{\textup{\textsc{sr}}}^2-E(W_{\textup{\textsc{sr}}}\tilde{Z}_{\textup{\textsc{sr}}}^2\mid C)\big\}E(W_{\textup{\textsc{sr}}}\tilde{U}_{\textup{\textsc{sr}}}\tilde{U}_{\textup{\textsc{sr}}}^{\T})\beta_{\textup{\textsc{sru,a}}}/E(W_{\textup{\textsc{sr}}}\tilde{Z}_{\textup{\textsc{sr}}}^2)\\
&\quad+E(W_{\textup{\textsc{sr}}}\tilde{U}_{\textup{\textsc{sr}}}\tilde{U}_{\textup{\textsc{sr}}}^{\T}\mid C)DE(W_{\textup{\textsc{sr}}}\tilde{U}_{\textup{\textsc{sr}}}\tilde{U}_{\textup{\textsc{sr}}}^{\T}\mid C)^{-1}W_{\textup{\textsc{sr}}}\zeta_{\textsc{a}}\tilde{U}_{\textup{\textsc{sr}}}\\
&\quad+W_{\textup{\textsc{sr}}}D\zeta_{\textsc{a}}\tilde{U}_{\textup{\textsc{sr}}},
\end{aligned}
\end{equation*}
with
$$D=\begin{pmatrix}-rI_{K}&0\\0&I_{L}\end{pmatrix},$$
and $\zeta_{\textsc{a}}$ being the population residual from the WLS regression of $Z$ on $(G, X, Y)$.

(ii) If $\tau_{\textup{\textsc{sr}}}=0$, we have $n\hat{\tau}_{\textup{\textsc{sr,a}}}^{\textup{c}}(r)\rightarrow \sum_{\ell=1}^{K+L}\lambda_{\textup{\textsc{sr,a}}\ell}\chi^2_{\ell}(1)$ in distribution as $n\rightarrow\infty$,  where $\lambda_{\textup{\textsc{sr,a}}\ell}\ (\ell=1,\ldots, K+L)$ are eigenvalues of
\begin{equation*}
\begin{aligned}
\Gamma_{\textup{\textsc{sr,a}}}&=\frac{1}{2}E(W_{\textsc{sr}}\tilde{Z}_{\textsc{sr}}^2)^{-1}\big\{E(W_{\textsc{sr}}\tilde{U}_{\textsc{sr}}\tilde{U}_{\textsc{sr}}^{\T})^{-1}D
+DE(W_{\textsc{sr}}\tilde{U}_{\textsc{sr}}\tilde{U}_{\textsc{sr}}^{\T})^{-1}\big\}^{1/2}\\
&\quad\times\bigg\{\sum_{s=1}^{S}\pi_sE[W_{\textup{\textsc{sr}}}^2\zeta_{\textsc{a}}^2\tilde{U}_{\textup{\textsc{sr}}}\tilde{U}_{\textup{\textsc{sr}}}^{\T}\mid C=s]\bigg\}
\big\{E(W_{\textsc{sr}}\tilde{U}_{\textsc{sr}}\tilde{U}_{\textsc{sr}}^{\T})^{-1}D
+DE(W_{\textsc{sr}}\tilde{U}_{\textsc{sr}}\tilde{U}_{\textsc{sr}}^{\T})^{-1}\big\}^{1/2}.
\end{aligned}
\end{equation*}
\end{theorem}

As in Theorem 2, the asymptotic distribution of $\hat{\tau}_{\textsc{sr,a}}^{\textup{c}}(r)-\tau_{\textsc{sr,a}}^{\textup{c}}$ depends on $\tau_{\textsc{sr}}$. When $\tau_{\textsc{sr}}\neq0$, the estimator $\hat{\tau}_{\textsc{sr,a}}^{\textup{c}}(r)$ converges to a normal distribution, and when $\tau_{\textsc{sr}}=0$, it converges to a weighted sum of chi-squared distributions.  We can construct confidence intervals for $\tau_{\textsc{sr,a}}^{\textup{c}}$ by replacing the asymptotic variance and eigenvalues with their sample analogues; we omit the technical details.

To estimate  the optimal coefficient $r_{\textup{opt}}$, we use Theorem \ref{AsyDis-BetaTau-SR-withX} to compute $\textup{var}(\hat{\tau}_{\textsc{sry,a}}^{\textup{c}}+\hat{\tau}_{\textsc{srx,a}}^{\textup{c}})$,  $\textup{var}(\hat{\tau}_{\textsc{sry,a}}^{\textup{c}}-\hat{\tau}_{\textsc{srx,a}}^{\textup{c}})$, and $\textup{var}(\hat{\tau}_{\textsc{sry,a}}^{\textup{c}})$ by setting $r=-1$, $1$, and $0$, respectively. We then obtain   $\textup{var}(\hat{\tau}_{\textsc{srx,a}}^{\textup{c}})=\textup{var}(\hat{\tau}_{\textsc{sry,a}}^{\textup{c}}+\hat{\tau}_{\textsc{srx,a}}^{\textup{c}})/2
+\textup{var}(\hat{\tau}_{\textsc{sry,a}}^{\textup{c}}-\hat{\tau}_{\textsc{srx,a}}^{\textup{c}})/2-\textup{var}(\hat{\tau}_{\textsc{sry,a}}^{\textup{c}})$. Replacing these population variances in the expression of $r_{\textup{opt}}$ with their sample analogues yields the estimator
$\hat{r}_{\textup{opt}}$. Deriving the asymptotic distribution of $\hat{\tau}_{\textsc{sr,a}}^{\textup{c}}(\hat{r}_{\textup{opt}})$ is cumbersome, so we leave it for future work.

\subsection{A real-data example}\label{sec-RealExample-SRE}

\cite{Chong2016} conducted a stratified randomized experiment with 219 students from a rural secondary school in Peru during the 2009 school year to evaluate the effect of iron supplementation on academic performance. The researchers first supplied the village clinic with iron supplements and trained local staff to distribute one free iron pill to any adolescent who requested it in person. Students were stratified by class level from 1--5. Within each level, they were randomly assigned in equal proportions to one of three video intervention arms: the first video showed a popular soccer player encouraging iron supplements to maximize energy (``soccer'' arm); the second video showed a doctor encouraging iron supplements for overall health (``physician'' arm); the third video did not mention iron at all (``control'' arm). Our analysis focuses on a subset of the original data, comparing only the ``physician''  and ``control'' arms.

\begin{figure}
\centering
    \includegraphics[width=0.45\linewidth]{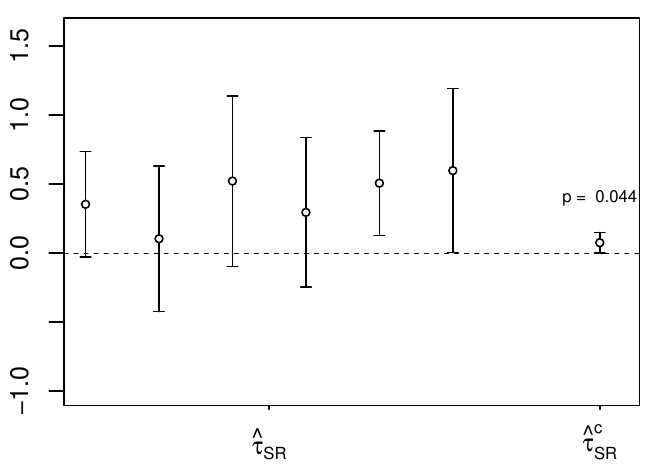}
\caption{Point estimates and 95\% confidence intervals for treatment effects on single outcomes ($\hat\tau_{\textsc{sr}}$) and the composite outcome ($\hat\tau_{\textsc{sr}}^{\textup{c}}$) for the example from \citet{Chong2016} analysed in Section \ref{sec-RealExample-SRE}. The $p$-value is from the test based on $\hat\tau_{\textsc{sr}}^{\textup{c}}$.}
\label{figS1-chong}
\end{figure}

Academic performance was measured using six outcomes: the average grade across math, English, social science, science, and communication in the first quarter of 2009, along with the grades in these subjects in the third and fourth quarters. Using the method in Section \ref{subsec-SREnoX-varweightATE}, we construct a composite outcome from these six measures.  Figure \ref{figS1-chong} displays point estimates and 95\% confidence intervals for the intervention effect on each single grade outcome and the composite outcome. The intervention shows a marginally significant effect on the composite outcome, suggesting that the physician video may improve overall academic performance. Among the single outcomes, only the social science grade exhibits a significant difference between the two arms.

\section{Extension to logistic regression}
\label{sec::logistic}

\subsection{Completely randomized experiments with multiple outcomes}
\label{subsec::logistic-CRE}

In this section, we extend the inverse linear regression framework to inverse logistic regression. We first consider a completely randomized experiment and then extend the results to an observational study.  The notation and definitions follow Section 2 of the main text. To construct a composite outcome from $Y$, we fit the inverse weighted logistic regression of $Z$ on $Y$:
\begin{equation}\label{logitmodel-CR}
\logit\big\{\pr(Z=1\mid Y)\big\}=\gamma_0+\gamma^{\T}Y,
\end{equation}
using the weights $W=Z/p+(1-Z)/(1-p)$, where $p=\pr(Z=1)$. Let $\gamma_{\textsc{f}}=(\gamma_0,\gamma^{\T})^{\T}$ denote the full parameter vector.

The observed data consist of $n$ independent copies of $(Z, Y)$, denoted by $(z_i, y_i)$ for $i=1,\ldots,n$. The weight assigned to unit $i$ is
$w_i=z_i/p+(1-z_i)/(1-p)$.  For $i=1,\ldots,n$, define $\pi(y_i,\gamma_{\textsc{f}})=\pr(z_i=1\mid y_i)=\exp(\gamma_0+\gamma^{\T}y_i)/\{1+\exp(\gamma_0+\gamma^{\T}y_i)\}$. The scaled weighted log-likelihood function for model \eqref{logitmodel-CR} is
\begin{equation*}
l(\gamma_{\textsc{f}})=n^{-1}\sum\limits_{i=1}^{n}w_i\big[z_i\log\{\pi(y_i,\gamma_{\textsc{f}})\}+(1-z_i)\log\{1-\pi(y_i,\gamma_{\textsc{f}})\}\big].
\end{equation*}
Differentiating $l(\gamma_{\textsc{f}})$ with respect to $\gamma_{\textsc{f}}$ yields the score function
\begin{equation*}
m(\gamma_{\textsc{f}})=\frac{\partial l(\gamma_{\textsc{f}})}{\partial \gamma_{\textsc{f}}}=n^{-1}\sum\limits_{i=1}^{n}w_i\{z_i-\pi(y_i,\gamma_{\textsc{f}})\}(1, y_i^{\T})^{\T}.
\end{equation*}
The Hessian matrix is
\begin{equation*}
H(\gamma_{\textsc{f}})=\frac{\partial^2 l(\gamma_{\textsc{f}})}{\partial \gamma_{\textsc{f}}\partial\gamma_{\textsc{f}}^{\T}}=-n^{-1}\sum_{i=1}^{n}w_i\pi(y_i,\gamma_{\textsc{f}})\{1-\pi(y_i,\gamma_{\textsc{f}})\}(1, y_i^{\T})^{\T}(1, y_i^{\T}).
\end{equation*}
Let $\hat\gamma_{\textsc{f}}=(\hat\gamma_0, \hat\gamma^{\T})^{\T}$ denote the weighted maximum likelihood estimator of $\gamma_{\textsc{f}}$.
The estimator $\hat\gamma_{\textsc{f}}$ maximizes $l(\gamma_{\textsc{f}})$ and satisfies the first-order condition
$m(\hat\gamma_{\textsc{f}})=n^{-1}\sum_{i=1}^{n}w_i\{z_i-\pi(y_i,\hat\gamma_{\textsc{f}})\}(1, y_i^{\T})^{\T}=0$. Let $\gamma_{\textsc{f}}^*=(\gamma_0^*,(\gamma^*)^{\T})^{\T}$ denote the probability limit of $\hat\gamma_{\textsc{f}}$, which satisfies $\gamma_{\textsc{f}}^*=\argmax_{\gamma_{\textsc{f}}}E\{l(\gamma_{\textsc{f}})\}$.
Based on the inverse regression coefficient of $\gamma^*$,  we construct the composite outcome:
\begin{equation*}
Y^{\textup{logit,c}} =(\gamma^*)^{\T}Y.
\end{equation*}
The corresponding average treatment effect is $\tau^{\textup{logit,c}}=(\gamma^*)^{\T}\tau$. We estimate $\tau^{\textup{logit,c}}$ by $\hat{\tau}^{\textup{logit,c}}=\hat\gamma^{\T}\hat\tau$, where $\hat{\tau}=(\phi_{zz}^{w})^{-1}\phi_{yz}^{w}$, $\phi_{zz}^{w}=S^{w}_{\smash{zz}}-S^{w}_{\smash{z1}}(S^{w}_{\smash{11}})^{-1}S^{w}_{\smash{1z}}$, and $\phi_{yz}^{w}=S^{w}_{\smash{yz}}-S^{w}_{\smash{y1}}(S^{w}_{\smash{11}})^{-1}S^{w}_{\smash{1z}}$.

Proposition \ref{Equiv-logit-CR-noX} below shows that, under the null hypothesis $\tau=E\{Y(1)\}-E\{Y(0)\}=0$, the weighted maximum likelihood estimator $\hat\gamma$ is asymptotically equivalent to the treatment effect estimator $\hat{\tau}$.

\begin{proposition}\label{Equiv-logit-CR-noX}
Assume complete randomization with $Z\ind\{Y(1),$ $Y(0)\}$.
Under $\tau=0$, we have $\hat{\gamma}=4\phi_{zz}^{w}(\phi_{yy}^{w})^{-1}\hat{\tau}+o_p(n^{-1/2})$.
\end{proposition}

Recall that Proposition 1 of the main text establishes the equivalence between the inverse linear regression coefficient $\hat{\beta}$ and the treatment effect estimator $\hat{\tau}$. Combining Proposition 1 with Proposition \ref{Equiv-logit-CR-noX} yields $\hat{\gamma}=4\hat{\beta}+o_p(n^{-1/2})$, that is,
the weighted maximum likelihood estimator $\hat{\gamma}$ from the inverse logistic regression  is asymptotically equivalent to the OLS estimator $\hat\beta$ from the inverse linear regression.

The asymptotic equivalence between $\hat\gamma$ and $\hat{\tau}$ motivates a Wald test based on $\hat\gamma$ for the null hypothesis $H_0: \tau= 0$. We define the Wald statistic as $\mathcal{W}^{\hat{\gamma}} = \hat{\gamma}^{\T}\hat{V}_{\gamma}^{-1}\hat{\gamma}$,
where
\begin{equation*}
\begin{aligned}
\hat{V}_{\gamma}&= 4n^{-1}(\phi_{yy}^{w})^{-1}\bigg(n^{-1}\sum_{i=1}^{n}\big[w_i^2\big\{y_i-(S_{11}^{w})^{-1}S_{y1}^{w}\big\}\big\{y_i-(S_{11}^{w})^{-1}S_{y1}^{w}\big\}^{\T}\big]\bigg)(\phi_{yy}^{w})^{-1}.
\end{aligned}
\end{equation*}
The following proposition establishes the asymptotic null distribution of $\mathcal{W}^{\text{logit}}$ for assessing statistical significance.

\begin{proposition}\label{asyDis-wald-logit-CR-noX}
Assume complete randomization with $Z\ind\{Y(1),Y(0)\}$. Under $\tau=0$, we have
$\mathcal{W}^{\hat{\gamma}}  = \hat{\gamma}^{\T}\hat{V}_{\gamma}^{-1}\hat{\gamma}\rightarrow \chi^2(L)$ in distribution as $n\rightarrow\infty$.
\end{proposition}

Under the null hypothesis $\tau=0$, Proposition \ref{Equiv-logit-CR-noX}, together with $\hat{\tau}\stackrel{p}\rightarrow\tau$, implies that
the average treatment effect estimator $\hat{\tau}^{\textup{logit,c}}=\hat\gamma^{\T}\hat\tau$ converges in probability to zero. The following theorem  establishes the asymptotic distribution of $\hat{\tau}^{\textup{logit,c}}$. The proof follows directly from the second part of Theorem 1 and is therefore omitted.

\begin{theorem}\label{AsyDis-GammaTau-CR-noX}
Assume complete randomization $Z\ind\{Y(1),Y(0)\}$. Under $\tau=0$, we have
$n\hat{\tau}^{\textup{logit,c}}\rightarrow\sum_{\ell=1}^{L}\lambda_{\ell}^{\textup{logit}}\chi^2_{\ell}(1)$  in distribution as $n\rightarrow\infty$,
where $\lambda_{\ell}^{\textup{logit}}(\ell=1,\ldots, L)$ are eigenvalues of
$$\Gamma^{\textup{logit}}=2(\Phi_{\textsc{yy}}^{\textsc{w}})^{-1/2}E\big[W^2\big\{Y-E(W)^{-1}E(WY)\big\}\big\{Y-E(W)^{-1}E(WY)\big\}^{\T}\big](\Phi_{\textsc{yy}}^{\textsc{w}})^{-1/2}.$$
\end{theorem}

\subsection{Observational studies with multiple outcomes}
\label{subsec::logistic-SRE}

The results in Section \ref{subsec::logistic-CRE} extend directly to observational studies under strong ignorability $Z\ind\{Y(1), Y(0)\} \mid X$  and overlap $0<e(X)<1$. Throughout the subsection, we retain the notation and definitions introduced in Section 3 of the main text. Recall that the propensity score is $e(X)=\pr(Z=1\mid X)$. We construct a composite outcome from $Y$ by fitting the inverse weighted logistic regression of $Z$ on $Y$: $\logit\{\pr(Z=1\mid Y)\}=\gamma_{\textsc{os0}}+\gamma_{\textsc{os}}^{\T}Y$, using the inverse propensity score weights $W=Z/e(X)+(1-Z)/\{1-e(X)\}$. The observed data are $(z_i, x_i, y_i)$ $(i=1,\ldots,n)$, which are independent copies of $(Z, X, Y)$. Let $\gamma_{\textsc{osf}}=(\gamma_{\textsc{os0}}, \gamma_{\textsc{os}}^{\T})^{\T}$ denote the full parameter vector, and let
$\hat\gamma_{\textsc{osf}}=(\hat\gamma_{\textsc{os0}}, \hat\gamma_{\textsc{os}}^{\T})^{\T}$ denote its weighted maximum likelihood estimator with weights
$\hat{w}_i=z_i/\hat{e}(x_i)+(1-z_i)/\{1-\hat{e}(x_i)\}$, where $\hat{e}(x_i)$ denotes the estimated propensity score for unit $i$.
Let $\gamma_{\textsc{osf}}^*=(\gamma_{\textsc{os0}}^*,(\gamma^*_{\textsc{os}})^{\T})^{\T}$ denote the probability limit of $\hat\gamma_{\textsc{osf}}$, which is the unique maximizer of the population weighted log-likelihood.
We use $\gamma^*_{\textsc{os}}$ to define the composite outcome:
\begin{equation*}
Y^{\textup{logit,c}}_{\textsc{os}} =(\gamma^*_{\textsc{os}})^{\T}Y.
\end{equation*}
The corresponding average treatment effect is $\tau^{\textup{logit,c}}_{\textsc{os}}=(\gamma^*_{\textsc{os}})^{\T}\tau$. We estimate $\tau^{\textup{logit,c}}_{\textsc{os}}$ by
$\hat{\tau}^{\textup{logit,c}}_{\textsc{os}}=\hat\gamma_{\textsc{os}}^{\T}\hat\tau_{\textsc{os}}$, where $\hat{\tau}_{\textsc{os}}=(\phi_{zz}^{\hat{w}})^{-1}\phi_{yz}^{\hat{w}}$ with $\phi_{zz}^{\hat{w}}=S^{\hat{w}}_{\smash{zz}}-S^{\hat{w}}_{\smash{z1}}(S^{\hat{w}}_{\smash{11}})^{-1}S^{\hat{w}}_{\smash{1z}}$ and $\phi_{yz}^{\hat{w}}=S^{\hat{w}}_{\smash{yz}}-S^{\hat{w}}_{\smash{y1}}(S^{\hat{w}}_{\smash{11}})^{-1}S^{\hat{w}}_{\smash{1z}}$.
Analogous to Proposition \ref{Equiv-logit-CR-noX}, we can establish an asymptotic equivalence between $\hat\gamma_{\textsc{os}}$ and $\hat{\tau}_{\textsc{os}}$ under $\tau=0$.

\begin{proposition}\label{Equiv-logit-OBS-noX}
Assume strong ignorability $Z\ind$ $\{Y(1),Y(0)\} \mid X$ and overlap $0<e(X)<1$.
Under $\tau=0$, we have $\hat{\gamma}_{\textup{\textsc{os}}}=4\phi_{zz}^{\hat{w}}(\phi_{yy}^{\hat{w}})^{-1}\hat{\tau}_{\textup{\textsc{os}}}+o_p(n^{-1/2})$.
\end{proposition}

The proof follows the same arguments as the proof of  Proposition \ref{Equiv-logit-CR-noX}, and is therefore omitted. Proposition \ref{Equiv-logit-OBS-noX} motivates a  Wald statistic $\mathcal{W}^{\hat\gamma_{\textsc{os}}}$, based on $\hat\gamma_{\textsc{os}}$, for testing the null hypothesis $H_0: \tau=0$. The construction parallels Proposition \ref{asyDis-wald-logit-CR-noX}, except that the asymptotic variance of $\hat\gamma_{\textsc{os}}$ must account for uncertainty from estimating the propensity score.

Similarly, extending the result in Theorem \ref{AsyDis-GammaTau-CR-noX} to observational studies yields the asymptotic distribution of $\hat{\tau}^{\textup{logit,c}}_{\textsc{os}}$. The asymptotic variance retains the same general structure as that of $\hat{\tau}^{\textup{logit,c}}$, while also accounting for propensity score estimation. Because the derivations of the derivations of the Wald statistic and the asymptotic distribution of $\hat{\tau}^{\textup{logit,c}}_{\textsc{os}}$ closely parallel Proposition \ref{asyDis-wald-logit-CR-noX} and Theorem \ref{AsyDis-GammaTau-CR-noX}, respectively, we omit the details.

\section{Proofs of the results in Section 2} \label{sec::proof-CRE-noX}

\subsection{Proof of Proposition 1}\label{subsec::prop1-proof}

\noindent(i) We first prove the result for generic weights $w_i$. Solving the normal equations of WLS yields
$\hat{\beta}=(n^{-1}\sum_{i=1}^nw_i\tilde{y}_i\tilde{y}_i^{\T})^{-1}(n^{-1}\sum_{i=1}^nw_i\tilde{y}_i\tilde{z}_i)$ and $\hat{\tau}=(n^{-1}\sum_{i=1}^nw_i\tilde{z}_i^2)^{-1}\{n^{-1}\sum_{i=1}^nw_i\tilde{y}_i\tilde{z}_i\}$, where
$\tilde{y}_i=y_i-\sum_{i=1}^{n}w_iy_i/(\sum_{i=1}^{n}w_i)$ and $\tilde{z}_i=z_i-\sum_{i=1}^{n}w_iz_i/(\sum_{i=1}^{n}w_i)$. Using the notation $S^{w}_{\smash{ab}}$, we write $\hat{\beta}=(\phi_{yy}^{w})^{-1}
\phi_{yz}^{w}$ and $\hat{\tau}=(\phi_{zz}^{w})^{-1}\phi_{yz}^{w}$, where $\phi_{yy}^{w}=S^{w}_{\smash{yy}}-S^{w}_{\smash{y1}}(S^{w}_{\smash{11}})^{-1}S^{w}_{\smash{1y}}$, $\phi_{zz}^{w}=S^{w}_{\smash{zz}}-S^{w}_{\smash{z1}}(S^{w}_{\smash{11}})^{-1}S^{w}_{\smash{1z}}$, and $\phi_{yz}^{w}=S^{w}_{\smash{yz}}-S^{w}_{\smash{y1}}(S^{w}_{\smash{11}})^{-1}S^{w}_{\smash{1z}}$.
Comparing the preceding expressions gives
$$\hat{\beta}=\phi_{zz}^{w}(\phi_{yy}^{w})^{-1}\hat{\tau}.$$

To further characterize the relationship between $\hat{\beta}$ and $\hat{\tau}$, we derive an explicit expression for $\phi_{zz}^{w}(\phi_{yy}^{w})^{-1}$.
Let $\bar{y}_w(1)=\sum_{i=1}^{n}w_iz_iy_{i}/(\sum_{i=1}^{n}w_iz_i)$ and $\bar{y}_w(0)=\sum_{i=1}^{n}w_i(1-z_i)y_{i}/\{\sum_{i=1}^{n}w_i(1-z_i)\}$ denote the weighted means in the treatment and control groups, respectively. Define $\hat{\textup{cov}}_{w}\{Y(1)\}=n^{-1}\sum_{i=1}^{n}w_iz_i\{y_i-\bar{y}_{w}(1)\}\{y_i-\bar{y}_{w}(1)\}^{\T}$  and $\hat{\textup{cov}}_{w}\{Y(0)\}=n^{-1}\sum_{i=1}^{n}w_i$ $(1-z_i)\{y_i-\bar{y}_{w}(0)\}\{y_i-\bar{y}_{w}(0)\}^{\T}$.
We will show that
$$(\phi_{zz}^{w})^{-1}\phi_{yy}^{w}=(\phi_{zz}^{w})^{-1}[\hat{\textup{cov}}_{w}\{Y(1)\}+\hat{\textup{cov}}_{w}\{Y(0)\}]+\hat{\tau}\hat{\tau}^{\T}.$$
To prove this identity, we first rewrite $\hat{\textup{cov}}_{w}\{Y(1)\}$ and $\hat{\textup{cov}}_{w}\{Y(0)\}$ as
\begin{equation*}
\begin{aligned}
\hat{\textup{cov}}_{w}\{Y(1)\}&=n^{-1}\sum_{i=1}^{n}w_iz_iy_iy_i^{\T}-S^{w}_{\smash{yz}}(S^{w}_{\smash{1z}})^{-1}S^{w}_{\smash{zy}},
\end{aligned}
\end{equation*}
and
\begin{equation*}
\begin{aligned}
\hat{\textup{cov}}_{w}\{Y(0)\}&=n^{-1}\sum_{i=1}^{n}w_i(1-z_i)y_iy_i^{\T}-(S^{w}_{\smash{y1}}-S^{w}_{\smash{yz}})
(S^{w}_{\smash{11}}-S^{w}_{\smash{1z}})^{-1}(S^{w}_{\smash{1y}}-S^{w}_{\smash{zy}}).
\end{aligned}
\end{equation*}
Summing the two covariance matrices yields
$$\hat{\textup{cov}}_{w}\{Y(1)\}+\hat{\textup{cov}}_{w}\{Y(0)\}=S^{w}_{\smash{yy}}-S^{w}_{\smash{yz}}(S^{w}_{\smash{1z}})^{-1}S^{w}_{\smash{zy}}
-(S^{w}_{\smash{y1}}-S^{w}_{\smash{yz}})(S^{w}_{\smash{11}}-S^{w}_{\smash{1z}})^{-1}(S^{w}_{\smash{1y}}-S^{w}_{\smash{zy}}).$$
Recall that $\phi_{zz}^{w}=S^{w}_{\smash{zz}}-S^{w}_{\smash{z1}}(S^{w}_{\smash{11}})^{-1}S^{w}_{\smash{1z}}$. Then, after simplification, we obtain
\begin{equation*}
\begin{aligned}
\phi_{zz}^{w}[\hat{\textup{cov}}_{w}\{Y(1)\}+\hat{\textup{cov}}_{w}\{Y(0)\}]&=\phi_{zz}^{w}S^{w}_{\smash{yy}}-(S^{w}_{\smash{11}}-S^{w}_{\smash{1z}})S^{w}_{\smash{yz}}(S^{w}_{\smash{11}})^{-1}S^{w}_{\smash{zy}}\\
&\quad-S^{w}_{\smash{1z}}(S^{w}_{\smash{y1}}-S^{w}_{\smash{yz}})(S^{w}_{\smash{11}})^{-1}(S^{w}_{\smash{1y}}-S^{w}_{\smash{zy}}),
\end{aligned}
\end{equation*}
and
\begin{equation*}
\begin{aligned}
(\phi_{zz}^{w})^2\hat{\tau}\hat{\tau}^{\T}&=S^{w}_{\smash{yz}}S^{w}_{\smash{zy}}-S^{w}_{\smash{1z}}(S^{w}_{\smash{11}})^{-1}S^{w}_{\smash{yz}}S^{w}_{\smash{1y}}
-S^{w}_{\smash{1z}}(S^{w}_{\smash{11}})^{-1}S^{w}_{\smash{y1}}S^{w}_{\smash{zy}}
+(S^{w}_{\smash{1z}})^2(S^{w}_{\smash{11}})^{-2}S^{w}_{\smash{y1}}S^{w}_{\smash{1y}}.
\end{aligned}
\end{equation*}
Adding the two expressions yields
\begin{equation*}
\begin{aligned}
\phi_{zz}^{w}[\hat{\textup{cov}}_{w}\{Y(1)\}+\hat{\textup{cov}}_{w}\{Y(0)\}]+(\phi_{zz}^{w})^2\hat{\tau}\hat{\tau}^{\T}
=\phi_{zz}^{w}\left\{S^{w}_{\smash{yy}}-S^{w}_{\smash{y1}}(S^{w}_{\smash{11}})^{-1}S^{w}_{\smash{1y}}\right\}=\phi_{zz}^{w}\phi_{yy}^{w}.
\end{aligned}
\end{equation*}
Therefore,
$$\hat{\beta}=\Big((\phi_{zz}^{w})^{-1}[\hat{\textup{cov}}_{w}\{Y(1)\}+\hat{\textup{cov}}_{w}\{Y(0)\}]+\hat{\tau}\hat{\tau}^{\T}\Big)^{-1}\hat{\tau}
=\big(\hat\Sigma+\hat{\tau}\hat{\tau}^{\T}\big)^{-1}\hat\tau.$$
Applying Lemma \ref{Woodburyformula} to $A=\hat\Sigma$, $\mu=\hat{\tau}$, and $\nu=\hat{\tau}$ gives
$$\hat{\beta}=
\Big(\hat\Sigma^{-1}-\frac{\hat\Sigma^{-1}\hat{\tau}\hat{\tau}^{\T}\hat\Sigma^{-1}}{1+\hat{\tau}^{\T}\hat\Sigma^{-1}\hat{\tau}}\Big)\hat\tau
=\frac{\hat\Sigma^{-1}\hat\tau}{1+\hat\tau^\T \hat\Sigma^{-1}\hat\tau}.$$
Under the weights $w_i=z_i/p+(1-z_i)/(1-p)$, the weighted means $\bar{y}_w(1)$ and $\bar{y}_w(0)$  and the weighted covariance matrices $\hat{\textup{cov}}_{w}\{Y(1)\}$ and $\hat{\textup{cov}}_{w}\{Y(0)\}$ reduce to $\bar y(1)$, $\bar y(0)$, $\hat{\textup{cov}}\{Y(1)\}$, and $\hat{\textup{cov}}\{Y(0)\}$, respectively. Substituting these quantities into the expression of $\hat{\beta}$ yields the result stated in the proposition.

\vspace{0.2cm}
\noindent(ii) Define $\tilde{Y}=Y-E(W)^{-1}E(WY)$ and $\tilde{Z}=Z-E(W)^{-1}E(WZ)$. By solving the normal equations of WLS, we obtain $\beta=E(W\tilde{Y}\tilde{Y}^{\T})^{-1}E(W\tilde{Y}\tilde{Z})$. A direct calculation gives
$$\beta=(\Phi_{\textsc{yy}}^{\textsc{w}})^{-1}
\big\{E(WZY)-E(WY)E(W)^{-1}E(WZ)\big\},$$
where $\Phi_{\textsc{yy}}^{\textsc{w}}=E(WYY^{\T})-E(WY)E(W)^{-1}E(WY^{\T})$. Recall that
the treatment effect $\tau$ is the coefficient of $Z$ from the element-wise WLS regression of $Y$ on $(1, Z)$, so
\begin{equation*}
\begin{aligned}
\tau&=(\Phi_{\textsc{zz}}^{\textsc{w}})^{-1}\big\{E(WZY)-E(WY)E(W)^{-1}E(WZ)\big\},
\end{aligned}
\end{equation*}
where $\Phi_{\textsc{zz}}^{\textsc{w}}=E(WZ^2)-E(WZ)E(W)^{-1}E(WZ)$. For the weight $W=Z/e(X)+(1-Z)/\{1-e(X)\}$, we have $\Phi_{\textsc{zz}}^{\textsc{w}}=1/2$. Comparing the expressions of $\beta$ and $\tau$ yields
$$\beta= (\Phi_{\textsc{yy}}^{\textsc{w}})^{-1}\tau/2.$$

Note that $E(WYY^{\T})=E\{Y(1)Y(1)^{\T}+Y(0)Y(0)^{\T}\}$, $E(WY)=E\{Y(1)+Y(0)\}$, and $E(W)=2$.
Substituting these identities into $\Phi_{\textsc{yy}}^{\textsc{w}}$ yields $\Phi_{\textsc{yy}}^{\textsc{w}}=\textup{cov}\{Y(1)\}+\textup{cov}\{Y(0)\}+\tau\tau^{\T}/2$.
Then we have
$$\beta=\big(\Sigma+\tau\tau^{\T}\big)^{-1}\tau.$$
Applying Lemma \ref{Woodburyformula} to $A=\Sigma$, $\mu=\tau$, and $\nu=\tau$, we obtain
\begin{equation*}
\begin{aligned}
\beta&=\Big(\Sigma^{-1}-\frac{\Sigma^{-1}\tau\tau^{\T}\Sigma^{-1}}{1+\tau^{\T}\Sigma^{-1}\tau}\Big)\tau
=\frac{\Sigma^{-1}\tau}{1+\tau^{\T}\Sigma^{-1}\tau}.
\end{aligned}
\end{equation*}

\subsection{Proof of Theorem 1}
\label{subsec::thm1-proof}

For each unit $i$, define the residual $\varepsilon_i = z_i-\beta_0-\beta^{\T}y_i$.  Substituting $z_i=\beta_0+\beta^{\T}y_i+\varepsilon_i$ into the expression of $\hat{\beta}$ gives
$$\hat{\beta}=\beta+(\phi_{yy}^{w})^{-1}
\{S^{w}_{\smash{y\varepsilon}}-S^{w}_{\smash{y1}}(S^{w}_{\smash{11}})^{-1}S^{w}_{\smash{1\varepsilon}}\}.$$
Proposition 1 gives $\hat{\beta}=\phi_{zz}^{w}(\phi_{yy}^{w})^{-1}\hat{\tau}$ and $\beta=\Phi_{\textsc{zz}}^{\textsc{w}}\big(\Phi_{\textsc{yy}}^{\textsc{w}}\big)^{-1}\tau$. Using these identities, we obtain
\begin{equation*}
\begin{aligned}
\hat{\beta}^{\T}\hat{\tau}-\beta^{\T}\tau&
=\beta^{\T}
\left\{(\phi_{zz}^{w})^{-1}\phi_{yy}^{w}-(\Phi_{\textsc{zz}}^{\textsc{w}})^{-1}\Phi_{\textsc{yy}}^{\textsc{w}}\right\}\beta+2\beta^{\T}(\phi_{zz}^{w})^{-1}\{S^{w}_{\smash{y\varepsilon}}-S^{w}_{\smash{y1}}(S^{w}_{\smash{11}})^{-1}S^{w}_{\smash{1\varepsilon}}\}\\
&\quad+(\phi_{zz}^{w})^{-1}\{S^{w}_{\smash{y\varepsilon}}-S^{w}_{\smash{y1}}(S^{w}_{\smash{11}})^{-1}S^{w}_{\smash{1\varepsilon}}\}^{\T}
(\phi_{yy}^{w})^{-1}\{S^{w}_{\smash{y\varepsilon}}-S^{w}_{\smash{y1}}(S^{w}_{\smash{11}})^{-1}S^{w}_{\smash{1\varepsilon}}\}.
\end{aligned}
\end{equation*}
We next derive the asymptotic distribution of $\hat{\beta}^{\T}\hat{\tau}-\beta^{\T}\tau$  in two cases: (i) $\tau\neq0$ and (ii) $\tau=0$.

\vspace{0.2cm}
\noindent(i) $\tau\neq0$. By the LLN, $\phi_{zz}^{w}=\Phi_{\textsc{zz}}^{\textsc{w}}+o_p(1)$. Therefore,
\begin{equation*}
\begin{aligned}
&\beta^{\T}
\big\{(\phi_{zz}^{w})^{-1}\phi_{yy}^{w}-(\Phi_{\textsc{zz}}^{\textsc{w}})^{-1}\Phi_{\textsc{yy}}^{\textsc{w}}\big\}\beta\\
&=(\Phi_{\textsc{zz}}^{\textsc{w}})^{-1}\beta^{\T}\big\{(\phi_{yy}^{w}-\Phi_{\textsc{yy}}^{\textsc{w}})\beta-(\phi_{zz}^{w}-\Phi_{\textsc{zz}}^{\textsc{w}})\Phi_{\textsc{yy}}^{\textsc{w}}\beta/\Phi_{\textsc{zz}}^{\textsc{w}})\big\}\times\{1+o_p(1)\}.
\end{aligned}
\end{equation*}
To obtain an asymptotic linear representation of the preceding expression, we first note that
\begin{equation*}
\begin{aligned}
\phi_{yy}^{w}-\Phi_{\textsc{yy}}^{\textsc{w}}&=n^{-1}\sum_{i=1}^{n}w_i\big\{y_i-(S^{w}_{\smash{11}})^{-1}S^{w}_{\smash{y1}}\big\}
\big\{y_i-(S^{w}_{\smash{11}})^{-1}S^{w}_{\smash{y1}}\big\}^{\T}-\Phi_{\textsc{yy}}^{\textsc{w}}\\
&=n^{-1}\sum_{i=1}^{n}\Big[w_i\big\{y_i-E(W)^{-1}E(WY)\big\}
\big\{y_i-E(W)^{-1}E(WY)\big\}^{\T}-\Phi_{\textsc{yy}}^{\textsc{w}}\Big]\\
&\quad-S^{w}_{\smash{11}}\big\{(S^{w}_{\smash{11}})^{-1}S^{w}_{\smash{y1}}-E(W)^{-1}E(WY)\big\}
\big\{(S^{w}_{\smash{11}})^{-1}S^{w}_{\smash{y1}}-E(W)^{-1}E(WY)\big\}^{\T}.
\end{aligned}
\end{equation*}
The LLN implies $S^{w}_{\smash{11}}=E(W)+o_p(1)$ and $S^{w}_{\smash{1y}}=E(WY)+o_p(1)$, while the CLT implies $n^{1/2}\{S^{w}_{\smash{11}}-E(W)\}=O_p(1)$ and
$n^{1/2}\{S^{w}_{\smash{y1}}-E(WY)\}=O_p(1)$. Consequently, $(S^{w}_{\smash{11}})^{-1}S^{w}_{\smash{y1}}-E(W)^{-1}E(WY)=O_p(n^{-1/2})$, which implies that the second term in the decomposition of $\phi_{yy}^{w}-\Phi_{\textsc{yy}}^{\textsc{w}}$ is $o_p(n^{-1/2})$. Hence,
$$n^{1/2}(\phi_{yy}^{w}-\Phi_{\textsc{yy}}^{\textsc{w}})=n^{-1/2}\sum_{i=1}^{n}\Big[w_i\big\{y_i-E(W)^{-1}E(WY)\big\}
\big\{y_i-E(W)^{-1}E(WY)\big\}^{\T}-\Phi_{\textsc{yy}}^{\textsc{w}}\Big]+o_p(1).$$
Applying analogous arguments yields
$$n^{1/2}(\phi_{zz}^{w}-\Phi_{\textsc{zz}}^{\textsc{w}})=n^{-1/2}\sum_{i=1}^{n}\Big[w_i\big\{z_i-E(W)^{-1}E(WZ)\big\}^2
-\Phi_{\textsc{zz}}^{\textsc{w}}\Big]+o_p(1).$$
Combining the two expansions, we obtain
\begin{equation*}
\begin{aligned}
&n^{1/2}\beta^{\T}
\big\{(\phi_{zz}^{w})^{-1}\phi_{yy}^{w}-(\Phi_{\textsc{zz}}^{\textsc{w}})^{-1}\Phi_{\textsc{yy}}^{\textsc{w}}\big\}\beta\\
&=\frac{\beta^{\T}}{n^{1/2}\Phi_{\textsc{zz}}^{\textsc{w}}}\sum_{i=1}^{n}\Big(\big[w_i\big\{y_i-E(W)^{-1}E(WY)\big\}
\big\{y_i-E(W)^{-1}E(WY)\big\}^{\T}-\Phi_{\textsc{yy}}^{\textsc{w}}\big]\beta\\
&\hspace{2.8cm}-\big[w_i\big\{z_i-E(W)^{-1}E(WZ)\big\}^2
-\Phi_{\textsc{zz}}^{\textsc{w}}\big]\Phi_{\textsc{yy}}^{\textsc{w}}\beta/\Phi_{\textsc{zz}}^{\textsc{w}}\Big)+o_p(1).
\end{aligned}
\end{equation*}

To analyze the second term in the expansion of  $\hat\beta^{\T}\hat\tau-\beta^{\T}\tau$, note that
\begin{equation}\label{CRnoX-asy-term2}
\begin{aligned}
&S^{w}_{\smash{y\varepsilon}}-S^{w}_{\smash{y1}}(S^{w}_{\smash{11}})^{-1}S^{w}_{\smash{1\varepsilon}}\\
&=n^{-1}\sum_{i=1}^{n}w_i\varepsilon_i\{y_i-E(W)^{-1}E(WY)\}
-\{(S^{w}_{\smash{11}})^{-1}S^{w}_{\smash{y1}}-E(W)^{-1}E(WY)\}\Big(n^{-1}\sum_{i=1}^{n}w_i\varepsilon_i\Big)\\
&=n^{-1}\sum_{i=1}^{n}w_i\varepsilon_i\{y_i-E(W)^{-1}E(WY)\}+o_p(n^{-1/2}),
\end{aligned}
\end{equation}
where the second equality follows because $n^{-1}\sum_{i=1}^{n}w_i\varepsilon_i=O_p(n^{-1/2})$ by the CLT  and
$S^{w}_{\smash{y1}}/S^{w}_{\smash{11}}-E(W)^{-1}E(WY)=o_p(1)$ by the LLN and Slutsky's theorem. Furthermore, substituting $\phi_{zz}^{w}=\Phi_{\textsc{zz}}^{\textsc{w}}+o_p(1)$ into the second term of $\hat\beta^{\T}\hat\tau-\beta^{\T}\tau$ gives
\begin{equation*}
\begin{aligned}
2\beta^{\T}(\phi_{zz}^{w})^{-1}\{S^{w}_{\smash{y\varepsilon}}-S^{w}_{\smash{y1}}(S^{w}_{\smash{11}})^{-1}S^{w}_{\smash{1\varepsilon}}\}
=\frac{2\beta^{\T}}{n\Phi_{\textsc{zz}}^{\textsc{w}}}\sum_{i=1}^{n}w_i\varepsilon_i\{y_i-E(W)^{-1}E(WY)\}
+o_p(n^{-1/2}).
\end{aligned}
\end{equation*}

Equation \eqref{CRnoX-asy-term2} and the CLT imply that $n^{1/2}\{S^{w}_{\smash{y\varepsilon}}-S^{w}_{\smash{y1}}(S^{w}_{\smash{11}})^{-1}S^{w}_{\smash{1\varepsilon}}\}=O_p(1)$. Consequently, the third term in the expansion of $\hat{\beta}^{\T}\hat{\tau}-\beta^{\T}\tau$ satisfies
$$(\phi_{zz}^{w})^{-1}\{S^{w}_{\smash{y\varepsilon}}-S^{w}_{\smash{y1}}(S^{w}_{\smash{11}})^{-1}S^{w}_{\smash{1\varepsilon}}\}^{\T}
(\phi_{yy}^{w})^{-1}\{S^{w}_{\smash{y\varepsilon}}-S^{w}_{\smash{y1}}(S^{w}_{\smash{11}})^{-1}S^{w}_{\smash{1\varepsilon}}\}=O_p(n^{-1}).$$
Combining the asymptotic expansions of the first, second, and third terms yields
\begin{equation*}
\begin{aligned}
&n^{1/2}(\hat{\beta}^{\T}\hat{\tau}-\beta^{\T}\tau)\\ &=\frac{\beta^{\T}}{n^{1/2}\Phi_{\textsc{zz}}^{\textsc{w}}}\sum_{i=1}^{n}\Big(\big[w_i\{y_i-E(W)^{-1}E(WY)\}
\{y_i-E(W)^{-1}E(WY)\}^{\T}-\Phi_{\textsc{yy}}^{\textsc{w}}\big]\beta\\
&\hspace{2.8cm}-\big[w_i\big\{z_i-E(W)^{-1}E(WZ)\big\}^2
-\Phi_{\textsc{zz}}^{\textsc{w}}\big]\Phi_{\textsc{yy}}^{\textsc{w}}\beta/\Phi_{\textsc{zz}}^{\textsc{w}}\\
&\hspace{2.8cm}+2w_i\varepsilon_i\{y_i-E(W)^{-1}E(WY)\}\Big)+o_p(1)
\end{aligned}
\end{equation*}
By the multivariate CLT and the delta method, we obtain
$$n^{1/2}(\hat{\tau}^{\textup{c}}-\tau^{\textup{c}})\stackrel{d}\rightarrow N\big(0, 4\beta^\T\textup{cov}\{R(Z,Y)\}\beta\big),$$
where
\begin{equation*}
\begin{aligned}
R(Z,Y)&=\big[W\{Y-E(W)^{-1}E(WY)\}\{Y-E(W)^{-1}E(WY)\}^{\T}-\Phi_{\textsc{yy}}^{\textsc{w}}\big]\beta_{\textup{\textsc{os}}}\\
&\quad-\big[W\{Z-E(W)^{-1}E(WZ)\}^2
-\Phi_{\textsc{zz}}^{\textsc{w}}\big]\Phi_{\textsc{yy}}^{\textsc{w}}\beta/\Phi_{\textsc{zz}}^{\textsc{w}}+2W\varepsilon \{Y-E(W)^{-1}E(WY)\}.
\end{aligned}
\end{equation*}
Using $\Phi_{\textsc{zz}}^{\textsc{w}}=1/2$ and $\{Z-E(W)^{-1}E(WZ)\}^2=1/4$, we can simplify $R(Z,Y)$ to the expression stated in the theorem

\vspace{0.2cm}
\noindent(ii) $\tau=0$.  In this case, the first two terms in the expansion of $\hat{\beta}^{\T}\hat{\tau}-\beta^{\T}\tau$ vanish, and therefore
\begin{equation*}
\begin{aligned}
n(\hat{\beta}^{\T}\hat{\tau}-\beta^{\T}\tau)&=n(\phi_{zz}^{w})^{-1}\{S^{w}_{\smash{y\varepsilon}}-S^{w}_{\smash{y1}}(S^{w}_{\smash{11}})^{-1}S^{w}_{\smash{1\varepsilon}}\}^{\T}
(\phi_{yy}^{w})^{-1}\{S^{w}_{\smash{y\varepsilon}}-S^{w}_{\smash{y1}}(S^{w}_{\smash{11}})^{-1}S^{w}_{\smash{1\varepsilon}}\}.
\end{aligned}
\end{equation*}
Define $M=n^{1/2}(\phi_{zz}^{w})^{-1/2}(\phi_{yy}^{w})^{-1/2}\{S^{w}_{\smash{y\varepsilon}}-S^{w}_{\smash{y1}}(S^{w}_{\smash{11}})^{-1}S^{w}_{\smash{1\varepsilon}}\}$. Then $n(\hat{\beta}^{\T}\hat{\tau}-\beta^{\T}\tau)=M^{\T}M$.
Equation \eqref{CRnoX-asy-term2} and the multivariate CLT imply
$$n^{1/2}\{S^{w}_{\smash{y\varepsilon}}-S^{w}_{\smash{y1}}(S^{w}_{\smash{11}})^{-1}S^{w}_{\smash{1\varepsilon}}\}\stackrel{d}\rightarrow N\big(0, E[W^2\varepsilon^2\{Y-E(W)^{-1}E(WY)\}\{Y-E(W)^{-1}E(WY)\}^{\T}]\big).$$
Furthermore, by the LLN,  $\phi_{zz}^{w}=\Phi_{\textsc{zz}}^{\textsc{w}}+o_p(1)$ and $\phi_{yy}^{w}=\Phi_{\textsc{yy}}^{\textsc{w}}+o_p(1)$. Combining these results with Slutsky's theorem gives  $M \stackrel{d}\rightarrow N\left(0, \Gamma\right)$,
where
$$\Gamma=(\Phi_{\textsc{zz}}^{\textsc{w}})^{-1}(\Phi_{\textsc{yy}}^{\textsc{w}})^{-1/2} E[W^2\varepsilon^2\{Y-E(W)^{-1}E(WY)\}\{Y-E(W)^{-1}E(WY)\}^{\T}](\Phi_{\textsc{yy}}^{\textsc{w}})^{-1/2}.$$
Let $\Gamma=Q^{\T}\Delta Q$ denote the eigendecomposition of $\Gamma$, where $\Delta=\diag(\lambda_1,\ldots, \lambda_L)$ contains the eigenvalues of
$\Gamma$ and  $Q$ is the orthogonal matrix whose columns are the corresponding eigenvectors.
Define $B=\Delta^{-1/2}QM$. Since $M \stackrel{d}\rightarrow N\left(0, \Gamma\right)$, it follows that $B\stackrel{d}\rightarrow N(0, I_L)$. Note that
$M^{\T}M=(Q^{\T}\Delta^{1/2}B)^{\T}(Q^{\T}\Delta^{1/2}B)=B^\T\Delta B.$
Writing $B=(B_{1},\ldots, B_{L})^\T$, we obtain
	\begin{equation*}
		n\hat\beta^\T\hat\tau=B^\T\Delta B=\sum_{\ell=1}^{L}\lambda_{\ell}B_{\ell}^2\stackrel{d}\rightarrow\sum_{\ell=1}^{L}\lambda_{\ell}\chi^2_{\ell}(1).
\end{equation*}

\section{Proofs of the results in Section 3}
\label{sec::proof-OBS-noX}

\subsection{Proof of Proposition 2}
\label{subsec::prop2-proof}

The proof follows the same arguments as those used in Proposition 1 after replacing $w_i=z_i/p+(1-z_i)/(1-p)$ and $W=Z/p+(1-Z)/(1-p)$ with $\hat{w}_i=z_i/\hat{e}(x_i)+(1-z_i)/{1-\hat{e}(x_i)}$ and $W=Z/e(X)+(1-Z)/\{1-e(X)\}$, respectively. We therefore omit the details.

\subsection{Proof of Lemma \ref{AsyDis-Beta-OBS}}
\label{subsec::lemmaS5-proofd}

Substituting $z_i=\beta_{0}+\beta^{\T}y_i+\varsigma_i$ into the expression of $\hat{\beta}_{\textsc{os}}$ yields
$$
\hat{\beta}_{\textsc{os}}=\beta+\big\{S^{\hat{w}}_{\smash{yy}}-S^{\hat{w}}_{\smash{y1}}(S^{\hat{w}}_{\smash{11}})^{-1}S^{\hat{w}}_{\smash{1y}}\big\}^{-1}
\big\{S^{\hat{w}}_{\smash{y\varsigma}}-S^{\hat{w}}_{\smash{y1}}(S^{\hat{w}}_{\smash{11}})^{-1}S^{\hat{w}}_{\smash{1\varsigma}}\big\}.
$$
Define $w(z,x;\alpha)=z/e(x; \alpha)+(1-z)/\{1-e(x; \alpha)\},$ then $\hat{w}_i=w(z,x;\hat\alpha)$.  Because the propensity score model $e(X;\alpha)$ is correctly specified,  $\hat{\alpha}$ converges in probability to $\alpha_0$. Consequently, $e(x;\hat\alpha)=e(x;\alpha_0)+o_p(1)$ for every $x$. Decompose the estimated weight as $w(z_i,x_i;\hat\alpha) =w(z_i,x_i;\alpha_0)+\{w(z_i,x_i;\hat{\alpha})-w(z_i,x_i;\alpha_0)\}$. A first-order Taylor expansion gives $$w(z_i,x_i;\hat{\alpha})-w(z_i,x_i;\alpha_0)=\nabla_{\alpha}w(z_i,x_i;\alpha_0)^{\T}(\hat{\alpha}-\alpha_0)+O_p(||\hat{\alpha}-\alpha_0||^2).$$
Standard  M-estimation theory \citep{van2000} implies that
$n^{1/2}(\hat{\alpha}-\alpha_0)=n^{-1/2}I(\alpha_0)^{-1}\sum_{i=1}^{n}$ $S(z_i,x_i;\alpha_0)+o_p(1).$
It follows that
\begin{equation}\label{OBSnoX-betaAsy-weight}
w(z_i,x_i;\hat{\alpha})-w(z_i,x_i;\alpha_0)=n^{-1}\nabla_{\alpha}w(z_i,x_i;\alpha_0)^{\T}I(\alpha_0)^{-1}\sum_{i=1}^{n}S(z_i,x_i;\alpha_0)+o_p(n^{-1/2}).
\end{equation}

Substituting \eqref{OBSnoX-betaAsy-weight} into the definition of $S^{\hat{w}}_{\smash{yy}}$ gives
\begin{equation*}
\begin{aligned}
S^{\hat{w}}_{\smash{yy}}&=n^{-1}\sum_{i=1}^{n}w(z_i,x_i;\alpha_0)y_iy_i^{\T}+n^{-1}\sum_{i=1}^{n}\big\{w(z_i,x_i;\hat{\alpha})-w(z_i,x_i;\alpha_0)\big\}y_iy_i^{\T}\\
&=E\{w(Z,X;\alpha_0)YY^{\T}\}+\bigg\{n^{-1}\sum_{i=1}^{n}\nabla_{\alpha}w(z_i,x_i;\alpha_0)^{\T}I(\alpha_0)^{-1}y_iy_i^{\T}\bigg\}
\bigg\{n^{-1}\sum_{i=1}^{n}S(z_i,x_i;\alpha_0)\bigg\}\\
&\quad+o_p(n^{-1/2}).
\end{aligned}
\end{equation*}
By the LLN, $n^{-1}\sum_{i=1}^{n}\nabla_{\alpha}w(z_i,x_i;\alpha_0)^{\T}I(\alpha_0)^{-1}y_iy_i^{\T}=E\{\nabla_{\alpha}w(Z,X;\alpha_0)^{\T}I(\alpha_0)^{-1}YY^{\T}\}$, and by the CLT, $n^{-1/2}\sum_{i=1}^{n}S(z_i,x_i;\alpha_0)=O_p(1)$. Therefore, $S^{\hat{w}}_{\smash{yy}}=E(WYY^{\T})+O_p(n^{-1/2})$, where $W=w(Z,X;\alpha_0)$.
Applying the same argument yields $S^{\hat{w}}_{\smash{y1}}=E(WY)+O_p(n^{-1/2})$ and $S^{\hat{w}}_{\smash{11}}=E(W)+O_p(n^{-1/2})$.
Combining these expansions with Slutsky's theorem gives
$$\phi_{yy}^{\hat{w}}=S^{\hat{w}}_{\smash{yy}}-S^{\hat{w}}_{\smash{y1}}(S^{\hat{w}}_{\smash{11}})^{-1}S^{\hat{w}}_{\smash{1y}}=E(WYY^{\T})-E(WY)E(W)^{-1}E(WY^{\T})+o_p(1)=\Phi_{\textsc{yy}}^{\textsc{w}}+o_p(1).$$
We now consider the term $S^{\hat{w}}_{\smash{y\varsigma}}-S^{\hat{w}}_{\smash{y1}}(S^{\hat{w}}_{\smash{11}})^{-1}S^{\hat{w}}_{\smash{1\varsigma}}$. Write
\begin{equation*}
\begin{aligned}
S^{\hat{w}}_{\smash{y\varsigma}}-S^{\hat{w}}_{\smash{y1}}(S^{\hat{w}}_{\smash{11}})^{-1}S^{\hat{w}}_{\smash{1\varsigma}}
=S^{\hat{w}}_{\smash{y\varsigma}}-E(WY)E(W)^{-1}S^{\hat{w}}_{\smash{1\varsigma}}
-\big\{S^{\hat{w}}_{\smash{y1}}(S^{\hat{w}}_{\smash{11}})^{-1}-E(WY)E(W)^{-1}\big\}S^{\hat{w}}_{\smash{1\varsigma}}.
\end{aligned}
\end{equation*}
Applying the same arguments used for $S^{\hat{w}}_{\smash{yy}}$ yields $S^{\hat{w}}_{\smash{1\varsigma}}=E(W\varsigma)+O_p(n^{-1/2})$, where $\varsigma=Z-\beta_{\textup{0}}-\beta^{\T}Y$ denotes the population residual from the WLS regression of $Z$ on $(1, Y)$ with weights $W$. Because the population WLS residual satisfies
$E(W\varsigma)=0$, then $S^{\hat{w}}_{\smash{1\varsigma}}=O_p(n^{-1/2})$. Moreover, $(S^{\hat{w}}_{\smash{11}})^{-1}S^{\hat{w}}_{\smash{y1}}-E(W)^{-1}E(WY)=o_p(1)$. Consequently,
\begin{equation*}
S^{\hat{w}}_{\smash{y\varsigma}}-S^{\hat{w}}_{\smash{y1}}(S^{\hat{w}}_{\smash{11}})^{-1}S^{\hat{w}}_{\smash{1\varsigma}}
=n^{-1}\sum_{i=1}^{n}\hat{w}_i\varsigma_i\big\{y_i-E(WY)E(W)^{-1}\big\}+o_p(n^{-1/2}).
\end{equation*}

Substituting \eqref{OBSnoX-betaAsy-weight} into
\begin{equation}\label{OBSnoX-betaAsy-resexpand}
\begin{aligned}
&S^{\hat{w}}_{\smash{y\varsigma}}-S^{\hat{w}}_{\smash{y1}}(S^{\hat{w}}_{\smash{11}})^{-1}S^{\hat{w}}_{\smash{1\varsigma}}\\
&=n^{-1}\sum_{i=1}^{n}w(z_i,x_i;\alpha_0)\varsigma_i\{y_i-E(WY)E(W)^{-1}\}\\
&\quad+n^{-1}\sum_{i=1}^{n}\varsigma_i\left\{y_i-E(WY)E(W)^{-1}\right\}\nabla_{\alpha}w(z_i,x_i;\alpha_0)^{\T}I(\alpha_0)^{-1}\Big\{n^{-1}\sum_{i=1}^{n}S(z_i,x_i;\alpha_0)\Big\}+o_p(n^{-1/2})\\
&=n^{-1}\sum_{i=1}^{n}\varsigma_iw(z_i,x_i;\alpha_0)\big\{y_i-E(WY)E(W)^{-1}\big\}\\
&\quad+n^{-1}\sum_{i=1}^{n}E\big[\varsigma\big\{Y-E(WY)E(W)^{-1}\big\}\nabla_{\alpha}w(Z,X;\alpha_0)^{\T}\big]I(\alpha_0)^{-1}S(z_i,x_i;\alpha_0)+o_p(n^{-1/2}),
\end{aligned}
\end{equation}
where the second equality follows from the LLN. By the CLT, we obtain
$$n^{1/2}\big\{S^{\hat{w}}_{\smash{y\varsigma}}-S^{\hat{w}}_{\smash{y1}}(S^{\hat{w}}_{\smash{11}})^{-1}S^{\hat{w}}_{\smash{1\varsigma}}\big\}\stackrel{d}\rightarrow N\big(0, E\{\psi_{\textsc{y}{\scriptscriptstyle\varsigma}}(Z,X,Y)\psi_{\textsc{y}{\scriptscriptstyle\varsigma}}(Z,X,Y)^{\T}\}\big),$$
where
\begin{equation*}
\begin{aligned}
\psi_{\textsc{y}{\scriptscriptstyle\varsigma}}(Z,X,Y)&=\varsigma_iw(Z,X;\alpha_0)\big\{Y-E(WY)E(W)^{-1}\big\}\\
&\quad+E\big[\varsigma\big\{Y-E(WY)E(W)^{-1}\big\}\nabla_{\alpha}w(Z,X;\alpha_0)^{\T}\big]I(\alpha_0)^{-1}S(Z,X;\alpha_0).
\end{aligned}
\end{equation*}
Finally, $S^{\hat{w}}_{\smash{yy}}-S^{\hat{w}}_{\smash{y1}}(S^{\hat{w}}_{\smash{11}})^{-1}S^{\hat{w}}_{\smash{1y}}=\Phi_{\textsc{yy}}^{\textsc{w}}+o_p(1)$ and Slutsky's theorem  yields the stated result.

\subsection{Proof of Theorem 2}
\label{subsec::thm2-proof}

Recall that $\hat{\beta}_{\textup{\textsc{os}}}=\phi_{zz}^{\hat{w}}\big(\phi_{yy}^{\hat{w}}\big)^{-1}
\hat{\tau}_{\textup{\textsc{os}}}$ and $\beta=\Phi_{\textsc{zz}}^{\textsc{w}}\big(\Phi_{\textsc{yy}}^{\textsc{w}}\big)^{-1}\tau$.
By the proof of  Lemma \ref{AsyDis-Beta-OBS},  $\hat{\beta}_{\textsc{os}}=\beta+(\phi_{yy}^{\hat{w}})^{-1}
\{S^{\hat{w}}_{\smash{y\varsigma}}-S^{\hat{w}}_{\smash{y1}}(S^{\hat{w}}_{\smash{11}})^{-1}S^{\hat{w}}_{\smash{1\varsigma}}\}$. Substituting this expression into $\hat{\beta}_{\textsc{os}}^{\T}\hat{\tau}_{\textsc{os}}-\beta^{\T}\tau$ yields
\begin{equation*}
\begin{aligned}
\hat{\beta}_{\textsc{os}}^{\T}\hat{\tau}_{\textsc{os}}-\beta^{\T}\tau&
=\beta^{\T}
\{(\phi_{zz}^{\hat{w}})^{-1}\phi_{yy}^{\hat{w}}-(\Phi_{\textsc{zz}}^{\textsc{w}})^{-1}\Phi_{\textsc{yy}}^{\textsc{w}}\}\beta
+2\beta^{\T}(\phi_{zz}^{\hat{w}})^{-1}\{S^{\hat{w}}_{\smash{y\varsigma}}-S^{\hat{w}}_{\smash{y1}}(S^{\hat{w}}_{\smash{11}})^{-1}S^{\hat{w}}_{\smash{1\varsigma}}\}\\
&\quad+(\phi_{zz}^{\hat{w}})^{-1}\{S^{\hat{w}}_{\smash{y\varsigma}}-S^{\hat{w}}_{\smash{y1}}(S^{\hat{w}}_{\smash{11}})^{-1}S^{\hat{w}}_{\smash{1\varsigma}}\}^{\T}
(\phi_{yy}^{\hat{w}})^{-1}\{S^{\hat{w}}_{\smash{y\varsigma}}-S^{\hat{w}}_{\smash{y1}}(S^{\hat{w}}_{\smash{11}})^{-1}S^{\hat{w}}_{\smash{1\varsigma}}\},
\end{aligned}
\end{equation*}
where $\Phi_{\textsc{zz}}^{\textsc{w}}=1/2.$ We derive the asymptotic distribution of $\hat{\beta}_{\textsc{os}}^{\T}\hat{\tau}_{\textsc{os}}-\beta^{\T}\tau$  in two cases: (i) $\tau\neq0$ and (ii) $\tau=0$.

\vspace{0.2cm}
\noindent(i) $\tau\neq0$. For the first term in the expansion,
\begin{equation*}
\begin{aligned}
\beta^{\T}
\big\{(\phi_{zz}^{\hat{w}})^{-1}\phi_{yy}^{\hat{w}}-(\Phi_{\textsc{zz}}^{\textsc{w}})^{-1}\Phi_{\textsc{yy}}^{\textsc{w}}\big\}\beta
=(\Phi_{\textsc{zz}}^{\textsc{w}})^{-1}\beta^{\T}\big\{(\phi_{yy}^{\hat{w}}-\Phi_{\textsc{yy}}^{\textsc{w}})\beta
-(\phi_{zz}^{\hat{w}}-\Phi_{\textsc{zz}}^{\textsc{w}})\Phi_{\textsc{yy}}^{\textsc{w}}\beta/\Phi_{\textsc{zz}}^{\textsc{w}})\big\}\times\{1+o_p(1)\},
\end{aligned}
\end{equation*}
where the second equality follows from $\phi_{zz}^{\hat{w}}=\Phi_{\textsc{zz}}^{\textsc{w}}+o_p(1)$.
We rewrite the first component as
\begin{equation*}
\begin{aligned}
\phi_{yy}^{\hat{w}}-\Phi_{\textsc{yy}}^{\textsc{w}}&=n^{-1}\sum_{i=1}^{n}\hat{w}_i\big\{y_i-(S^{\hat{w}}_{\smash{11}})^{-1}S^{\hat{w}}_{\smash{y1}}\big\}
\big\{y_i-(S^{\hat{w}}_{\smash{11}})^{-1}S^{\hat{w}}_{\smash{y1}}\big\}^{\T}-\Phi_{\textsc{yy}}^{\textsc{w}}\\
&=n^{-1}\sum_{i=1}^{n}\big[\hat{w}_i\{y_i-E(W)^{-1}E(WY)\}
\{y_i-E(W)^{-1}E(WY)\}^{\T}-\Phi_{\textsc{yy}}^{\textsc{w}}\big]\\
&\quad-S^{\hat{w}}_{\smash{11}}\big\{(S^{\hat{w}}_{\smash{11}})^{-1}S^{\hat{w}}_{\smash{y1}}-E(W)^{-1}E(WY)\big\}
\big\{(S^{\hat{w}}_{\smash{11}})^{-1}S^{\hat{w}}_{\smash{y1}}-E(W)^{-1}E(WY)\big\}^{\T}.
\end{aligned}
\end{equation*}
The proof of Lemma \ref{AsyDis-Beta-OBS} implies that $S^{\hat{w}}_{\smash{y1}}=E(WY)+O_p(n^{-1/2})$ and $S^{\hat{w}}_{\smash{11}}=E(W)+O_p(n^{-1/2})$. It follows that $n^{1/2}\{(S^{\hat{w}}_{\smash{11}})^{-1}S^{\hat{w}}_{\smash{y1}}-E(W)^{-1}E(WY)\}=O_p(1)$. Consequently, the quadratic remainder term is $o_p(n^{-1/2})$ and
\begin{equation*}
\begin{aligned}
\phi_{yy}^{\hat{w}}-\Phi_{\textsc{yy}}^{\textsc{w}}=n^{-1}\sum_{i=1}^{n}\big[\hat{w}_i\{y_i-E(W)^{-1}E(WY)\}
\{y_i-E(W)^{-1}E(WY)\}^{\T}-\Phi_{\textsc{yy}}^{\textsc{w}}\big]+o_p(n^{-1/2}).
\end{aligned}
\end{equation*}
Then, we have
\begin{equation*}
\begin{aligned}
&(\Phi_{\textsc{zz}}^{\textsc{w}})^{-1}\beta^{\T}(\phi_{yy}^{\hat{w}}-\Phi_{\textsc{yy}}^{\textsc{w}})\beta\\
&=2n^{-1}\beta^{\T}\sum_{i=1}^{n}\Big[w(z_i,x_i;\alpha_0)\big\{y_i-E(W)^{-1}E(WY)\big\}
\big\{y_i-E(W)^{-1}E(WY)\big\}^{\T}-\Phi_{\textsc{yy}}^{\textsc{w}}\Big]\beta\\
&\quad+2n^{-1}\beta^{\T}\sum_{i=1}^{n}\{w(z_i,x_i;\hat\alpha)-w(z_i,x_i;\alpha_0)\}\big\{y_i-E(W)^{-1}E(WY)\big\}
\big\{y_i-E(W)^{-1}E(WY)\big\}^{\T}\beta\\
&\quad+o_p(n^{-1/2})\\
&=2n^{-1}\beta^{\T}\sum_{i=1}^{n}\Big[w(z_i,x_i;\alpha_0)\big\{y_i-E(W)^{-1}E(WY)\big\}
\big\{y_i-E(W)^{-1}E(WY)\big\}^{\T}-\Phi_{\textsc{yy}}^{\textsc{w}}\Big]\beta\\
&\quad+2n^{-1}\beta^{\T}\sum_{i=1}^{n}E\Big[\big\{Y-E(W)^{-1}E(WY)\big\}
\big\{Y-E(W)^{-1}E(WY)\big\}^{\T}\beta\nabla_{\alpha}w(Z,X;\alpha_0)^{\T}I(\alpha_0)^{-1}\Big]\\
&\hspace{2.5cm}\times S(z_i,x_i;\alpha_0)+o_p(n^{-1/2}).
\end{aligned}
\end{equation*}
Applying the same argument to $\phi_{zz}^{\hat{w}}-\Phi_{\textsc{zz}}^{\textsc{w}}$ gives
\begin{equation*}
\begin{aligned}
&(\Phi_{\textsc{zz}}^{\textsc{w}})^{-2}\beta^{\T}(\phi_{zz}^{\hat{w}}-\Phi_{\textsc{zz}}^{\textsc{w}})\Phi_{\textsc{yy}}^{\textsc{w}}\beta\\
&=4n^{-1}\beta^{\T}\sum_{i=1}^{n}\big[w(z_i,x_i;\alpha_0)
\{z_i-E(W)^{-1}E(WZ)\}^2-\Phi_{\textsc{zz}}^{\textsc{w}}\big]\Phi_{\textsc{yy}}^{\textsc{w}}\beta\\
&\quad+4n^{-1}\beta^{\T}\sum_{i=1}^{n}E\big[\{Z-E(W)^{-1}E(WZ)\}^2
\Phi_{\textsc{yy}}^{\textsc{w}}\beta\nabla_{\alpha}w(Z,X;\alpha_0)^{\T}I(\alpha_0)^{-1}\big]S(z_i,x_i;\alpha_0)+o_p(n^{-1/2})\\
&=n^{-1}\beta^{\T}\sum_{i=1}^{n}\big[\{w(z_i,x_i;\alpha_0)
-2\}+E\{\nabla_{\alpha}w(Z,X;\alpha_0)^{\T}\}I(\alpha_0)^{-1}S(z_i,x_i;\alpha_0)\big]\Phi_{\textsc{yy}}^{\textsc{w}}\beta+o_p(n^{-1/2}).\hspace{3cm}
\end{aligned}
\end{equation*}

Next, consider the second term in the expansion of $\hat{\beta}_{\textsc{os}}^{\T}\hat{\tau}_{\textsc{os}}-\beta^{\T}\tau$. Using \eqref{OBSnoX-betaAsy-resexpand}, we write
\begin{equation*}
\begin{aligned}
&2\beta^{\T}(\phi_{zz}^{\hat{w}})^{-1}\big\{S^{\hat{w}}_{\smash{y\varsigma}}-S^{\hat{w}}_{\smash{y1}}(S^{\hat{w}}_{\smash{11}})^{-1}S^{\hat{w}}_{\smash{1\varsigma}}\big\}\\
&=4n^{-1}\sum_{i=1}^{n}\beta^{\T}w(z_i,x_i;\alpha_0)\varsigma_i\{y_i-E(W)^{-1}E(WY)\}\\
&\quad+4n^{-1}\sum_{i=1}^{n}\beta^{\T}E\big[\varsigma\{Y-E(W)^{-1}E(WY)\}\nabla_{\alpha}w(Z,X;\alpha_0)^{\T}\big]I(\alpha_0)^{-1}S(z_i,x_i;\alpha_0) +o_p(n^{-1/2}).
\end{aligned}
\end{equation*}
Moreover, since $n^{1/2}\{S^{\hat{w}}_{\smash{y\varsigma}}-S^{\hat{w}}_{\smash{y1}}(S^{\hat{w}}_{\smash{11}})^{-1}S^{\hat{w}}_{\smash{1\varsigma}}\}=O_p(1)$, the third term in the expansion of $\hat{\beta}_{\textsc{os}}^{\T}\hat{\tau}_{\textsc{os}}-\beta^{\T}\tau$ is $O_p(n^{-1})$ and therefore asymptotically negligible.
Combining all terms yields
\begin{equation*}
\begin{aligned}
&n^{1/2}(\hat{\beta}_{\textsc{os}}^{\T}\hat{\tau}_{\textsc{os}}-\beta^{\T}\tau)\\ &=n^{-1/2}\beta^{\T}\sum_{i=1}^{n}\bigg\{2\Big(\big[w(z_i,x_i;\alpha_0)\{y_i-E(W)^{-1}E(WY)\}\{y_i-E(W)^{-1}E(WY)\}^{\T}-\Phi_{\textsc{yy}}^{\textsc{w}}\big]\beta\\
&\hspace{3.3cm}+E\big[\{Y-E(W)^{-1}E(WY)\}\{Y-E(W)^{-1}E(WY)\}^{\T}\beta\nabla_{\alpha}w(Z,X;\alpha_0)^{\T}\big]\\
&\hspace{3.7cm}\times I(\alpha_0)^{-1}S(z_i,x_i;\alpha_0)\Big)\\
&\hspace{2.7cm}-\big[\{w(z_i,x_i;\alpha_0)-2\}+E\{\nabla_{\alpha}w(Z,X;\alpha_0)^{\T}\}I(\alpha_0)^{-1}S(z_i,x_i;\alpha_0)\big]\Phi_{\textsc{yy}}^{\textsc{w}}\beta\\
&\hspace{2.7cm}+4\Big(w(z_i,x_i;\alpha_0)\varsigma_i\{y_i-E(WY)E(W)^{-1}\}\\
&\hspace{3.2cm}+E\big[\varsigma\{Y-E(WY)E(W)^{-1}\}\nabla_{\alpha}w(Z,X;\alpha_0)^{\T}\big]I(\alpha_0)^{-1}S(z_i,x_i;\alpha_0)\Big)\bigg\}+o_p(1).
\end{aligned}
\end{equation*}
By the multivariate CLT and the delta method, we have
$$n^{1/2}\big(\hat{\tau}_{\textsc{os}}^{\textup{c}}-\tau^{\textup{c}}\big)\stackrel{d}\rightarrow N\big(0, 4\beta^\T\textup{cov}\{R_{\textup{\textsc{os}}}(Z,X,Y)\}\beta\big),$$
where $R_{\textup{\textsc{os}}}(Z,X,Y)$ is defined in the theorem.

\vspace{0.2cm}
\noindent(ii) $\tau=0$.  In this case, the first two terms in the decomposition vanish and
\begin{equation*}
\begin{aligned}
n(\hat{\beta}_{\textsc{os}}^{\T}\hat{\tau}_{\textsc{os}}-\beta^{\T}\tau)&=n(\phi_{zz}^{\hat{w}})^{-1}\big\{S^{\hat{w}}_{\smash{y\varsigma}}
-S^{\hat{w}}_{\smash{y1}}(S^{\hat{w}}_{\smash{11}})^{-1}S^{\hat{w}}_{\smash{1\varsigma}}\big\}^{\T}
(\phi_{yy}^{\hat{w}})^{-1}\big\{S^{\hat{w}}_{\smash{y\varsigma}}-S^{\hat{w}}_{\smash{y1}}(S^{\hat{w}}_{\smash{11}})^{-1}S^{\hat{w}}_{\smash{1\varsigma}}\big\}.
\end{aligned}
\end{equation*}
Define $M_{\textsc{os}}=n^{1/2}(\phi_{zz}^{\hat{w}})^{-1/2}(\phi_{yy}^{\hat{w}})^{-1/2}\{S^{\hat{w}}_{\smash{y\varsigma}}-S^{\hat{w}}_{\smash{y1}}
(S^{\hat{w}}_{\smash{11}})^{-1}S^{\hat{w}}_{\smash{1\varsigma}}\}$. Then $n(\hat{\beta}_{\textsc{os}}^{\T}\hat{\tau}_{\textsc{os}}-\beta^{\T}\tau)=M_{\textsc{os}}^{\T}M_{\textsc{os}}$.
It follows from the proof of Lemma \ref{AsyDis-Beta-OBS} that $\phi_{zz}^{\hat{w}}=\Phi_{\textsc{zz}}^{\textsc{w}}+o_p(1)$, $\phi_{yy}^{\hat{w}}=\Phi_{\textsc{yy}}^{\textsc{w}}+o_p(1)$, and
$$n^{1/2}\big\{S^{\hat{w}}_{\smash{y\varsigma}}-S^{\hat{w}}_{\smash{y1}}(S^{\hat{w}}_{\smash{11}})^{-1}S^{\hat{w}}_{\smash{1\varsigma}}\big\}\stackrel{d}\rightarrow N\big(0, E\{\psi_{\textsc{y}{\scriptscriptstyle\varsigma}}(Z,X,Y)\psi_{\textsc{y}{\scriptscriptstyle\varsigma}}(Z,X,Y)^{\T}\}\big).$$
The LLN and Slutsky's theorem therefore imply $M_{\textsc{os}}\stackrel{d}\rightarrow N(0, \Gamma_{\textsc{os}})$,
where
$$\Gamma_{\textsc{os}}=(\Phi_{\textsc{zz}}^{\textsc{w}})^{-1}(\Phi_{\textsc{yy}}^{\textsc{w}})^{-1/2} E\{\psi_{\textsc{y}{\scriptscriptstyle\varsigma}}(Z,X,Y)\psi_{\textsc{y}{\scriptscriptstyle\varsigma}}(Z,X,Y)^{\T}\}(\Phi_{\textsc{yy}}^{\textsc{w}})^{-1/2}.$$
Let
$\Gamma_{\textsc{os}}=Q_{\textsc{os}}^{\T}\Delta_{\textsc{os}} Q_{\textsc{os}},$
where $Q_{\textsc{os}}$ is an orthogonal matrix and $\Delta_{\textsc{os}}=\text{diag}(\lambda_{{\textsc{os}}1},\ldots,\lambda_{{\textsc{os}}L})$ is a diagonal matrix. Define $B_{\textsc{os}}=\Delta_{\textsc{os}}^{-1/2}Q_{\textsc{os}}M_{\textsc{os}}=(B_{{\textsc{os}}1},\ldots, B_{{\textsc{os}}L})^{\T}$. Then  $B_{\textsc{os}}\stackrel{d}\rightarrow N(0, I_L)$. Moreover,
$M_{\textsc{os}}^{\T}M_{\textsc{os}}=(Q_{\textsc{os}}^{\T}\Delta_{\textsc{os}}^{1/2}B_{\textsc{os}})^{\T}(Q_{\textsc{os}}^{\T}\Delta_{\textsc{os}}^{1/2}B_{\textsc{os}})
=\sum_{\ell=1}^{L}\lambda_{{\textsc{os}}\ell}B_{{\textsc{os}}\ell}^2$. Therefore,
\begin{equation*}
n\hat{\beta}_{\textsc{os}}^{\T}\hat{\tau}_{\textsc{os}}=B_{\textsc{os}}^{\T}\Delta_{\textsc{os}} B_{\textsc{os}}=\sum_{\ell=1}^{L}\lambda_{{\textsc{os}}\ell}B_{{\textsc{os}}\ell}^2
\stackrel{d}\rightarrow\sum_{\ell=1}^{L}\lambda_{{\textsc{os}}\ell}\chi^2_{\ell}(1).
\end{equation*}

\section{Proofs of results in Sections \ref{sec::CRE-OBS-withX}-\ref{sec::logistic} }
\label{sec::secSupp-proof}

\subsection{Proof of Proposition \ref{Equiv-CR-withX}}\label{subsec-propS1-proof}

(i) A direct calculation shows that
\begin{equation*}
\check{y}^{\T}\Lambda z=y^{\T}\Lambda\check{z}
=y^{\T}\Lambda z-(y^{\T}\Lambda\mathring{x})(\mathring{x}^{\T}\Lambda\mathring{x})^{-1}
(\mathring{x}^{\T}\Lambda z).
\end{equation*}
Comparing the expressions of $\hat{\beta}_{\textsc{a}}$ and $\hat{\tau}_{\textsc{a}}$ yields $\hat{\beta}_{\textup{\textsc{a}}}=(\check{z}^{\T}\Lambda\check{z})(\check{y}^{\T}\Lambda\check{y})^{-1}\hat{\tau}_{\textup{\textsc{a}}}$.

We next further characterize the relationship between $\hat{\beta}_{\textsc{a}}$ and $\hat{\tau}_{\textsc{a}}$. We derive an explicit expression for $(\check{z}^{\T}\Lambda\check{z})(\check{y}^{\T}\Lambda\check{y})^{-1}$ for generic weights $w_i$. Recall that
$\bar{y}_w(1)=\sum_{i=1}^{n}w_iz_iy_{i}/(\sum_{i=1}^{n}w_iz_i)$, $\bar{y}_w(0)=\sum_{i=1}^{n}w_i(1-z_i)y_{i}/\{\sum_{i=1}^{n}w_i(1-z_i)\}$, $\hat{\textup{cov}}_{w}\{Y(1)\}=n^{-1}\sum_{i=1}^{n}w_iz_i\{y_i-\bar{y}_{w}(1)\}\{y_i-\bar{y}_{w}(1)\}^{\T}$,  and $\hat{\textup{cov}}_{w}\{Y(0)\}=n^{-1}\sum_{i=1}^{n}w_i(1-z_i)\{y_i-\bar{y}_{w}(0)\}\{y_i-\bar{y}_{w}(0)\}^{\T}$. We similarly define
$\bar{x}_{w1}=\sum_{i=1}^{n}w_iz_ix_{i}/(\sum_{i=1}^{n}w_iz_i)$, $\bar{x}_{w0}=\sum_{i=1}^{n}w_i(1-z_i)x_{i}/\{\sum_{i=1}^{n}w_i(1-z_i)\}$, $\hat{\textup{cov}}_{w1}(X)=$ $n^{-1}\sum_{i=1}^{n}w_iz_i(x_i-\bar{x}_{w1})(x_i-\bar{x}_{w1})^{\T}$, and $\hat{\textup{cov}}_{w0}(X)=n^{-1}\sum_{i=1}^{n}w_i(1-z_i)(x_i-\bar{x}_{w0})(x_i-\bar{x}_{w0})^{\T}$. Morevoer, we define $\hat{\textup{cov}}_{w}\{Y(1),X\}=n^{-1}\sum_{i=1}^{n}w_iz_i\{y_i-\bar{y}_w(1)\}(x_i-\bar{x}_{w1})^{\T}$,
$\hat{\textup{cov}}_{w}\{Y(0), X\}=n^{-1}\sum_{i=1}^{n}w_i(1-z_i)\{y_i-\bar{y}_w(0)\}(x_i-\bar{x}_{w0})^{\T}$.

We decompose $\check{y}^{\T}\Lambda \check{y}$ as
\begin{equation*}
\begin{aligned}
\check{y}^{\T}\Lambda \check{y}&=\check{y}^{\T}\Lambda \check{z}(\check{z}^{\T}\Lambda \check{z})^{-1}\check{z}^{\T}\Lambda \check{y}
+\check{y}^{\T}\Lambda\{I_{n}-\check{z}(\check{z}^{\T}\Lambda \check{z})^{-1}\check{z}^{\T}\Lambda\}\check{y}\\
&=(\check{z}^{\T}\Lambda \check{z})\hat{\tau}_{\textsc{a}}\hat{\tau}_{\textsc{a}}^{\T}+\check{y}^{\T}\Lambda\{I_{n}-\check{z}(\check{z}^{\T}\Lambda \check{z})^{-1}\check{z}^{\T}\Lambda\}\check{y}.
\end{aligned}
\end{equation*}
Next we analyze $\check{y}^{\T}\Lambda\{I_{n}-\check{z}(\check{z}^{\T}\Lambda \check{z})^{-1}\check{z}^{\T}\Lambda\}\check{y}$.  By straightforward algebra, we have
\begin{equation*}
\begin{aligned}
n^{-1}\check{y}^{\T}\Lambda \check{y}&=n^{-1}y^{\T}\Lambda y-(n^{-1}y^{\T}\Lambda\mathring{x})(n^{-1}\mathring{x}^{\T}\Lambda\mathring{x})^{-1}
(n^{-1}\mathring{x}^{\T}\Lambda y)\\
&=\{S_{yy}^w-S_{y1}^w(S_{11}^w)^{-1}S_{1y}^w\}\\
&\quad-\{S_{yx}^w-S_{y1}^w(S_{11}^w)^{-1}S_{1x}^w\}\{S_{xx}^w-S_{x1}^w(S_{11}^w)^{-1}S_{1x}^w\}^{-1}\{S_{xy}^w-S_{x1}^w(S_{11}^w)^{-1}S_{1y}^w\}\\
&=\phi_{yy}^{w}-\phi_{yx}^{w}(\phi_{xx}^{w})^{-1}\phi_{xy}^{w}.
\end{aligned}
\end{equation*}
$n^{-1}\check{z}^{\T}\Lambda \check{z}=\phi_{zz}^{w}-\phi_{zx}^{w}(\phi_{xx}^{w})^{-1}\phi_{xz}^{w}$ and
$n^{-1}y^{\T}\Lambda \check{z}=\phi_{yz}^{w}-\phi_{yx}^{w}(\phi_{xx}^{w})^{-1}\phi_{xz}^{w}$.
Combining these results yields
\begin{equation*}
\begin{aligned}
\check{y}^{\T}\Lambda\big\{I_{n}-\check{z}(\check{z}^{\T}\Lambda \check{z})^{-1}\check{z}^{\T}\Lambda\big\}\check{y}
=\phi_{yy}^{w}-(\phi_{yx}^{w}, \phi_{yz}^{w})\left(\begin{matrix}
\phi_{xx}^{w}  & \phi_{xz}^{w}\\
\phi_{zx}^{w} & \phi_{zz}^{w}
\end{matrix}\right)^{-1}\left(\begin{matrix}
\phi_{xy}^{w}\\
\phi_{zy}^{w}
\end{matrix}\right).
\end{aligned}
\end{equation*}

On the other hand, recall from the proof of Proposition 1 that
\begin{equation*}
\begin{aligned}
\hat{\textup{cov}}_{w}\{Y(1)\}&=n^{-1}\sum_{i=1}^{n}w_iz_iy_iy_i^{\T}-S^{w}_{\smash{yz}}(S^{w}_{\smash{1z}})^{-1}S^{w}_{\smash{zy}},\\
\hat{\textup{cov}}_{w}\{Y(0)\}&=n^{-1}\sum_{i=1}^{n}w_i(1-z_i)y_iy_i^{\T}-(S^{w}_{\smash{y1}}-S^{w}_{\smash{yz}})
(S^{w}_{\smash{11}}-S^{w}_{\smash{1z}})^{-1}(S^{w}_{\smash{1y}}-S^{w}_{\smash{zy}}).
\end{aligned}
\end{equation*}
Adding these two expressions yields
\begin{equation*}
\begin{aligned}
\hat{\textup{cov}}_{w}\{Y(1)\}+\hat{\textup{cov}}_{w}\{Y(0)\}&=S^{w}_{\smash{yy}}-S^{w}_{\smash{yz}}(S^{w}_{\smash{1z}})^{-1}S^{w}_{\smash{zy}}
-(S^{w}_{\smash{y1}}-S^{w}_{\smash{yz}})(S^{w}_{\smash{11}}-S^{w}_{\smash{1z}})^{-1}(S^{w}_{\smash{1y}}-S^{w}_{\smash{zy}})\\
&=\phi_{yy}^{w}-\phi_{yz}^{w}(\phi_{zz}^{w})^{-1}\phi_{zy}^{w}.
\end{aligned}
\end{equation*}
Similarly,
\begin{equation*}
\begin{aligned}
\hat{\textup{cov}}_{w1}(X)+\hat{\textup{cov}}_{w0}(X)&=S^{w}_{\smash{xx}}-S^{w}_{\smash{xz}}(S^{w}_{\smash{1z}})^{-1}S^{w}_{\smash{zx}}
-(S^{w}_{\smash{x1}}-S^{w}_{\smash{xz}})(S^{w}_{\smash{11}}-S^{w}_{\smash{1z}})^{-1}(S^{w}_{\smash{1x}}-S^{w}_{\smash{zx}})\\
&=\phi_{xx}^{w}-\phi_{xz}^{w}(\phi_{zz}^{w})^{-1}\phi_{zx}^{w}.
\end{aligned}
\end{equation*}
and
\begin{equation*}
\begin{aligned}
\hat{\textup{cov}}_{w}\{Y(1), X\}+\hat{\textup{cov}}_{w}\{Y(0), X\}&=S^{w}_{\smash{yx}}-S^{w}_{\smash{yz}}(S^{w}_{\smash{1z}})^{-1}S^{w}_{\smash{zx}}
-(S^{w}_{\smash{y1}}-S^{w}_{\smash{yz}})(S^{w}_{\smash{11}}-S^{w}_{\smash{1z}})^{-1}(S^{w}_{\smash{1x}}-S^{w}_{\smash{zx}})\\
&=\phi_{yx}^{w}-\phi_{yz}^{w}(\phi_{zz}^{w})^{-1}\phi_{zx}^{w}.
\end{aligned}
\end{equation*}
Combining these expressions, together with
\[ \begin{aligned} &\left( \begin{matrix} \phi_{xx}^{w} & \phi_{xz}^{w}\\ \phi_{zx}^{w} & \phi_{zz}^{w} \end{matrix} \right)^{-1}\\
&\quad= (\phi_{zz}^{w})^{-1}\left( \begin{matrix} \phi_{zz}^{w}\{\phi_{xx}^{w}-\phi_{xz}^{w}(\phi_{zz}^{w})^{-1}\phi_{zx}^{w}\}^{-1} & -\{\phi_{xx}^{w}-\phi_{xz}^{w}(\phi_{zz}^{w})^{-1}\phi_{zx}^{w}\}^{-1}\phi_{xz}^{w} \\[2mm] -\phi_{zx}^{w}\{\phi_{xx}^{w}-\phi_{xz}^{w}(\phi_{zz}^{w})^{-1}\phi_{zx}^{w}\}^{-1} & 1 + \phi_{zx}^{w} \{\phi_{xx}^{w}-\phi_{xz}^{w}(\phi_{zz}^{w})^{-1}\phi_{zx}^{w}\}^{-1} \phi_{xz}^{w}
 (\phi_{zz}^{w})^{-1}\end{matrix} \right), \end{aligned} \]
we obtain
\begin{equation*}
\begin{aligned}
&\hat{\textup{cov}}_{w}\{Y(1)\}+\hat{\textup{cov}}_{w}\{Y(0)\}-[\hat{\textup{cov}}_{w}\{Y(1), X\}+\hat{\textup{cov}}_{w}\{Y(0), X\}]
[\hat{\textup{cov}}_{w1}(X)+\hat{\textup{cov}}_{w0}(X)]^{-1}\\
&\hspace{4.5cm}\times
[\hat{\textup{cov}}_{w}\{Y(1), X\}+\hat{\textup{cov}}_{w}\{Y(0), X\}]^{\T}\\
&=\phi_{yy}^{w}-(\phi_{yx}^{w}, \phi_{yz}^{w})\left(\begin{matrix}
\phi_{xx}^{w}  & \phi_{xz}^{w}\\
\phi_{zx}^{w} & \phi_{zz}^{w}
\end{matrix}\right)^{-1}\left(\begin{matrix}
\phi_{xy}^{w}\\
\phi_{zy}^{w}
\end{matrix}\right).
\end{aligned}
\end{equation*}
It follows that
\begin{equation*}
\begin{aligned}
n^{-1}\check{y}^{\T}\Lambda\{I_{n}-\check{z}(\check{z}^{\T}\Lambda \check{z})^{-1}\check{z}^{\T}\Lambda\}\check{y}
=(\check{z}^{\T}\Lambda \check{z})\hat{\Sigma}_{\textsc{a}}.
\end{aligned}
\end{equation*}
Thus,
$(\check{z}^{\T}\Lambda \check{z})^{-1}\check{y}^{\T}\Lambda \check{y}=\hat{\Sigma}_{\textsc{a}}+\hat{\tau}_{\textsc{a}}\hat{\tau}_{\textsc{a}}^{\T}$ and
$\hat{\beta}_{\textsc{a}}=\big(\hat{\Sigma}_{\textsc{a}}+\hat{\tau}_{\textup{\textsc{a}}}\hat{\tau}_{\textup{\textsc{a}}}^{\T}\big)^{-1}\hat{\tau}_{\textup{\textsc{a}}}.$
Applying Lemma \ref{Woodburyformula} to $A=\hat{\Sigma}_{\textsc{a}}$, $\mu=\hat{\tau}_{\textsc{a}}$, and $\nu=\hat{\tau}_{\textsc{a}}$ gives
$$\hat{\beta}_{\textup{\textsc{a}}}
=\frac{\hat{\Sigma}_{\textsc{a}}^{-1}\hat{\tau}_{\textsc{a}}}{1+\hat{\tau}_{\textsc{a}}^{\T}\Sigma_{\textsc{a}}^{-1}\hat{\tau}_{\textsc{a}}}.$$
Under the weights $w_i=z_i/p+(1-z_i)/(1-p)$, the weighted means $\bar{y}_w(1)$, $\bar{y}_w(0)$  $\bar{x}_{w1}$, and $\bar{x}_{w0}$ reduce to $\bar{y}(1)$, $\bar{y}(0)$, $\bar{x}_1$, and $\bar{x}_0$. Moreover, the weighted covariance matrices $\hat{\textup{cov}}_{w}\{Y(1)\}$,  $\hat{\textup{cov}}_{w}\{Y(0)\}$, $\hat{\textup{cov}}_{w1}(X)$, $\hat{\textup{cov}}_{w0}(X)$, $\hat{\textup{cov}}_{w}\{Y(1), X\}$, and $\hat{\textup{cov}}_{w}\{Y(0), X\}$ reduce to $\hat{\textup{cov}}\{Y(1)\}$, $\hat{\textup{cov}}\{Y(0)\}$, $\hat{\textup{cov}}_{1}(X)$, $\hat{\textup{cov}}_{0}(X)$, $\hat{\textup{cov}}\{Y(1), X\}$, and $\hat{\textup{cov}}\{Y(0), X\}$, respectively. Substituting these quantities into the expression of $\hat{\beta}$ yields the result stated in the proposition.

\vspace{0.2cm}
\noindent(ii) Substituting $\check{Y}=Y-E(WY\mathring{X}^{\T})E(W \mathring{X}\mathring{X}^{\T})^{-1}\mathring{X}$ into $E(W\check{Y}Z)$ gives
\begin{equation*}
\begin{aligned}
E(W\check{Y}Z)=E(WYZ)-E(WY\mathring{X}^{\T})E(W\mathring{X}\mathring{X}^{\T})^{-1}E(W\mathring{X}Z).
\end{aligned}
\end{equation*}
Straightforward algebra yields $E(WYZ)=E\{Y(1)\}$, $E(WY\mathring{X}^{\T})=(E\{Y(1)+Y(0)\}, E[\{Y(1)+Y(0)\}X^{\T}])$, and
$E(W\mathring{X}Z)=(1, E(X^{\T}))^{\T}$. Furthermore, applying the inverse formula of a $2\times2$  block matrix gives
\begin{equation*}
E(W\mathring{X}\mathring{X}^{\T})^{-1}=\frac{1}{2}\left(\begin{matrix}
1+E(X^{\T})\textup{cov}(X)^{-1}E(X) \quad & -E(X^{\T})\textup{cov}(X)^{-1}\\
-\textup{cov}(X)^{-1}E(X) \quad& \textup{cov}(X)^{-1}
\end{matrix}\right).
\end{equation*}
Combining these identities yields
$E(W\check{Y}Z)=E\{Y(1)-Y(0)\}/2=\tau/2.$
Moreover, since $E(WZ^2)=1$, we have
$$E(W\check{Z}^2)=E(WZ^2)-E(WZ\mathring{X}^{\T})E(W\mathring{X}\mathring{X}^{\T})^{-1}E(WZ\mathring{X})=1/2.$$
It follows that $\tau=E(W\check{Z}^2)^{-1}E(W\check{Y}Z)$. Hence, $\beta_{\textup{\textsc{a}}}=E(W\check{Z}^2)E(W\check{Y}\check{Y}^{\T})^{-1}\tau
$.

Substituting $\check{Y}=Y-E(WY\mathring{X}^{\T})E(W \mathring{X}\mathring{X}^{\T})^{-1}\mathring{X}$ into $E(W\check{Y}\check{Y}^{\T})$ gives
$$E(W\check{Y}\check{Y}^{\T})=E(WYY^{\T})-E(WY\mathring{X}^{\T})E(W\mathring{X}\mathring{X}^{\T})^{-1}E(W\mathring{X}Y^{\T}).$$
Note that $E(WYY^{\T})=E\{Y(1)Y(1)^{\T}+Y(0)Y(0)^{\T}\}$ and $E(WY\mathring{X}^{\T})=E[\{Y(1)+Y(0)\}\mathring{X}^{\T}]$. Thus,
\begin{equation*}
\begin{aligned}
&2E(WY\mathring{X}^{\T})E(W\mathring{X}\mathring{X}^{\T})^{-1}E(W\mathring{X}Y^{\T})\\
&=E\{Y(1)+Y(0)\}E\{Y(1)^{\T}+Y(0)^{\T}\}
+\textup{cov}\{Y(1)+Y(0), X\}\textup{cov}(X)^{-1}\textup{cov}\{X, Y(1)+Y(0)\}.
\end{aligned}
\end{equation*}
Combining these identities into $\Phi_{\textsc{yy}}^{\textsc{w}}$ yields
\begin{equation*}
\begin{aligned}
E(W\check{Z}^2)^{-1}E(W\check{Y}\check{Y}^{\T})=2E(W\check{Y}\check{Y}^{\T})=\Sigma_{\textsc{a}}+\tau\tau^{\T}.
\end{aligned}
\end{equation*}
Then $\beta_{\textup{\textsc{a}}}=(\Sigma_{\textsc{a}}+\tau\tau^{\T})^{-1}\tau.$
Applying Lemma \ref{Woodburyformula} to $A=\Sigma_{\textsc{a}}$, $\mu=\tau$, and $\nu=\tau$, we obtain
\begin{equation*}
\begin{aligned}
\beta_{\textup{\textsc{a}}}&=\Big(\Sigma_{\textsc{a}}^{-1}-\frac{\Sigma_{\textsc{a}}^{-1}\tau\tau^{\T}\Sigma_{\textsc{a}}^{-1}}{1+\tau^{\T}\Sigma_{\textsc{a}}^{-1}\tau}\Big)\tau
=\frac{\Sigma_{\textsc{a}}^{-1}\tau}{1+\tau^{\T}\Sigma_{\textsc{a}}^{-1}\tau}.
\end{aligned}
\end{equation*}
This completes the proof.

\subsection{Proof of Theorem \ref{AsyDis-BetaTau-CR-withX}}\label{subsec-thmS2-proof}

Let $(\beta_{\textsc{a}\mathring{\textsc{x}}},\beta_{\textsc{a}})$ denote the coefficient vector from the WLS regression of $Z$ on $(\mathring{X}, Y)$, that is,
$(\beta_{\textsc{a}\mathring{\textsc{x}}},\beta_{\textsc{a}})=\argmin_{b_{\mathring{\textsc{x}}},b} E\{W(Z-b_{\mathring{\textsc{x}}}^{\T}X-b^{\T}Y)^2\}$.
Define $\varepsilon_{\textsc{a}i}=z_i-\beta_{\textsc{a}\mathring{\textsc{x}}}^{\T}\mathring{x}_i-\beta_{\textsc{a}}^{\T}y_i$ and $\varepsilon_{\textsc{a}}^{\textup{v}}=(\varepsilon_{\textsc{a}1},\ldots, \varepsilon_{\textsc{a}n})^{\T}$. Substituting $z=\varepsilon_{\textsc{a}}^{\textup{v}}+\mathring{x}\beta_{\textsc{a}\mathring{\textsc{x}}}+y\beta_{\textsc{a}}$
into the expressions of $\hat{\beta}_{\textsc{a}}$ and $\hat{\tau}_{\textsc{a}}$ yields
\begin{equation*}
\hat{\beta}_{\textup{\textsc{a}}}=\beta_{\textsc{a}}+(\check{y}^{\T}\Lambda\check{y})^{-1}
(\check{y}^{\T}\Lambda \varepsilon_{\textsc{a}}^{\textup{v}}).
\end{equation*}
Recall that $\hat{\beta}_{\textup{\textsc{a}}}=(\check{z}^{\T}\Lambda\check{z})(\check{y}^{\T}\Lambda\check{y})^{-1}\hat{\tau}_{\textup{\textsc{a}}}$
and $\beta_{\textup{\textsc{a}}}=E(W\check{Z}^2)E(W\check{Y}\check{Y}^{\T})^{-1}\tau$. Therefore,  we have the decomposition:
\begin{equation*}
\begin{aligned}
\hat{\beta}_{\textsc{a}}^{\T}\hat{\tau}_{\textsc{a}}-\beta_{\textsc{a}}^{\T}\tau
&=\beta_{\textsc{a}}^{\T}\big\{(n^{-1}\check{z}^{\T}\Lambda\check{z})^{-1}(n^{-1}\check{y}^{\T}\Lambda\check{y})-E(W\check{Z}^2)^{-1}E(W\check{Y}\check{Y}^{\T})\big\}\beta_{\textsc{a}}
+2(n^{-1}\check{z}^{\T}\Lambda\check{z})^{-1}\beta_{\textsc{a}}^{\T}(n^{-1}\check{y}^{\T}\Lambda \varepsilon_{\textsc{a}}^{\textup{v}})\\
&\quad+(n^{-1}\check{z}^{\T}\Lambda\check{z})^{-1}(n^{-1}\check{y}^{\T}\Lambda \varepsilon_{\textsc{a}}^{\textup{v}})^{\T}(n^{-1}\check{y}^{\T}\Lambda\check{y})^{-1}(n^{-1}\check{y}^{\T}\Lambda \varepsilon_{\textsc{a}}^{\textup{v}}).
\end{aligned}
\end{equation*}
In what follows, we derive the asymptotic distribution of $\hat{\beta}_{\textsc{a}}^{\T}\hat{\tau}_{\textsc{a}}-\beta_{\textsc{a}}^{\T}\tau$ separately in two cases: (i) $\tau\neq0$ and (ii) $\tau=0$.

\vspace{0.2cm}
\noindent(i) $\tau\neq0$. We begin by analyzing the first term in the expansion of $\hat{\beta}_{\textsc{a}}^{\T}\hat{\tau}_{\textsc{a}}-\beta_{\textsc{a}}^{\T}\tau$. Recall that $\check{z}=z-\mathring{x}(\mathring{x}^{\T}\Lambda\mathring{x})^{-1}(\mathring{x}^{\T}\Lambda z)$. We obtain
\begin{equation*}
\begin{aligned}
n^{-1}\check{z}^{\T}\Lambda\check{z}
&=n^{-1}z^{\T}\Lambda z-(n^{-1}z^{\T}\Lambda\mathring{x})(n^{-1}\mathring{x}^{\T}\Lambda\mathring{x})^{-1}
(n^{-1}\mathring{x}^{\T}\Lambda z)\\
&=n^{-1}\sum_{i=1}^{n}w_iz_i^2-\Big(n^{-1}\sum_{i=1}^{n}w_iz_i\mathring{x}_i^{\T}\Big)\Big(n^{-1}\sum_{i=1}^{n}w_i\mathring{x}_i^{\T}\mathring{x}_i\Big)^{-1}
\Big(n^{-1}\sum_{i=1}^{n}w_iz_i\mathring{x}_i\Big).
\end{aligned}
\end{equation*}
By the LLN, $n^{-1}\check{z}^{\T}\Lambda\check{z}=E(WZ^2)-E(WZ\mathring{X}^{\T})E(W\mathring{X}\mathring{X}^{\T})^{-1}E(WZ\mathring{X})+o_p(1)=E(W\check{Z}^2)+o_p(1).$
Consequently,
\begin{equation*}
\begin{aligned}
&\beta_{\textsc{a}}^{\T}\big\{(n^{-1}\check{z}^{\T}\Lambda\check{z})^{-1}(n^{-1}\check{y}^{\T}\Lambda\check{y})-E(W\check{Z}^2)^{-1}E(W\check{Y}\check{Y}^{\T})\big\}\beta_{\textsc{a}}\\
&=E(W\check{Z}^2)^{-1}\beta_{\textsc{a}}^{\T}\big[\{n^{-1}\check{y}^{\T}\Lambda\check{y}-E(W\check{Y}\check{Y}^{\T})\}
-\{n^{-1}\check{z}^{\T}\Lambda\check{z}-E(W\check{Z}^2)\}E(W\check{Y}\check{Y}^{\T})/E(W\check{Z}^2)\big]\beta_{\textsc{a}}\\
&\quad\times\{1+o_p(1)\}.
\end{aligned}
\end{equation*}
To derive an asymptotic linear representation of this expression, we next consider $n^{-1}\check{y}^{\T}\Lambda\check{y}$. Write
\begin{equation*}
\begin{aligned}
n^{-1}\check{y}^{\T}\Lambda\check{y}
&=n^{-1}\sum_{i=1}^{n}w_iy_iy_i^{\T}-\Big(n^{-1}\sum_{i=1}^{n}w_iy_i\mathring{x}_i^{\T}\Big)\Big(n^{-1}\sum_{i=1}^{n}w_i\mathring{x}_i^{\T}\mathring{x}_i\Big)^{-1}
\Big(n^{-1}\sum_{i=1}^{n}w_i\mathring{x}_iy_i^{\T}\Big)\\
&=n^{-1}\sum_{i=1}^{n}w_i\big\{y_i-(n^{-1}y^{\T}\Lambda\mathring{x})(n^{-1}\mathring{x}^{\T}\Lambda\mathring{x})^{-1}\mathring{x}_i\big\}
\big\{y_i-(n^{-1}y^{\T}\Lambda\mathring{x})(n^{-1}\mathring{x}^{\T}\Lambda\mathring{x})^{-1}\mathring{x}_i\big\}^{\T}\\
&=n^{-1}\sum_{i=1}^{n}w_i\big\{y_i-E(WY\mathring{X}^{\T})E(W\mathring{X}\mathring{X}^{\T})^{-1}\mathring{x}_i\big\}
\big\{y_i-E(WY\mathring{X}^{\T})E(W\mathring{X}\mathring{X}^{\T})^{-1}\mathring{x}_i\big\}^{\T}\\
&\quad-\big\{(n^{-1}y^{\T}\Lambda\mathring{x})(n^{-1}\mathring{x}^{\T}\Lambda\mathring{x})^{-1}-E(WY\mathring{X}^{\T})E(W\mathring{X}\mathring{X}^{\T})^{-1}\big\}(n^{-1}\mathring{x}^{\T}\Lambda\mathring{x})
\\
&\qquad\times\big\{(n^{-1}y^{\T}\Lambda\mathring{x})(n^{-1}\mathring{x}^{\T}\Lambda\mathring{x})^{-1}-E(WY\mathring{X}^{\T})E(W\mathring{X}\mathring{X}^{\T})^{-1}\big\}^{\T}.
\end{aligned}
\end{equation*}
The LLN implies that $n^{-1}y^{\T}\Lambda\mathring{x}=E(WY\mathring{X}^{\T})+o_p(1)$ and $n^{-1}\mathring{x}^{\T}\Lambda\mathring{x}=E(W\mathring{X}\mathring{X}^{\T})+o_p(1)$, while the CLT implies that
$n^{1/2}\{n^{-1}y^{\T}\Lambda\mathring{x}-E(WY\mathring{X}^{\T})\}=O_p(1)$ and $n^{1/2}\{n^{-1}\mathring{x}^{\T}\Lambda\mathring{x}-E(W\mathring{X}\mathring{X}^{\T})\}=O_p(1)$. Hence, $(n^{-1}y^{\T}\Lambda\mathring{x})(n^{-1}\mathring{x}^{\T}\Lambda\mathring{x})^{-1}-E(WY\mathring{X}^{\T})E(W\mathring{X}\mathring{X}^{\T})^{-1}=O_p(n^{-1/2})$. It follows that
\begin{equation*}
\begin{aligned}
&n^{1/2}\big\{n^{-1}\check{y}^{\T}\Lambda\check{y}-E(W\check{Y}\check{Y}^{\T})\big\}\\
&=n^{-1/2}\sum_{i=1}^{n}\big[w_i\big\{y_i-E(WY\mathring{X}^{\T})E(W\mathring{X}\mathring{X}^{\T})^{-1}\mathring{x}_i\big\}
\big\{y_i-E(WY\mathring{X}^{\T})E(W\mathring{X}\mathring{X}^{\T})^{-1}\mathring{x}_i\big\}^{\T}
-E(W\check{Y}\check{Y}^{\T})\big]\\
&\quad+o_p(1).
\end{aligned}
\end{equation*}
Applying the same argument, we can obtain similar expansion for $n^{1/2}\{n^{-1}\check{z}^{\T}\Lambda\check{z}-E(W\check{Z}^2)\}$.
Combining the two expansions, we obtain
\begin{equation*}
\begin{aligned}
&n^{1/2}\beta_{\textsc{a}}^{\T}\big\{(n^{-1}\check{z}^{\T}\Lambda\check{z})^{-1}(n^{-1}\check{y}^{\T}\Lambda\check{y})-E(W\check{Z}^2)^{-1}E(W\check{Y}\check{Y}^{\T})\big\}\beta_{\textsc{a}}\\
&=\frac{n^{-1/2}\beta_{\textsc{a}}^{\T}}{E(W\check{Z}^2)}\sum_{i=1}^{n}\bigg(\Big[w_i\big\{y_i-E(WY\mathring{X}^{\T})E(W\mathring{X}\mathring{X}^{\T})^{-1}\mathring{x}_i\big\}
\big\{y_i-E(WY\mathring{X}^{\T})E(W\mathring{X}\mathring{X}^{\T})^{-1}\mathring{x}_i\big\}^{\T}
-E(W\check{Y}\check{Y}^{\T})\Big]\\
&\hspace{2.8cm}-\Big[w_i\big\{z_i-E(WZ\mathring{X}^{\T})E(W\mathring{X}\mathring{X}^{\T})^{-1}\mathring{x}_i\big\}^2
-E(W\check{Z}^2)\Big]\frac{E(W\check{Y}\check{Y}^{\T})}{E(W\check{Z}^2)}\bigg)\beta_{\textsc{a}}+o_p(1).
\end{aligned}
\end{equation*}

We next analyze the second term in the expansion of $\hat{\beta}_{\textsc{a}}^{\T}\hat{\tau}_{\textsc{a}}-\beta_{\textsc{a}}^{\T}\tau$. We observe that
\begin{equation}\label{CRwithX-asy-term2}
\begin{aligned}
n^{-1}\check{y}^{\T}\Lambda \varepsilon_{\textsc{a}}^{\textup{v}}
&=n^{-1}\sum_{i=1}^{n}w_i\varepsilon_{\textsc{a}i}\big\{y_i-E(WY\mathring{X}^{\T})E(W\mathring{X}\mathring{X}^{\T})^{-1}\mathring{x}_i\big\}\\
&\quad-\big\{(n^{-1}y^{\T}\Lambda\mathring{x})(n^{-1}\mathring{x}^{\T}\Lambda\mathring{x})^{-1}-E(WY\mathring{X}^{\T})E(W\mathring{X}\mathring{X}^{\T})^{-1}\big\}\Big(n^{-1}\sum_{i=1}^{n}w_i\varepsilon_{\textsc{a}i}\Big)\\
&=n^{-1}\sum_{i=1}^{n}w_i\varepsilon_{\textsc{a}i}\big\{y_i-E(WY\mathring{X}^{\T})E(W\mathring{X}\mathring{X}^{\T})^{-1}\mathring{x}_i\big\}+o_p(n^{-1/2}).
\end{aligned}
\end{equation}
The second equality follows because $n^{-1}\sum_{i=1}^{n}w_i\varepsilon_{\textsc{a}i}=O_p(n^{-1/2})$ by the CLT,  and
$(n^{-1}y^{\T}\Lambda\mathring{x})(n^{-1}\mathring{x}^{\T}\Lambda\mathring{x})^{-1}-E(WY\mathring{X}^{\T})E(W\mathring{X}\mathring{X}^{\T})^{-1}=o_p(1)$ by the LLN and Slutsky's theorem. Moreover, since $n^{-1}\check{z}^{\T}\Lambda\check{z}=E(W\check{Z}^2)+o_p(1)$, we have
\begin{equation*}
\begin{aligned}
2(n^{-1}\check{z}^{\T}\Lambda\check{z})^{-1}\beta_{\textsc{a}}^{\T}(n^{-1}\check{y}^{\T}\Lambda \varepsilon_{\textsc{a}}^{\textup{v}})
=\frac{2\beta_{\textsc{a}}^{\T}}{nE(W\check{Z}^2)}\sum_{i=1}^{n}w_i\varepsilon_{\textsc{a}i}\big\{y_i-E(WY\mathring{X}^{\T})E(W\mathring{X}\mathring{X}^{\T})^{-1}\mathring{x}_i\big\}
+o_p(n^{-1/2}).
\end{aligned}
\end{equation*}

Finally, we show that the third term is asymptotically negligible. By Equation \eqref{CRwithX-asy-term2} and the CLT,  $n^{-1}\check{y}^{\T}\Lambda \varepsilon_{\textsc{a}}^{\textup{v}}=O_p(n^{-1/2})$. Consequently,
$$(n^{-1}\check{z}^{\T}\Lambda\check{z})^{-1}(n^{-1}\check{y}^{\T}\Lambda \varepsilon_{\textsc{a}}^{\textup{v}})^{\T}(n^{-1}\check{y}^{\T}\Lambda\check{y})^{-1}(n^{-1}\check{y}^{\T}\Lambda \varepsilon_{\textsc{a}}^{\textup{v}})=O_p(n^{-1}).$$
Combining the asymptotic linear representation of the three terms yields
\begin{equation*}
\begin{aligned}
&n^{1/2}(\hat{\beta}_{\textsc{a}}^{\T}\hat{\tau}_{\textsc{a}}-\beta_{\textsc{a}}^{\T}\tau)\\ &=\frac{n^{-1/2}\beta_{\textsc{a}}^{\T}}{E(W\check{Z}^2)}\sum_{i=1}^{n}\bigg(\Big[w_i\big\{y_i-E(WY\mathring{X}^{\T})E(W\mathring{X}\mathring{X}^{\T})^{-1}\mathring{x}_i\big\}
\big\{y_i-E(WY\mathring{X}^{\T})E(W\mathring{X}\mathring{X}^{\T})^{-1}\mathring{x}_i\big\}^{\T}\\
&\hspace{3.3cm}
-E(W\check{Y}\check{Y}^{\T})\Big]\beta_{\textsc{a}}\\
&\hspace{3cm}-\Big[w_i\big\{z_i-E(WZ\mathring{X}^{\T})E(W\mathring{X}\mathring{X}^{\T})^{-1}\mathring{x}_i\big\}^2
-E(W\check{Z}^2)\Big]\frac{E(W\check{Y}\check{Y}^{\T})\beta_{\textsc{a}}}{E(W\check{Z}^2)}\\
&\hspace{3cm}+2w_i\varepsilon_{\textsc{a}i}\big\{y_i-E(WY\mathring{X}^{\T})E(W\mathring{X}\mathring{X}^{\T})^{-1}\mathring{x}_i\big\}\bigg)+o_p(1).
\end{aligned}
\end{equation*}
By the multivariate CLT and the delta method, we obtain
$$n^{1/2}(\hat{\tau}_{\textup{\textsc{a}}}^{\textup{c}}-\tau_{\textup{\textsc{a}}}^{\textup{c}})\rightarrow N\big(0, 4\beta_{\textup{\textsc{a}}}^\T\textup{cov}\{R_{\textup{\textsc{a}}}(Z,X,Y)\}\beta_{\textup{\textsc{a}}}\big),$$
where
\begin{equation*}
\begin{aligned}
R_{\textup{\textsc{a}}}(Z,X,Y)&=\big[W\{Y-E(WY\mathring{X}^{\T})E(W \mathring{X}\mathring{X}^{\T})^{-1}\mathring{X}\}\{Y-E(WY\mathring{X}^{\T})E(W \mathring{X}\mathring{X}^{\T})^{-1}\mathring{X}\}^{\T}-E(W\check{Y}\check{Y}^{\T})\big]\beta_{\textup{\textsc{a}}}\\
&\quad-\big[W\{Z-E(WZ\mathring{X}^{\T})E(W \mathring{X}\mathring{X}^{\T})^{-1}\mathring{X}\}^2
-E(W\check{Z}^2)\big]E(W\check{Y}\check{Y}^{\T})\beta_{\textsc{a}}/E(W\check{Z}^2)\\
&\quad+2W\varepsilon_{\textsc{a}} \{Y-E(WY\mathring{X}^{\T})E(W \mathring{X}\mathring{X}^{\T})^{-1}\mathring{X}\}.
\end{aligned}
\end{equation*}
Using $E(W\check{Z}^2)=1/2$ and $E(WZ\mathring{X}^{\T})E(W \mathring{X}\mathring{X}^{\T})^{-1}\mathring{X}=1/2$, we further simplify $R_{\textup{\textsc{a}}}(Z,X,Y)$ to the expression stated in the theorem.

\vspace{0.2cm}
\noindent(ii) $\tau=0$. In this case, the expression of $\hat{\beta}_{\textsc{a}}^{\T}\hat{\tau}_{\textsc{a}}-\beta_{\textsc{a}}^{\T}\tau$ simplifies to
\begin{equation*}
\begin{aligned} \hat{\beta}_{\textsc{a}}^{\T}\hat{\tau}_{\textsc{a}}-\beta_{\textsc{a}}^{\T}\tau=(n^{-1}\check{z}^{\T}\Lambda\check{z})^{-1}(n^{-1}\check{y}^{\T}\Lambda \varepsilon_{\textsc{a}}^{\textup{v}})^{\T}(n^{-1}\check{y}^{\T}\Lambda\check{y})^{-1}(n^{-1}\check{y}^{\T}\Lambda \varepsilon_{\textsc{a}}^{\textup{v}}).
\end{aligned}
\end{equation*}
Define $M_{\textsc{a}}=(n^{-1}\check{z}^{\T}\Lambda\check{z})^{-1/2}(n^{-1}\check{y}^{\T}\Lambda\check{y})^{-1/2}
(n^{-1/2}\check{y}^{\T}\Lambda \varepsilon_{\textsc{a}}^{\textup{v}}).$
Then $n(\hat{\beta}_{\textsc{a}}^{\T}\hat{\tau}_{\textsc{a}}-\beta_{\textsc{a}}^{\T}\tau)=M_{\textsc{a}}^{\T}M_{\textsc{a}}$.
Applying the CLT to \eqref{CRwithX-asy-term2} yields
$$n^{-1/2}\check{y}^{\T}\Lambda \varepsilon_{\textsc{a}}^{\textup{v}}\stackrel{d}\rightarrow N\big(0,E(W^2\varepsilon_{\textup{\textsc{a}}}^2\check{Y}\check{Y}^{\T})\big).$$
Since $n^{-1}\check{z}^{\T}\Lambda\check{z}= E(W\check{Z}^2)+o_p(1)$ and $n^{-1}\check{y}^{\T}\Lambda\check{y}=E(W\check{Y}\check{Y}^{\T})+o_p(1)$, by Slutsky's theorem, we obtain
$M_{\textsc{a}}
\stackrel{d}\rightarrow N(0, \Gamma_{\textup{\textsc{a}}})$, where
$\Gamma_{\textup{\textsc{a}}}=E(W\check{Z}^2)^{-1}E(W\check{Y}\check{Y}^{\T})^{-1/2} E(W^2\varepsilon_{\textup{\textsc{a}}}^2\check{Y}\check{Y}^{\T})
E(W\check{Y}\check{Y}^{\T})^{-1/2}.$
Let
$\Gamma_{\textup{\textsc{a}}}=Q_{\textsc{a}}^{\T}\Delta_{\textsc{a}} Q_{\textsc{a}}$ be the eigendecomposition of $\Gamma_{\textup{\textsc{a}}}$, where
$\Delta_{\textsc{a}}=\diag\{\lambda_{\textsc{a}1},\ldots,\lambda_{\textsc{a}L}\}$ contains the eigenvalues of $\Gamma_{\textup{\textsc{a}}}$ and
$Q_{\textsc{a}}$ is the orthogonal matrix whose columns are the corresponding eigenvectors.
Define $B_{\textsc{a}}=\Delta_{\textsc{a}}^{-1/2}Q_{\textsc{a}}M_{\textsc{a}}=(B_{\textsc{a}1},\dots, B_{\textsc{a}L})$. Then $B_{\textsc{a}}\stackrel{d}\rightarrow N(0, I_L)$. Therefore,
$n\hat{\beta}_{\textsc{a}}^{\T}\hat{\tau}_{\textsc{a}}=M_{\textsc{a}}^{\T}M_{\textsc{a}}= B_{\textsc{a}}^{\T}\Delta_{\textsc{a}}B_{\textsc{a}}=\sum_{\ell=1}^{L}\lambda_{{\textsc{a}}\ell}B_{{\textsc{a}}\ell}^2\stackrel{d}\rightarrow\sum_{\ell=1}^{L}\lambda_{{\textsc{a}}\ell}\chi^2_{\ell}(1).
$

\subsection{Proof of Proposition \ref{SREnoX-equiv}}
\label{subsec-propS3-proof}

\noindent(i) Recall that $\hat{\beta}_{\textsc{sr}}=(\tilde{y}_{\textsc{sr}}^{\T}\Lambda_{\textsc{sr}}\tilde{y}_{\textsc{sr}})^{-1}
\tilde{y}_{\textsc{sr}}^{\T}\Lambda_{\textsc{sr}}z$ and $\hat{\tau}_{\textsc{sr}}=(\tilde{z}_{\textsc{sr}}^{\T}\Lambda_{\textsc{sr}}\tilde{z}_{\textsc{sr}})^{-1}
y^{\T}\Lambda_{\textsc{sr}}\tilde{z}_{\textsc{sr}}$, where
$\tilde{y}_{\textsc{sr}}=y-g(g^{\T}\Lambda_{\textsc{sr}}g)^{-1}(g^{\T}\Lambda_{\textsc{sr}}y)$ and $\tilde{z}_{\textsc{sr}}=z-g(g^{\T}\Lambda_{\textsc{sr}}g)^{-1}(g^{\T}\Lambda_{\textsc{sr}}z)$. A direct calculation shows that
$$\tilde{y}_{\textsc{sr}}^{\T}\Lambda_{\textsc{sr}}z=y^{\T}\Lambda_{\textsc{sr}}\tilde{z}_{\textsc{sr}}=y^{\T}\Lambda_{\textsc{sr}}z-(y^{\T}\Lambda_{\textsc{sr}}g)(g^{\T}\Lambda_{\textsc{sr}}g)^{-1}(g^{\T}\Lambda_{\textsc{sr}}z).$$
Therefore,
$\hat{\beta}_{\textup{\textsc{sr}}}=(\tilde{z}_{\textup{\textsc{sr}}}^{\T}\Lambda_{\textup{\textsc{sr}}}\tilde{z}_{\textup{\textsc{sr}}})
(\tilde{y}_{\textup{\textsc{sr}}}^{\T}\Lambda_{\textup{\textsc{sr}}}\tilde{y}_{\textup{\textsc{sr}}})^{-1}\hat{\tau}_{\textup{\textsc{sr}}}.
$

Next, we derive a more explicit characterization of the relationship between $\hat{\beta}_{\textsc{sr}}$ and $\hat{\tau}_{\textsc{sr}}$. Substituting the expressions of $\tilde{z}$ and $\tilde{y}$, we obtain
\begin{equation*}
\begin{aligned}
\tilde{z}_{\textup{\textsc{sr}}}^{\T}\Lambda_{\textup{\textsc{sr}}}\tilde{z}_{\textup{\textsc{sr}}}&=z^{\T}\Lambda_{\textsc{sr}}z-(z^{\T}\Lambda_{\textsc{sr}}g)
(g^{\T}\Lambda_{\textsc{sr}}g)^{-1}(g^{\T}
\Lambda_{\textsc{sr}}z),\\
\tilde{y}_{\textup{\textsc{sr}}}^{\T}\Lambda_{\textup{\textsc{sr}}}\tilde{y}_{\textup{\textsc{sr}}}&=y^{\T}\Lambda_{\textsc{sr}}y-(y^{\T}\Lambda_{\textsc{sr}}g)
(g^{\T}\Lambda_{\textsc{sr}}g)^{-1}(g^{\T}
\Lambda_{\textsc{sr}}y).
\end{aligned}
\end{equation*}
We observe that
\begin{equation*}
\begin{aligned}
g^{\T}\Lambda_{\textsc{sr}}g=\sum_{i=1}^{n}w_{\textsc{sr}i}g_ig_i^{\T}=\diag\bigg\{\sum_{i=1}^{n}w_{\textsc{sr}i}I(c_i=1),\ldots, \sum_{i=1}^{n}w_{\textsc{sr}i}I(c_i=S)\bigg\}.
\end{aligned}
\end{equation*}
By straightforward algebra, we obtain $\sum_{i=1}^{n}w_{\textsc{sr}i}I(c_i=s)=n_{1s}(1-p_{[s]})+n_{0s}p_{[s]}$. Substituting this identity into the expression of $\tilde{z}_{\textup{\textsc{sr}}}^{\T}\Lambda_{\textup{\textsc{sr}}}\tilde{z}_{\textup{\textsc{sr}}}$ yields
\begin{equation*}
\begin{aligned}
\tilde{z}_{\textup{\textsc{sr}}}^{\T}\Lambda_{\textup{\textsc{sr}}}\tilde{z}_{\textup{\textsc{sr}}}
&=\sum_{i=1}^{n}w_{\textsc{sr}i}z_i^2-\Big(\sum_{i=1}^{n}w_{\textsc{sr}i}z_ig_i^{\T}\Big)\Big(\sum_{i=1}^{n}w_{\textsc{sr}i}g_ig_i^{\T}\Big)^{-1}
\Big(\sum_{i=1}^{n}w_{\textsc{sr}i}z_ig_i\Big)\\
&=\sum_{i=1}^{n}w_{\textsc{sr}i}z_i^2-\sum_{s=1}^{S}\bigg\{\sum_{i=1}^{n}\sum_{j=1}^{n}w_{\textsc{sr}i}w_{\textsc{sr}j}z_iz_jI(c_i=s)I(c_j=s)\bigg\}/\bigg\{\sum_{i=1}^{n}w_{\textsc{sr}i}I(c_i=s)\bigg\}\\
&=\sum_{s=1}^Sn_{1s}n_{0s}p_{[s]}(1-p_{[s]})/\{n_{1s}(1-p_{[s]})+n_{0s}p_{[s]}\}.
\end{aligned}
\end{equation*}
For notational convenience, define $\kappa_s=n_{1s}n_{0s}p_{[s]}(1-p_{[s]})/\{n_{1s}(1-p_{[s]})+n_{0s}p_{[s]}\}$. Then, $\tilde{z}_{\textup{\textsc{sr}}}^{\T}\Lambda_{\textup{\textsc{sr}}}\tilde{z}_{\textup{\textsc{sr}}}=\sum_{s=1}^{S}\kappa_s.$
Applying the same decomposition to $\tilde{y}_{\textup{\textsc{sr}}}^{\T}\Lambda_{\textup{\textsc{sr}}}\tilde{y}_{\textup{\textsc{sr}}}$ gives
\begin{equation*}
\begin{aligned}
\tilde{y}_{\textup{\textsc{sr}}}^{\T}\Lambda_{\textup{\textsc{sr}}}\tilde{y}_{\textup{\textsc{sr}}}
&=\sum_{s=1}^S\bigg[(1-p_{[s]})\sum_{i=1}^nz_iI(c_i=s)\{y_i-\bar{y}_{[s]}(1)\}\{y_i-\bar{y}_{[s]}(1)\}^{\T}\\
&\quad\qquad+p_{[s]}\sum_{i=1}^n(1-z_i)I(c_i=s)\{y_i-\bar{y}_{[s]}(0)\}\{y_i-\bar{y}_{[s]}(0)\}^{\T}
+\kappa_s\hat{\tau}_{[s]}\hat{\tau}_{[s]}^{\T}\bigg].
\end{aligned}
\end{equation*}
Similarly,
$\tilde{y}_{\textup{\textsc{sr}}}^{\T}\Lambda_{\textup{\textsc{sr}}}\tilde{z}_{\textup{\textsc{sr}}}=\sum_{s=1}^{S}\kappa_s\hat{\tau}_{[s]}$.
Consequently,
$\hat\tau_{\textsc{sr}}=(\sum_{s=1}^{S}\kappa_s)^{-1}(\sum_{s=1}^{S}\kappa_s\hat{\tau}_{[s]})$. Furthermore,
\begin{equation*}
\begin{aligned}
\frac{\sum_{s=1}^S\kappa_s(\hat{\tau}_{[s]}-\hat\tau_{\textsc{sr}})(\hat{\tau}_{[s]}-\hat\tau_{\textsc{sr}})^{\T}}{\sum_{s=1}^{S}\kappa_s}
=\frac{\sum_{s=1}^S\kappa_s\hat{\tau}_{[s]}\hat{\tau}_{[s]}^{\T}}{\sum_{s=1}^{S}\kappa_s}-\hat\tau_{\textsc{sr}}\hat\tau_{\textsc{sr}}^{\T}.
\end{aligned}
\end{equation*}
Combining this identity with the preceding expressions of $\tilde{z}_{\textup{\textsc{sr}}}^{\T}\Lambda_{\textup{\textsc{sr}}}\tilde{z}_{\textup{\textsc{sr}}}$
and $\tilde{y}_{\textup{\textsc{sr}}}^{\T}\Lambda_{\textup{\textsc{sr}}}\tilde{y}_{\textup{\textsc{sr}}}$ yields
\begin{equation*}
\begin{aligned}
(\tilde{z}_{\textup{\textsc{sr}}}^{\T}\Lambda_{\textup{\textsc{sr}}}\tilde{z}_{\textup{\textsc{sr}}})
(\tilde{y}_{\textup{\textsc{sr}}}^{\T}\Lambda_{\textup{\textsc{sr}}}\tilde{y}_{\textup{\textsc{sr}}})^{-1}=(\hat{\Sigma}_{\textsc{sr}}+\hat\tau_{\textsc{sr}}\hat\tau_{\textsc{sr}}^{\T})^{-1}.
\end{aligned}
\end{equation*}
Applying  Lemma \ref{Woodburyformula} to $A=\hat\Sigma_{\textsc{sr}}$, $\mu=\hat\tau_{\textsc{sr}}$, and $\nu=\hat\tau_{\textsc{sr}}$ gives
	\begin{equation*}
		\begin{aligned}
(\tilde{z}_{\textup{\textsc{sr}}}^{\T}\Lambda_{\textup{\textsc{sr}}}\tilde{z}_{\textup{\textsc{sr}}})
(\tilde{y}_{\textup{\textsc{sr}}}^{\T}\Lambda_{\textup{\textsc{sr}}}\tilde{y}_{\textup{\textsc{sr}}})^{-1}=\hat\Sigma_{\textsc{sr}}^{-1}-\frac{\hat\Sigma_{\textsc{sr}}^{-1}\hat\tau_{\textsc{sr}}
\hat\tau_{\textsc{sr}}^{\T}\hat\Sigma_{\textsc{sr}}^{-1}}{1+\hat\tau_{\textsc{sr}}^{\T}\hat\Sigma_{\textsc{sr}}^{-1}\hat\tau_{\textsc{sr}}}.
		\end{aligned}
	\end{equation*}
It follows that
$$\hat{\beta}_{\textsc{sr}}=\frac{\hat\Sigma_{\textsc{sr}}^{-1}\hat\tau_{\textsc{sr}}}{1+\hat\tau_{\textsc{sr}}^{\T}\hat\Sigma_{\textsc{sr}}^{-1}\hat\tau_{\textsc{sr}}}.$$

\vspace{0.5cm}
\noindent(ii) Recall that
\begin{equation*}
\begin{aligned}
\tau_{\textsc{sr}}&=E(W_{\textsc{sr}}\tilde{Z}_{\textsc{sr}}^2)^{-1}E(W_{\textsc{sr}}\tilde{Y}_{\textsc{sr}}\tilde{Z}_{\textsc{sr}})=E(W_{\textsc{sr}}\tilde{Z}_{\textsc{sr}}^2)^{-1}E(W_{\textsc{sr}}\tilde{Z}_{\textsc{sr}}Y),\\
\beta_{\textsc{sr}}&=E(W_{\textsc{sr}}\tilde{Y}_{\textsc{sr}}\tilde{Y}_{\textsc{sr}}^{\T})^{-1}E(W_{\textsc{sr}}\tilde{Y}_{\textsc{sr}}\tilde{Z}_{\textsc{sr}})=E(W_{\textsc{sr}}\tilde{Y}_{\textsc{sr}}\tilde{Y}_{\textsc{sr}}^{\T})^{-1}E(W_{\textsc{sr}}\tilde{Y}_{\textsc{sr}}Z),
\end{aligned}
\end{equation*}
where $\tilde{Y}_{\textsc{sr}}=Y-E(W_{\textsc{sr}} YG^{\T})E(W_{\textsc{sr}} GG^{\T})^{-1}G=Y-\{E(W_{\textsc{sr}}\mid C)\}^{-1}E(W_{\textsc{sr}}Y\mid C)$ and $\tilde{Z}_{\textsc{sr}}=Z-E(W_{\textsc{sr}} ZG^{\T})E(W_{\textsc{sr}} GG^{\T})^{-1}G=Z-\{E(W_{\textsc{sr}}\mid C)\}^{-1}E(W_{\textsc{sr}}Z\mid C)$. Comparing the expressions of $\tau_{\textsc{sr}}$ and $\beta_{\textsc{sr}}$ immediately gives
\begin{equation*}
\beta_{\textsc{sr}}=E(W_{\textsc{sr}}\tilde{Z}_{\textsc{sr}}^2)E(W_{\textsc{sr}}\tilde{Y}_{\textsc{sr}}\tilde{Y}_{\textsc{sr}}^{\T})^{-1}\tau_{\textsc{sr}}.
\end{equation*}

Next, we derive a more explicit characterization of the relationship between $\beta_{\textsc{sr}}$ and $\tau_{\textsc{sr}}$. Write
$$E(W_{\textsc{sr}}\tilde{Y}_{\textsc{sr}}\tilde{Y}_{\textsc{sr}}^{\T})=E(W_{\textsc{sr}}YY^{\T})-E(W_{\textsc{sr}}YG^{\T})E(W_{\textsc{sr}}GG^{\T})^{-1}E(W_{\textsc{sr}}GY^{\T}),$$
where  $E(W_{\textsc{sr}}YG^{\T})=(E\{W_{\textsc{sr}}YI(C=1)\},\ldots, E\{W_{\textsc{sr}}YI(C=S)\})$ and $E(W_{\textsc{sr}}GG^{\T})=\diag\{E\{$ $W_{\textsc{sr}}I(C=1)\}, \ldots, E\{W_{\textsc{sr}}I(C=S)\}\}$.
Recall that $W_{\textsc{sr}}=Z(1-p_{[C]})+(1-Z)p_{[C]}$. Direct calculations give
\begin{equation*}
\begin{aligned}
&E\{W_{\textsc{sr}}I(C=s)\}=2\pi_sp_{[s]}(1-p_{[s]}),\\
&E(W_{\textsc{sr}}ZI(C=s))=\pi_sp_{[s]}(1-p_{[s]}),\\
&E(W_{\textsc{sr}}YI(C=s))=\pi_sp_{[s]}(1-p_{[s]})E\{Y(1)+Y(0)\mid C=s\}.
\end{aligned}
\end{equation*}
Substituting these expressions into the previous display yields
\begin{equation*}
\begin{aligned}
&E(W_{\textsc{sr}}YG^{\T})E(W_{\textsc{sr}}GG^{\T})^{-1}E(W_{\textsc{sr}}GY^{\T})\\
&=\sum_{s=1}^{S}E(W_{\textsc{sr}}YI(C=s))E(W_{\textsc{sr}}Y^{\T}I(C=s))/E\{W_{\textsc{sr}}I(C=s)\}\\
&=\frac{1}{2}\sum_{s=1}^{S}\pi_sp_{[s]}(1-p_{[s]})E\{Y(1)+Y(0)\mid C=s\}E\{Y(1)^{\T}+Y(0)^{\T}\mid C=s\}.
\end{aligned}
\end{equation*}
In addition,
$E(W_{\textsc{sr}}YY^{\T})=\sum_{s=1}^{S}\pi_sp_{[s]}(1-p_{[s]})E\{Y(1)Y(1)^{\T}+Y(0)Y(0)^{\T}\mid C=s\}.$
Combining these two identities yields
\begin{equation*}
\begin{aligned}
E(W_{\textsc{sr}}\tilde{Y}_{\textsc{sr}}\tilde{Y}_{\textsc{sr}}^{\T})=\frac{1}{2}\sum_{s=1}^{S}\pi_sp_{[s]}(1-p_{[s]})\big[\tau_{[s]}\tau_{[s]}^{\T}+2\textup{cov}\{Y(1)\mid C=s\}
+2\textup{cov}\{Y(0)\mid C=s\}\big],
\end{aligned}
\end{equation*}
where $\tau_{[s]}=E\{Y(1)-Y(0)\mid C=s\}$.
Similarly,
\begin{equation*}
\begin{aligned}
E(W_{\textsc{sr}}\tilde{Z}_{\textsc{sr}}^2)=E(W_{\textsc{sr}}Z^2)-E(W_{\textsc{sr}}ZG^{\T})E(W_{\textsc{sr}}GG^{\T})^{-1}E(W_{\textsc{sr}}GZ)=\frac{1}{2}\sum_{s=1}^{S}\pi_sp_{[s]}(1-p_{[s]}).
\end{aligned}
\end{equation*}
Since
\begin{equation*}
\begin{aligned}
\frac{\sum_{s=1}^{S}\pi_sp_{[s]}(1-p_{[s]})(\tau_{[s]}-\tau_{\textsc{sr}})(\tau_{[s]}-\tau_{\textsc{sr}})^{\T}}{\sum_{s=1}^{S}\pi_sp_{[s]}(1-p_{[s]})}
&=\frac{\sum_{s=1}^{S}\pi_sp_{[s]}(1-p_{[s]})\tau_{[s]}\tau_{[s]}^{\T}}{\sum_{s=1}^{S}\pi_sp_{[s]}(1-p_{[s]})}-\tau_{\textsc{sr}}\tau_{\textsc{sr}}^{\T},
\end{aligned}
\end{equation*}
we can write
$E(W_{\textsc{sr}}\tilde{Z}_{\textsc{sr}}^2)^{-1}E(W_{\textsc{sr}}\tilde{Y}_{\textsc{sr}}\tilde{Y}_{\textsc{sr}}^{\T})=\Sigma_{\textsc{sr}}+\tau_{\textsc{sr}}\tau_{\textsc{sr}}^{\T}.$
Applying  Lemma \ref{Woodburyformula} to $A=\Sigma_{\textsc{sr}}$, $\mu=\tau_{\textsc{sr}}$, and $\nu=\tau_{\textsc{sr}}$ yields
	\begin{equation*}
		\begin{aligned}
E(W_{\textsc{sr}}\tilde{Z}_{\textsc{sr}}^2)E(W_{\textsc{sr}}\tilde{Y}_{\textsc{sr}}\tilde{Y}_{\textsc{sr}}^{\T})^{-1}=\Sigma_{\textsc{sr}}^{-1}-\frac{\Sigma_{\textsc{sr}}^{-1}\tau_{\textsc{sr}}
\tau_{\textsc{sr}}^{\T}\Sigma_{\textsc{sr}}^{-1}}{1+\tau_{\textsc{sr}}^{\T}\Sigma_{\textsc{sr}}^{-1}\tau_{\textsc{sr}}}.
		\end{aligned}
	\end{equation*}
Substituting the preceding expression into the formula of $\beta_{\textsc{sr}}$, we obtain
\begin{equation*}
\begin{aligned}
\beta_{\textsc{sr}}=\frac{\Sigma_{\textsc{sr}}^{-1}\tau_{\textsc{sr}}}{1+\tau_{\textsc{sr}}^{\T}\Sigma_{\textsc{sr}}^{-1}\tau_{\textsc{sr}}}.
\end{aligned}
\end{equation*}

\subsection{Proof of Theorem \ref{SREnoX-compATE-asy}}
\label{subsec-thmS4-proof}

For each unit $i$, define $\zeta_i=z_i-\beta_{\textsc{srg}}^{\T}g_i-\beta_{\textsc{sr}}^{\T}y_i$. Let $\zeta^{\textup{v}}=(\zeta_1,\ldots,\zeta_n)^{\T}$.
Substituting $z=\zeta^{\textup{v}}+g\beta_{\textsc{srg}}+y\beta_{\textsc{sr}}$ into the expression of $\hat{\beta}_{\textsc{sr}}=(\tilde{y}_{\textsc{sr}}^{\T}\Lambda_{\textsc{sr}}\tilde{y}_{\textsc{sr}})^{-1}
\tilde{y}_{\textsc{sr}}^{\T}\Lambda_{\textsc{sr}}z$ yields
$$\hat\beta_{\textsc{sr}}=(\tilde{y}_{\textsc{sr}}^{\T}\Lambda_{\textsc{sr}}\tilde{y}_{\textsc{sr}})^{-1}(\tilde{y}_{\textsc{sr}}^{\T}
\Lambda_{\textsc{sr}}\zeta^{\textup{v}})+
\beta_{\textsc{sr}}.
$$
Recall that
$\hat{\tau}_{\textsc{sr}}=(\tilde{z}_{\textsc{sr}}^{\T}\Lambda_{\textsc{sr}}\tilde{z}_{\textsc{sr}})^{-1}(\tilde{y}_{\textsc{sr}}^{\T}\Lambda_{\textsc{sr}}\tilde{y}_{\textsc{sr}})\hat{\beta}_{\textsc{sr}}$
and
$\beta_{\textup{\textsc{sr}}}=E(W_{\textup{\textsc{sr}}}\tilde{Z}_{\textup{\textsc{sr}}}^2)E(W_{\textup{\textsc{sr}}}\tilde{Y}_{\textup{\textsc{sr}}}
\tilde{Y}_{\textup{\textsc{sr}}}^{\T})^{-1}\tau_{\textup{\textsc{sr}}}$.
Then we have
\begin{equation*}
\begin{aligned}
\hat{\beta}_{\textsc{sr}}^{\T}\hat{\tau}_{\textsc{sr}}-\beta_{\textsc{sr}}^{\T}\tau_{\textsc{sr}} &=\beta_{\textsc{sr}}^{\T}\Big\{\big(n^{-1}\tilde{z}_{\textsc{sr}}^{\T}\Lambda_{\textsc{sr}}\tilde{z}_{\textsc{sr}}\big)^{-1}
\big(n^{-1}\tilde{y}_{\textsc{sr}}^{\T}\Lambda_{\textsc{sr}}\tilde{y}_{\textsc{sr}}\big)
-E(W_{\textup{\textsc{sr}}}\tilde{Z}_{\textup{\textsc{sr}}}^2)^{-1}E(W_{\textup{\textsc{sr}}}\tilde{Y}_{\textup{\textsc{sr}}}
\tilde{Y}_{\textup{\textsc{sr}}}^{\T})\Big\}\beta_{\textsc{sr}}\\
&\quad+2\big(n^{-1}\tilde{z}_{\textsc{sr}}^{\T}\Lambda_{\textsc{sr}}\tilde{z}_{\textsc{sr}}\big)^{-1}\hat{\beta}_{\textsc{sr}}^{\T}
\big(n^{-1}\tilde{y}_{\textsc{sr}}^{\T}\Lambda_{\textsc{sr}}\zeta^{\textup{v}}\big)
\\
&\quad+\big(n^{-1}\tilde{z}_{\textsc{sr}}^{\T}\Lambda_{\textsc{sr}}\tilde{z}_{\textsc{sr}}\big)^{-1}
\big(n^{-1}\tilde{y}_{\textsc{sr}}^{\T}\Lambda_{\textsc{sr}}\zeta^{\textup{v}}\big)^{\T}
\big(n^{-1}\tilde{y}_{\textsc{sr}}^{\T}\Lambda_{\textsc{sr}}\tilde{y}_{\textsc{sr}}\big)^{-1}
\big(n^{-1}\tilde{y}_{\textsc{sr}}^{\T}\Lambda_{\textsc{sr}}\zeta^{\textup{v}}\big).
\end{aligned}
\end{equation*}
In what follows, we derive the asymptotic distribution of $\hat{\beta}_{\textsc{sr}}^{\T}\hat{\tau}_{\textsc{sr}}-\beta_{\textsc{sr}}^{\T}\tau_{\textsc{sr}}$ in two cases: (i) $\tau_{\textsc{sr}}\neq0$ and (ii) $\tau_{\textsc{sr}}=0$.

\vspace{0.2cm}
\noindent\textit{(i) $\tau_{\textup{\textsc{sr}}}\neq0$.}  We begin by analyzing the first term in the expansion of
$\hat{\beta}_{\textsc{sr}}^{\T}\hat{\tau}_{\textsc{sr}}-\beta_{\textsc{sr}}^{\T}\tau_{\textsc{sr}}$. Recall that $\bar{z}_{w,[s]}=\sum_{i=1}^{n}w_{\textsc{sr}i}z_iI(c_i=s)/\{\sum_{i=1}^{n}w_{\textsc{sr}i}I(c_i=s)\}$. We write
\begin{equation}\label{SRnoX-asy-Zquar}
\begin{aligned}
n^{-1}\tilde{z}_{\textsc{sr}}^{\T}\Lambda_{\textsc{sr}}\tilde{z}_{\textsc{sr}}
&=n^{-1}\sum_{s=1}^{S}\sum_{i=1}^{n}w_{\textsc{sr}i}\big(z_i-\bar{z}_{w,[s]}\big)^2 I(c_i=s)\\
&=n^{-1}\sum_{s=1}^{S}\sum_{i=1}^{n}w_{\textsc{sr}i}\Big\{z_i-\frac{E(W_{\textsc{sr}}Z\mid C=s)}{E(W_{\textsc{sr}}\mid C=s)}\Big\}^2 I(c_i=s)\\
&\quad-n^{-1}\sum_{s=1}^{S}\sum_{i=1}^{n}w_{\textsc{sr}i}\Big\{\bar{z}_{w,[s]}-\frac{E(W_{\textsc{sr}}Z\mid C=s)}{E(W_{\textsc{sr}}\mid C=s)}\Big\}^2 I(c_i=s).
\end{aligned}
\end{equation}
Define $n_s=\sum_{i=1}^nI(c_i=s)$.
By the LLN,   $n_s^{-1}\sum_{i=1}^{n}w_{\textsc{sr}i}z_iI(c_i=s)\stackrel{p}\rightarrow E\{W_{\textsc{sr}}Z I(C=s)\}$ and $n_s^{-1}\sum_{i=1}^{n}w_{\textsc{sr}i}I(c_i=s)\stackrel{p}\rightarrow E\{W_{\textsc{sr}}I(C=s)\}$ as $n_s\rightarrow\infty$. Applying Slutsky's theorem therefore yields
$\bar{z}_{w,[s]}\stackrel{p}\rightarrow E(W_{\textsc{sr}}Z \mid C=s)/E(W_{\textsc{sr}}\mid C=s)$. Consequently, the second term in \eqref{SRnoX-asy-Zquar} is $o_p(1)$.
Moreover, by the LLN, as $n\rightarrow\infty$, we have
\begin{equation*}
\begin{aligned}
&n^{-1}\sum_{s=1}^{S}\sum_{i=1}^{n}w_{\textsc{sr}i}\big\{z_i-E(W_{\textsc{sr}}Z\mid C=s)/E(W_{\textsc{sr}}\mid C=s)\big\}^2 I(c_i=s)\\
&\stackrel{p}\rightarrow \sum_{s=1}^{S}\pi_s E\big[W_{\textsc{sr}}\big\{Z-E(W_{\textsc{sr}}Z\mid C=s)/E(W_{\textsc{sr}}\mid C=s)\big\}^2\mid C=s\big]=E(W_{\textsc{sr}}\tilde{Z}_{\textsc{sr}}^2),
\end{aligned}
\end{equation*}
where $\tilde{Z}_{\textsc{sr}}=Z-E(W_{\textsc{sr}}Z\mid C)/E(W_{\textsc{sr}}\mid C)$. Combining the preceding convergence results with \eqref{SRnoX-asy-Zquar} yields $n^{-1}\tilde{z}_{\textsc{sr}}^{\T}\Lambda_{\textsc{sr}}\tilde{z}_{\textsc{sr}}=E(W_{\textsc{sr}}\tilde{Z}_{\textsc{sr}}^2)+o_p(1)$.
Then we can write
\begin{equation*}
		\begin{aligned}
			&\big(n^{-1}\tilde{z}_{\textsc{sr}}^{\T}\Lambda_{\textsc{sr}}\tilde{z}_{\textsc{sr}}\big)^{-1}
\big(n^{-1}\tilde{y}_{\textsc{sr}}^{\T}\Lambda_{\textsc{sr}}\tilde{y}_{\textsc{sr}}\big)
-E(W_{\textup{\textsc{sr}}}\tilde{Z}_{\textup{\textsc{sr}}}^2)^{-1}E(W_{\textup{\textsc{sr}}}\tilde{Y}_{\textup{\textsc{sr}}}
\tilde{Y}_{\textup{\textsc{sr}}}^{\T})\\
&=E(W_{\textup{\textsc{sr}}}\tilde{Z}_{\textup{\textsc{sr}}}^2)^{-1}\big[\big\{n^{-1}\tilde{y}_{\textsc{sr}}^{\T}\Lambda_{\textsc{sr}}\tilde{y}_{\textsc{sr}}-E(W_{\textup{\textsc{sr}}}\tilde{Y}_{\textup{\textsc{sr}}}
\tilde{Y}_{\textup{\textsc{sr}}}^{\T})\big\}\\
&\hspace{2.5cm}-\big\{n^{-1}\tilde{z}_{\textsc{sr}}^{\T}\Lambda_{\textsc{sr}}\tilde{z}_{\textsc{sr}}
-E(W_{\textup{\textsc{sr}}}\tilde{Z}_{\textup{\textsc{sr}}}^2
)\big\}E(W_{\textup{\textsc{sr}}}\tilde{Y}_{\textup{\textsc{sr}}}
\tilde{Y}_{\textup{\textsc{sr}}}^{\T})/E(W_{\textup{\textsc{sr}}}\tilde{Z}_{\textup{\textsc{sr}}}^2)\big]\times\{1+o_p(1)\}.
		\end{aligned}
	\end{equation*}

Applying the same decomposition used in \eqref{SRnoX-asy-Zquar} gives
\begin{equation*}
\begin{aligned}
&n^{-1}\tilde{y}_{\textsc{sr}}^{\T}\Lambda_{\textsc{sr}}\tilde{y}_{\textsc{sr}}\\
&=n^{-1}\sum_{s=1}^{S}\sum_{i=1}^{n}w_{\textsc{sr}i}\Big\{y_i-\frac{E(W_{\textsc{sr}}Y\mid C=s)}{E(W_{\textsc{sr}}\mid C=s)}\Big\}
\Big\{y_i-\frac{E(W_{\textsc{sr}}Y\mid C=s)}{E(W_{\textsc{sr}}\mid C=s)}\Big\}^{\T} I(c_i=s)\\
&\quad-n^{-1}\sum_{s=1}^{S}\sum_{i=1}^{n}w_{\textsc{sr}i}\Big\{\bar{y}_{w,[s]}-\frac{E(W_{\textsc{sr}}Y\mid C=s)}{E(W_{\textsc{sr}}\mid C=s)}\Big\}\Big\{\bar{y}_{w,[s]}-\frac{E(W_{\textsc{sr}}Y\mid C=s)}{E(W_{\textsc{sr}}\mid C=s)}\Big\}^{\T} I(c_i=s),
\end{aligned}
\end{equation*}
where $\bar{y}_{w,[s]}=\sum_{i=1}^{n}w_{\textsc{sr}i}y_iI(c_i=s)/\{\sum_{i=1}^{n}w_{\textsc{sr}i}I(c_i=s)\}$.
The CLT implies that  $n_s^{1/2}\{\bar{y}_{w,[s]}-E(W_{\textsc{sr}}Y\mid C=s)/E(W_{\textsc{sr}}\mid C=s)\}=O_p(1)$
and the LLN implies $n_s^{-1}\sum_{i=1}^{n}w_{\textsc{sr}i}I(c_i=s)=E(W_{\textsc{sr}}\mid C=s)$. These results together imply
$$n^{-1/2}\sum_{s=1}^{S}\sum_{i=1}^{n}w_{\textsc{sr}i}\Big\{\bar{y}_{w,[s]}-\frac{E(W_{\textsc{sr}}Y\mid C=s)}{E(W_{\textsc{sr}}\mid C=s)}\Big\}\Big\{\bar{y}_{w,[s]}-\frac{E(W_{\textsc{sr}}Y\mid C=s)}{E(W_{\textsc{sr}}\mid C=s)}\Big\}^{\T} I(c_i=s)=o_p(1).$$
Therefore,
	\begin{equation*}
		\begin{aligned}
			&n^{1/2}\big\{n^{-1}\tilde{y}_{\textsc{sr}}^{\T}\Lambda_{\textsc{sr}}\tilde{y}_{\textsc{sr}}-E(W_{\textsc{sr}}\tilde{Y}_{\textsc{sr}}\tilde{Y}_{\textsc{sr}}^{\T})\big\}\\
			&=n^{-1/2}\sum_{s=1}^{S}\sum_{i=1}^{n}\bigg[w_{\textsc{sr}i}\Big\{y_i-\frac{E(W_{\textsc{sr}}Y\mid C=s)}{E(W_{\textsc{sr}}\mid C=s)}\Big\}
\Big\{y_i-\frac{E(W_{\textsc{sr}}Y\mid C=s)}{E(W_{\textsc{sr}}\mid C=s)}\Big\}^{\T}-E(W_{\textsc{sr}}\tilde{Y}_{\textsc{sr}}\tilde{Y}_{\textsc{sr}}^{\T}\mid C=s)\bigg]\\
&\hspace{2.8cm} \times I(c_i=s)+o_p(1).
		\end{aligned}
	\end{equation*}
	Analogously, we obtain
	\begin{equation*}
		\begin{aligned}
			&n^{1/2}\big\{n^{-1}\tilde{z}_{\textsc{sr}}^{\T}\Lambda_{\textsc{sr}}\tilde{z}_{\textsc{sr}}-E(W_{\textsc{sr}}\tilde{Z}_{\textsc{sr}}^2)\big\}\\
			&=n^{-1/2}\sum_{s=1}^{S}\sum_{i=1}^{n}\bigg[w_{\textsc{sr}i}\Big\{z_i-\frac{E(W_{\textsc{sr}}Z\mid C=s)}{E(W_{\textsc{sr}}\mid C=s)}\Big\}^2-E(W_{\textsc{sr}}\tilde{Z}_{\textsc{sr}}^2\mid C=s)\bigg] I(c_i=s)+o_p(1).
		\end{aligned}
	\end{equation*}
	Combining the preceding expressions yields
	\begin{equation*}
		\begin{aligned}
			&n^{1/2}\Big\{\big(n^{-1}\tilde{z}_{\textsc{sr}}^{\T}\Lambda_{\textsc{sr}}\tilde{z}_{\textsc{sr}}\big)^{-1}
\big(n^{-1}\tilde{y}_{\textsc{sr}}^{\T}\Lambda_{\textsc{sr}}\tilde{y}_{\textsc{sr}}\big)
-E(W_{\textup{\textsc{sr}}}\tilde{Z}_{\textup{\textsc{sr}}}^2)^{-1}E(W_{\textup{\textsc{sr}}}\tilde{Y}_{\textup{\textsc{sr}}}
\tilde{Y}_{\textup{\textsc{sr}}}^{\T})\Big\}\\
			&=\frac{n^{-1/2}}{E(W_{\textsc{sr}}\tilde{Z}_{\textsc{sr}}^2)}\sum_{s=1}^{S}\sum_{i=1}^{n}\bigg(\bigg[w_{\textsc{sr}i}\Big\{y_i-\frac{E(W_{\textsc{sr}}Y\mid C=s)}{E(W_{\textsc{sr}}\mid C=s)}\Big\}
\Big\{y_i-\frac{E(W_{\textsc{sr}}Y\mid C=s)}{E(W_{\textsc{sr}}\mid C=s)}\Big\}^{\T}\\
&\hspace{3.6cm}-E(W_{\textsc{sr}}\tilde{Y}_{\textsc{sr}}\tilde{Y}_{\textsc{sr}}^{\T}\mid C=s)\bigg]\\
			&\hspace{3.3cm}-\bigg[w_{\textsc{sr}i}\Big\{z_i-\frac{E(W_{\textsc{sr}}Z\mid C=s)}{E(W_{\textsc{sr}}\mid C=s)}\Big\}^2-E(W_{\textsc{sr}}\tilde{Z}_{\textsc{sr}}^2\mid C=s)\bigg]\frac{E(W_{\textsc{sr}}\tilde{Y}_{\textsc{sr}}\tilde{Y}_{\textsc{sr}}^{\T})}{E(W_{\textsc{sr}}\tilde{Z}_{\textsc{sr}}^2)}\bigg)\\
&\hspace{3.7cm}\times I(c_i=s)+o_p(1).
		\end{aligned}
	\end{equation*}

For the second term in the expansion of  $\hat{\beta}_{\textsc{sr}}^{\T}\hat{\tau}_{\textsc{sr}}-\beta_{\textsc{sr}}^{\T}\tau_{\textsc{sr}}$, we have
\begin{equation}\label{SRnoX-eq-Yres}
\begin{aligned}
n^{-1}\tilde{y}_{\textsc{sr}}^{\T}\Lambda_{\textsc{sr}}\zeta^{\textup{v}}
&=n^{-1}y^{\T}\Lambda_{\textsc{sr}}\zeta^{\textup{v}}-(n^{-1}y^{\T}\Lambda_{\textsc{sr}}g)(n^{-1}g^{\T}\Lambda_{\textsc{sr}}g)^{-1}(n^{-1}g^{\T}\Lambda_{\textsc{sr}}\zeta^{\textup{v}})\\
&=n^{-1}\sum_{s=1}^{S}\sum_{i=1}^{n}w_{\textsc{sr}i}\big(y_i-\bar{y}_{w,[s]}\big)\Big\{\zeta_i-\frac{\sum_{i=1}^{n}w_{\textsc{sr}i}\zeta_iI(c_i=s)}{\sum_{i=1}^{n}w_{\textsc{sr}i}I(c_i=s)}\Big\} I(c_i=s)\\
&=n^{-1}\sum_{s=1}^{S}\sum_{i=1}^{n}w_{\textsc{sr}i}\Big\{y_i-\frac{E(W_{\textsc{sr}}Y\mid C=s)}{E(W_{\textsc{sr}}\mid C=s)}\Big\}\zeta_iI(c_i=s)\\
&\quad-n^{-1}\sum_{s=1}^{S}\sum_{i=1}^{n}w_{\textsc{sr}i}\Big\{\bar{y}_{w,[s]}-\frac{E(W_{\textsc{sr}}Y\mid C=s)}{E(W_{\textsc{sr}}\mid C=s)}\Big\}\frac{\sum_{i=1}^{n}w_{\textsc{sr}i}\zeta_iI(c_i=s)}{\sum_{i=1}^{n}w_{\textsc{sr}i}I(c_i=s)} I(c_i=s)\\
&=n^{-1}\sum_{s=1}^{S}\sum_{i=1}^{n}w_{\textsc{sr}i}\Big\{y_i-\frac{E(W_{\textsc{sr}}Y\mid C=s)}{E(W_{\textsc{sr}}\mid C=s)}\Big\}\zeta_iI(c_i=s)+o_p(n^{-1/2}),
\end{aligned}
\end{equation}
where the fourth equality follows because $n_s^{1/2}\{\bar{y}_{w,[s]}-E(W_{\textsc{sr}}Y\mid C=s)/E(W_{\textsc{sr}}\mid C=s)\}=O_p(1)$,
$\sum_{i=1}^{n}w_{\textsc{sr}i}\zeta_iI(c_i=s)/n=E\{W_{\textsc{sr}}\zeta I(C=s)\}+o_p(1)$ by the LLN, and $E\{W_{\textsc{sr}}\zeta I(C=s)\}=0$ by
the property of the population WLS.
Using \eqref{SRnoX-eq-Yres} together with  $n^{-1}\tilde{z}_{\textsc{sr}}^{\T}\Lambda_{\textsc{sr}}\tilde{z}_{\textsc{sr}}=E\{W_{\textsc{sr}}\tilde{Z}_{\textsc{sr}}^2\}+o_p(1)$, we obtain
	\begin{equation*}
		\begin{aligned}
			&2\big(n^{-1}\tilde{z}_{\textsc{sr}}^{\T}\Lambda_{\textsc{sr}}\tilde{z}_{\textsc{sr}}\big)^{-1}\hat{\beta}_{\textsc{sr}}^{\T}
\big(n^{-1}\tilde{y}_{\textsc{sr}}^{\T}\Lambda_{\textsc{sr}}\zeta^{\textup{v}}\big)\\
			&=2E(W_{\textsc{sr}}\tilde{Z}_{\textsc{sr}}^2)^{-1}\beta_{\textsc{sr}}^{\T}\big(n^{-1}\tilde{y}_{\textsc{sr}}^{\T}\Lambda_{\textsc{sr}}\zeta^{\textup{v}}\big)\times\{1+o_p(1)\}\\
			&=\frac{2\beta_{\textsc{sr}}^{\T}}{E(W_{\textsc{sr}}\tilde{Z}_{\textsc{sr}}^2)}\bigg[n^{-1}\sum_{s=1}^{S}\sum_{i=1}^{n}w_{\textsc{sr}i}\Big\{y_i-\frac{E(W_{\textsc{sr}}Y\mid C=s)}{E(W_{\textsc{sr}}\mid C=s)}\Big\}\zeta_iI(c_i=s)+o_p(n^{-1/2})\bigg]\times\{1+o_p(1)\}.
		\end{aligned}
	\end{equation*}

Next, applying the CLT to \eqref{SRnoX-eq-Yres} yields $n^{-1/2}\tilde{y}_{\textsc{sr}}^{\T}\Lambda_{\textsc{sr}}\zeta^{\textup{v}}=O_p(1)$. Consequently,
the third term in the expansion of  $\hat{\beta}_{\textsc{sr}}^{\T}\hat{\tau}_{\textsc{sr}}-\beta_{\textsc{sr}}^{\T}\tau_{\textsc{sr}}$ satisfies
\begin{equation*}
\begin{aligned}
\big(n^{-1}\tilde{z}_{\textsc{sr}}^{\T}\Lambda_{\textsc{sr}}\tilde{z}_{\textsc{sr}}\big)^{-1}
\big(n^{-1}\tilde{y}_{\textsc{sr}}^{\T}\Lambda_{\textsc{sr}}\zeta^{\textup{v}}\big)^{\T}
\big(n^{-1}\tilde{y}_{\textsc{sr}}^{\T}\Lambda_{\textsc{sr}}\tilde{y}_{\textsc{sr}}\big)^{-1}
\big(n^{-1}\tilde{y}_{\textsc{sr}}^{\T}\Lambda_{\textsc{sr}}\zeta^{\textup{v}}\big)=o_p(n^{-1/2}).
\end{aligned}
	\end{equation*}
Combining the above results, we obtain
	\begin{equation*}
		\begin{aligned}
			&n^{1/2}\big(\hat{\beta}_{\textsc{sr}}^{\T}\hat{\tau}_{\textsc{sr}}-\beta_{\textsc{sr}}^{\T}\tau_{\textsc{sr}}\big)\\
			&=\frac{n^{-1/2}\beta_{\textsc{sr}}^{\T}}{E(W_{\textsc{sr}}\tilde{Z}_{\textsc{sr}}^2)}\sum_{s=1}^{S}\sum_{i=1}^{n}\bigg(\bigg[w_{\textsc{sr}i}\Big\{y_i-\frac{E(W_{\textsc{sr}}Y\mid C=s)}{E(W_{\textsc{sr}}\mid C=s)}\Big\}
\Big\{y_i-\frac{E(W_{\textsc{sr}}Y\mid C=s)}{E(W_{\textsc{sr}}\mid C=s)}\Big\}^{\T}\\
&\hspace{3.6cm}-E(W_{\textsc{sr}}\tilde{Y}_{\textsc{sr}}\tilde{Y}_{\textsc{sr}}^{\T}\mid C=s)\bigg]\beta_{\textsc{sr}}\\
			&\hspace{3.3cm}-\bigg[w_{\textsc{sr}i}\Big\{z_i-\frac{E(W_{\textsc{sr}}Z\mid C=s)}{E(W_{\textsc{sr}}\mid C=s)}\Big\}^2-E(W_{\textsc{sr}}\tilde{Z}_{\textsc{sr}}^2\mid C=s)\bigg]\frac{E(W_{\textsc{sr}}\tilde{Y}_{\textsc{sr}}\tilde{Y}_{\textsc{sr}}^{\T})\beta_{\textsc{sr}}}{E(W_{\textsc{sr}}\tilde{Z}_{\textsc{sr}}^2)}\\
			&\hspace{3.3cm}+2w_{\textsc{sr}i}\Big\{y_i-\frac{E(W_{\textsc{sr}}Y\mid C=s)}{E(W_{\textsc{sr}}\mid C=s)}\Big\}\zeta_i\bigg)I(c_i=s)+o_p(1).\\
		\end{aligned}
	\end{equation*}
Finally, applying the CLT to the leading term  and then applying the delta method yields
	\begin{equation*}
		n^{1/2}\big(\hat{\tau}^{\textup{c}}_{\textsc{sr}}-\tau^{\textup{c}}_{\textsc{sr}}\big)\stackrel{d}\rightarrow N\big(0, E(W_{\textsc{sr}}\tilde{Z}_{\textsc{sr}}^2)^{-2}\beta_{\textup{\textsc{sr}}}^{\T}E[\textup{cov}\{R_{\textup{\textsc{sr}}}(Z,C,Y)\mid C\}]\beta_{\textup{\textsc{sr}}}\big),
	\end{equation*}
	where $R_{\textup{\textsc{sr}}}(Z,C,Y)$ is defined in the theorem.

\vspace{0.2cm}
\noindent\textit{(ii) $\tau_{\textup{\textsc{sr}}}=0$.} In this case, the expression of $n(\hat{\beta}_{\textsc{sr}}^{\T}\hat{\tau}_{\textsc{sr}}-\beta_{\textsc{sr}}^{\T}\tau_{\textsc{sr}})$ simplifies to
\begin{equation*}
\begin{aligned}
n(\hat{\beta}_{\textsc{sr}}^{\T}\hat{\tau}_{\textsc{sr}}-\beta_{\textsc{sr}}^{\T}\tau_{\textsc{sr}})
&=\big(n^{-1}\tilde{z}_{\textsc{sr}}^{\T}\Lambda_{\textsc{sr}}\tilde{z}_{\textsc{sr}}\big)^{-1}
\big(n^{-1/2}\tilde{y}_{\textsc{sr}}^{\T}\Lambda_{\textsc{sr}}\zeta^{\textup{v}}\big)^{\T}
\big(n^{-1}\tilde{y}_{\textsc{sr}}^{\T}\Lambda_{\textsc{sr}}\tilde{y}_{\textsc{sr}}\big)^{-1}
\big(n^{-1/2}\tilde{y}_{\textsc{sr}}^{\T}\Lambda_{\textsc{sr}}\zeta^{\textup{v}}\big).
\end{aligned}
\end{equation*}
Let $M_{\textsc{sr}}=(n^{-1}\tilde{z}_{\textsc{sr}}^{\T}\Lambda_{\textsc{sr}}\tilde{z}_{\textsc{sr}})^{-1/2}
(n^{-1}\tilde{y}_{\textsc{sr}}^{\T}\Lambda_{\textsc{sr}}\tilde{y}_{\textsc{sr}})^{-1/2}
(n^{-1/2}\tilde{y}_{\textsc{sr}}^{\T}\Lambda_{\textsc{sr}}\zeta^{\textup{v}}).$
Applying the CLT to \eqref{SRnoX-eq-Yres}, we have
$n^{-1/2}\tilde{y}_{\textsc{sr}}^{\T}\Lambda_{\textsc{sr}}\zeta^{\textup{v}}\stackrel{d}\rightarrow N(0, \sum_{s=1}^{S}\pi_sE\{W_{\textsc{sr}}^2\zeta^2\tilde{Y}_{\textsc{sr}}
\tilde{Y}_{\textsc{sr}}^{\T}\mid C=s\}),$
Further applying Slutsky's theorem, we obtain $M_{\textsc{sr}}\stackrel{d}\rightarrow N(0, \Gamma_{\textsc{sr}})$, where $\Gamma_{\textsc{sr}}$ is defined in the theorem.
Let $\Gamma_{\textsc{sr}}=Q_{\textsc{sr}}^{\T}\Delta_{\textsc{sr}} Q_{\textsc{sr}}$ be the eigendecomposition of $\Gamma_{\textsc{sr}}$,
where $Q_{\textsc{sr}}$ is an orthogonal matrix and $\Delta_{\textsc{sr}}=\text{diag}(\lambda_{{\textsc{sr}}1},\ldots,\lambda_{{\textsc{sr}}L})$ is a diagonal matrix. Define $B_{\textsc{sr}}=\Delta_{\textsc{sr}}^{-1/2}Q_{\textsc{sr}}M_{\textsc{sr}}=(B_{{\textsc{sr}}1},\ldots, B_{{\textsc{sr}}L})^{\T}$. It then follows that  $B_{\textsc{sr}}\stackrel{d}\rightarrow N(0, I_L)$. Since
$M_{\textsc{sr}}^{\T}M_{\textsc{sr}}=(Q_{\textsc{sr}}^{\T}\Delta_{\textsc{sr}}^{1/2}B_{\textsc{sr}})^{\T}(Q_{\textsc{sr}}^{\T}\Delta_{\textsc{sr}}^{1/2}B_{\textsc{sr}})
=B_{\textsc{sr}}^{\T}\Delta_{\textsc{sr}} B_{\textsc{sr}}$, we have
$n(\hat\beta_{\textsc{sr}}^{\T}\hat\tau_{\textsc{sr}}-\beta_{\textsc{sr}}^{\T}\tau_{\textsc{sr}})= \sum_{\ell=1}^{L}\lambda_{{\textsc{sr}}\ell}B_{{\textsc{sr}}\ell}^2\stackrel{d}\rightarrow \sum_{\ell=1}^{L}\lambda_{{\textsc{sr}}\ell}\chi^2_{\ell}(1).$

\subsection{Proof of Theorem \ref{AsyDis-BetaTau-SR-withX}}
\label{subsec-thmS5-proof}

Let $(\beta_{\textsc{srg,a}}, \beta_{\textsc{srx,a}},\beta_{\textsc{sr,a}}) = \argmin_{b_{\textsc{g}},b_{\textsc{x}},b} E\{W_{\textsc{sr}}(Z-b_{\textsc{g}}^{\T}G-b_{\textsc{x}}^{\T}X-b^{\T}Y)^2\}$ denote the population WLS coefficients from the regression of $Z$ on $(G, X, Y)$. Define $\zeta_{\textsc{a}i} =z_i-\beta_{\textsc{srg,a}}^{\T}g_i-\beta_{\textsc{srx,a}}^{\T}x_i-\beta_{\textsc{sr,a}}^{\T}y_i$ and $\zeta_{\textsc{a}}^{\textup{v}}=(\zeta_{\textsc{a}1},\ldots,\zeta_{\textsc{a}n})^{\T}$. Substituting $z=\zeta_{\textsc{a}}^{\textup{v}}+g\beta_{\textsc{srg,a}}+u\beta_{\textsc{sru,a}}$ into the expressions of $\hat\beta_{\textup{\textsc{sru,a}}}=(\tilde{u}_{\textsc{sr}}^{\T}\Lambda_{\textsc{sr}}\tilde{u}_{\textsc{sr}})^{-1}
\tilde{u}_{\textsc{sr}}^{\T}\Lambda_{\textsc{sr}}z$ yields
$$\hat{\beta}_{\textsc{sru,a}}=(\tilde{u}_{\textsc{sr}}^{\T}\Lambda_{\textsc{sr}}\tilde{u}_{\textsc{sr}})^{-1}(\tilde{u}_{\textsc{sr}}^{\T}
\Lambda_{\textsc{sr}}\zeta_{\textsc{a}}^{\textup{v}})+
\beta_{\textsc{sru,a}}.
$$
Proposition \ref{Equiv-SR-withX} gives $\hat{\tau}_{\textup{\textsc{sru}}}=(\tilde{z}_{\textup{\textsc{sr}}}^{\T}\Lambda_{\textup{\textsc{sr}}}\tilde{z}_{\textup{\textsc{sr}}})^{-1}
(\tilde{u}_{\textup{\textsc{sr}}}^{\T}\Lambda_{\textup{\textsc{sr}}}\tilde{u}_{\textup{\textsc{sr}}})\hat\beta_{\textup{\textsc{sru,a}}}$ and $
\beta_{\textup{\textsc{sru,a}}}=E(W_{\textup{\textsc{sr}}}\tilde{Z}_{\textup{\textsc{sr}}}^2)E(W_{\textup{\textsc{sr}}}\tilde{U}_{\textup{\textsc{sr}}}$ $
\tilde{U}_{\textup{\textsc{sr}}}^{\T})^{-1}\tau_{\textup{\textsc{sru}}}$.
Using these identities, we decompose
\begin{equation*}
\begin{aligned}
\hat{\tau}_{\textsc{sr,a}}^{\textup{c}}(r)-\tau_{\textsc{sr,a}}^{\textup{c}}
&=\hat{\tau}_{\textsc{sr,a}y}^{\textup{c}}-r\hat{\tau}_{\textsc{sr,a}x}^{\textup{c}}-\beta_{\textsc{sr,a}}^{\T}\tau_{\textsc{sr}}
=\hat{\beta}_{\textsc{sru,a}}^{\T}D\hat{\tau}_{\textsc{sru}}-\beta_{\textsc{sru,a}}^{\T}D\tau_{\textsc{sru}},\\ &=\beta_{\textsc{sru,a}}^{\T}D\Big\{\big(n^{-1}\tilde{z}_{\textsc{sr}}^{\T}\Lambda_{\textsc{sr}}\tilde{z}_{\textsc{sr}}\big)^{-1}
\big(n^{-1}\tilde{u}_{\textsc{sr}}^{\T}\Lambda_{\textsc{sr}}\tilde{u}_{\textsc{sr}}\big)
-E(W_{\textup{\textsc{sr}}}\tilde{Z}_{\textup{\textsc{sr}}}^2)^{-1}E(W_{\textup{\textsc{sr}}}\tilde{U}_{\textup{\textsc{sr}}}
\tilde{U}_{\textup{\textsc{sr}}}^{\T})\Big\}\beta_{\textsc{sru,a}}\\
&\quad+\big(n^{-1}\tilde{z}_{\textsc{sr}}^{\T}\Lambda_{\textsc{sr}}\tilde{z}_{\textsc{sr}}\big)^{-1}\beta_{\textsc{sru,a}}^{\T}
\big(n^{-1}\tilde{u}_{\textsc{sr}}^{\T}\Lambda_{\textsc{sr}}\tilde{u}_{\textsc{sr}}\big)D\big(n^{-1}\tilde{u}_{\textsc{sr}}^{\T}\Lambda_{\textsc{sr}}\tilde{u}_{\textsc{sr}}\big)^{-1}
\big(n^{-1}\tilde{u}_{\textsc{sr}}^{\T}\Lambda_{\textsc{sr}}\zeta_{\textsc{a}}^{\textup{v}}\big)\\
&\quad+\big(n^{-1}\tilde{z}_{\textsc{sr}}^{\T}\Lambda_{\textsc{sr}}\tilde{z}_{\textsc{sr}}\big)^{-1}\beta_{\textsc{sru,a}}^{\T}D
\big(n^{-1}\tilde{u}_{\textsc{sr}}^{\T}\Lambda_{\textsc{sr}}\zeta_{\textsc{a}}^{\textup{v}}\big)
\\
&\quad+\big(n^{-1}\tilde{z}_{\textsc{sr}}^{\T}\Lambda_{\textsc{sr}}\tilde{z}_{\textsc{sr}}\big)^{-1}
\big(n^{-1}\tilde{u}_{\textsc{sr}}^{\T}\Lambda_{\textsc{sr}}\zeta_{\textsc{a}}^{\textup{v}}\big)^{\T}
\big(n^{-1}\tilde{u}_{\textsc{sr}}^{\T}\Lambda_{\textsc{sr}}\tilde{u}_{\textsc{sr}}\big)^{-1}
D\big(n^{-1}\tilde{u}_{\textsc{sr}}^{\T}\Lambda_{\textsc{sr}}\zeta_{\textsc{a}}^{\textup{v}}\big).
\end{aligned}
\end{equation*}
In what follows, we derive the asymptotic distribution of $\hat{\tau}_{\textsc{sr,a}}^{\textup{c}}(r)-\tau_{\textsc{sr,a}}^{\textup{c}}$  in two cases: (i) $\tau_{\textsc{sr}}\neq0$ and (ii) $\tau_{\textsc{sr}}=0$.

\vspace{0.2cm}
\noindent(i) $\tau_{\textup{\textsc{sr}}}\neq0$.  Using arguments analogous to those in the proof of Theorem 3, we obtain $n^{-1}\tilde{z}_{\textsc{sr}}^{\T}\Lambda_{\textsc{sr}}\tilde{z}_{\textsc{sr}}=E(W_{\textup{\textsc{sr}}}\tilde{Z}_{\textup{\textsc{sr}}}^2)+o_p(1)$ and $n^{-1}\tilde{u}_{\textsc{sr}}^{\T}\Lambda_{\textsc{sr}}\tilde{u}_{\textsc{sr}}=E\{W_{\textsc{sr}}\tilde{U}_{\textsc{sr}}\tilde{U}_{\textsc{sr}}^{\T}\}+o_p(1)$. The same arguments further give
\begin{equation*}
		\begin{aligned}
			&n^{1/2}\Big\{\big(n^{-1}\tilde{z}_{\textsc{sr}}^{\T}\Lambda_{\textsc{sr}}\tilde{z}_{\textsc{sr}}\big)^{-1}
\big(n^{-1}\tilde{u}_{\textsc{sr}}^{\T}\Lambda_{\textsc{sr}}\tilde{u}_{\textsc{sr}}\big)
-E(W_{\textup{\textsc{sr}}}\tilde{Z}_{\textup{\textsc{sr}}}^2)^{-1}E(W_{\textup{\textsc{sr}}}\tilde{U}_{\textup{\textsc{sr}}}
\tilde{U}_{\textup{\textsc{sr}}}^{\T})\Big\}\\
			&=\frac{n^{-1/2}}{E(W_{\textsc{sr}}\tilde{Z}_{\textsc{sr}}^2)}\sum_{s=1}^{S}\sum_{i=1}^{n}\bigg[\big\{w_{\textsc{sr}i}(u_i-\tilde{U}_{\textsc{sr}})
(u_i-\tilde{U}_{\textsc{sr}})^{\T}-E(W_{\textsc{sr}}\tilde{U}_{\textsc{sr}}\tilde{U}_{\textsc{sr}}^{\T}\mid C=s)\big\}\\
			&\hspace{3.3cm}-\big\{w_{\textsc{sr}i}(z_i-\tilde{Z}_{\textsc{sr}})^2-E(W_{\textsc{sr}}\tilde{Z}_{\textsc{sr}}^2\mid C=s)\big\}\frac{E(W_{\textsc{sr}}\tilde{U}_{\textsc{sr}}\tilde{U}_{\textsc{sr}}^{\T})}{E(W_{\textsc{sr}}\tilde{Z}_{\textsc{sr}}^2)}\bigg] I(c_i=s)+o_p(1),
		\end{aligned}
	\end{equation*}
and
\begin{equation}\label{SRwithX-eq-Yres}
\begin{aligned}
n^{-1}\tilde{u}_{\textsc{sr}}^{\T}\Lambda_{\textsc{sr}}\zeta_{\textsc{a}}^{\textup{v}}=n^{-1}\sum_{s=1}^{S}\sum_{i=1}^{n}w_{\textsc{sr}i}
(u_i-\tilde{U}_{\textsc{sr}})\zeta_{\textsc{a}i}I(c_i=s)+o_p(n^{-1/2}).
\end{aligned}
\end{equation}
By the CLT, $n^{-1/2}\tilde{u}_{\textsc{sr}}^{\T}\Lambda_{\textsc{sr}}\zeta_{\textsc{a}}^{\textup{v}}=O_p(1)$. Consequently,
the fourth term in the expansion of  $\hat{\tau}_{\textsc{sr,a}}^{\textup{c}}(r)-\tau_{\textsc{sr,a}}^{\textup{c}}$ satisfies
\begin{equation*}
\begin{aligned}
\big(n^{-1}\tilde{z}_{\textsc{sr}}^{\T}\Lambda_{\textsc{sr}}\tilde{z}_{\textsc{sr}}\big)^{-1}
\big(n^{-1}\tilde{u}_{\textsc{sr}}^{\T}\Lambda_{\textsc{sr}}\zeta_{\textsc{a}}^{\textup{v}}\big)^{\T}
\big(n^{-1}\tilde{u}_{\textsc{sr}}^{\T}\Lambda_{\textsc{sr}}\tilde{u}_{\textsc{sr}}\big)^{-1}
D\big(n^{-1}\tilde{u}_{\textsc{sr}}^{\T}\Lambda_{\textsc{sr}}\zeta_{\textsc{a}}^{\textup{v}}\big)=o_p(n^{-1/2}).
\end{aligned}
	\end{equation*}
Combining the preceding expressions yields
\begin{equation*}
\begin{aligned}
&n^{1/2}\big\{\hat{\tau}_{\textsc{sr,a}}^{\textup{c}}(r)-\tau_{\textsc{sr,a}}^{\textup{c}}\big\}\\
&=\frac{n^{-1/2}\beta_{\textsc{sru,a}}^{\T}}{E(W_{\textsc{sr}}\tilde{Z}_{\textsc{sr}}^2)}\sum_{s=1}^{S}\sum_{i=1}^{n}\bigg[D\big\{w_{\textsc{sr}i}(u_i-\tilde{U}_{\textsc{sr}})
(u_i-\tilde{U}_{\textsc{sr}})^{\T}-E(W_{\textsc{sr}}\tilde{U}_{\textsc{sr}}\tilde{U}_{\textsc{sr}}^{\T}\mid C=s)\big\}\beta_{\textsc{sru,a}}\\
			&\hspace{3.5cm}-D\big\{w_{\textsc{sr}i}(z_i-\tilde{Z}_{\textsc{sr}})^2-E(W_{\textsc{sr}}\tilde{Z}_{\textsc{sr}}^2\mid C=s)\big\}
\frac{E(W_{\textsc{sr}}\tilde{U}_{\textsc{sr}}\tilde{U}_{\textsc{sr}}^{\T})}{E(W_{\textsc{sr}}\tilde{Z}_{\textsc{sr}}^2)}\beta_{\textsc{sru,a}}\\
&\hspace{3.6cm}+E(W_{\textsc{sr}}\tilde{U}_{\textsc{sr}}\tilde{U}_{\textsc{sr}}^{\T}\mid C=s)DE(W_{\textsc{sr}}\tilde{U}_{\textsc{sr}}\tilde{U}_{\textsc{sr}}^{\T}\mid C=s)^{-1} w_{\textsc{sr}i}(u_i-\tilde{U}_{\textsc{sr}})\zeta_{\textsc{a}i}\\
&\hspace{3.6cm}+w_{\textsc{sr}i}D(u_i-\tilde{U}_{\textsc{sr}})\zeta_{\textsc{a}i}\bigg]I(c_i=s)+o_p(1).
\end{aligned}
\end{equation*}
The conclusion follows by applying the multivariate CLT and the delta method.

\vspace{0.2cm}
\noindent(ii) $\tau_{\textup{\textsc{sr}}}=0$. In this case, the expression of $\hat{\tau}_{\textsc{sr,a}}^{\textup{c}}(r)-\tau_{\textsc{sr,a}}^{\textup{c}}$ simplifies to
\begin{equation*}
\begin{aligned}
n\big\{\hat{\tau}_{\textsc{sr,a}}^{\textup{c}}(r)-\tau_{\textsc{sr,a}}^{\textup{c}}\big\}
&=\big(n^{-1}\tilde{z}_{\textsc{sr}}^{\T}\Lambda_{\textsc{sr}}\tilde{z}_{\textsc{sr}}\big)^{-1}
\big(n^{-1/2}\tilde{u}_{\textsc{sr}}^{\T}\Lambda_{\textsc{sr}}\zeta_{\textsc{a}}^{\textup{v}}\big)^{\T}
\big(n^{-1}\tilde{u}_{\textsc{sr}}^{\T}\Lambda_{\textsc{sr}}\tilde{u}_{\textsc{sr}}\big)^{-1}
D\big(n^{-1/2}\tilde{u}_{\textsc{sr}}^{\T}\Lambda_{\textsc{sr}}\zeta_{\textsc{a}}^{\textup{v}}\big)\\
&=\frac{1}{2}\big(n^{-1}\tilde{z}_{\textsc{sr}}^{\T}\Lambda_{\textsc{sr}}\tilde{z}_{\textsc{sr}}\big)^{-1}
\big(n^{-1/2}\tilde{u}_{\textsc{sr}}^{\T}\Lambda_{\textsc{sr}}\zeta_{\textsc{a}}^{\textup{v}}\big)^{\T}\\
&\quad\times\Big\{\big(n^{-1}\tilde{u}_{\textsc{sr}}^{\T}\Lambda_{\textsc{sr}}\tilde{u}_{\textsc{sr}}\big)^{-1}D
+D\big(n^{-1}\tilde{u}_{\textsc{sr}}^{\T}\Lambda_{\textsc{sr}}\tilde{u}_{\textsc{sr}}\big)^{-1}\Big\}
\big(n^{-1/2}\tilde{u}_{\textsc{sr}}^{\T}\Lambda_{\textsc{sr}}\zeta_{\textsc{a}}^{\textup{v}}\big),
\end{aligned}
\end{equation*}
where the second equality follows from the property of quadratic forms. Let $$M_{\textsc{sr,a}}=2^{-1/2}\big(n^{-1}\tilde{z}_{\textsc{sr}}^{\T}\Lambda_{\textsc{sr}}\tilde{z}_{\textsc{sr}}\big)^{-1/2}
\Big\{\big(n^{-1}\tilde{u}_{\textsc{sr}}^{\T}\Lambda_{\textsc{sr}}\tilde{u}_{\textsc{sr}}\big)^{-1}D
+D\big(n^{-1}\tilde{u}_{\textsc{sr}}^{\T}\Lambda_{\textsc{sr}}\tilde{u}_{\textsc{sr}}\big)^{-1}\Big\}^{1/2}
\big(n^{-1/2}\tilde{u}_{\textsc{sr}}^{\T}\Lambda_{\textsc{sr}}\zeta_{\textsc{a}}^{\textup{v}}\big).$$
Applying the CLT to \eqref{SRwithX-eq-Yres} yields
$n^{-1/2}\tilde{u}_{\textsc{sr}}^{\T}\Lambda_{\textsc{sr}}\zeta_{\textsc{a}}^{\textup{v}}\stackrel{d}\rightarrow N(0, \sum_{s=1}^{S}\pi_sE\{W_{\textsc{sr}}^2\zeta_{\textsc{a}}^2\tilde{U}_{\textsc{sr}}\tilde{U}_{\textsc{sr}}^{\T}\mid C=s\})$.  Further applying Slutsky's theorem, we obtain $M_{\textsc{sr,a}}\stackrel{d}\rightarrow N(0, \Gamma_{\textsc{sr,a}})$, where $\Gamma_{\textup{\textsc{sr,a}}}$ is defined in the theorem.
Let $\Gamma_{\textsc{sr,a}}=Q_{\textsc{sr,a}}^{\T}\Delta_{\textsc{sr,a}} Q_{\textsc{sr,a}}$ be the eigendecomposition of $\Gamma_{\textsc{sr,a}}$,
where $\Delta_{\textsc{sr,a}}=\text{diag}(\lambda_{{\textsc{sr,a}}1},\ldots,$ $\lambda_{{\textsc{sr,a}}L})$ contains the eigenvalues of $\Gamma_{\textsc{sr,a}}$
and $Q_{\textsc{sr,a}}$ is the orthogonal matrix whose columns are the corresponding eigenvectors.
Define $B_{\textsc{sr,a}}=\Delta_{\textsc{sr,a}}^{-1/2}Q_{\textsc{sr,a}}M_{\textsc{sr,a}}=(B_{{\textsc{sr,a}}1},\ldots, B_{{\textsc{sr,a}}L})^{\T}$. Then  $B_{\textsc{sr,a}}\stackrel{d}\rightarrow N(0, I_L)$. Therefore,
\begin{equation*}
n\big\{\hat{\tau}_{\textsc{sr,a}}^{\textup{c}}(r)-\tau_{\textsc{sr,a}}^{\textup{c}}\big\}=M_{\textsc{sr,a}}^{\T}M_{\textsc{sr,a}}=
B_{\textsc{sr,a}}^{\T}\Delta_{\textsc{sr,a}} B_{\textsc{sr,a}}= \sum_{\ell=1}^{L}\lambda_{{\textsc{sr,a}}\ell}B_{{\textsc{sr,a}}\ell}^2\stackrel{d}\rightarrow \sum_{\ell=1}^{L}\lambda_{{\textsc{sr,a}}\ell}\chi^2_{\ell}(1).
\end{equation*}

\subsection{Proof of Proposition \ref{Equiv-logit-CR-noX}} \label{subsec-propS5-proof}

The proof proceeds in two steps. First, we establish the convergence of $\hat\gamma_{\textsc{f}}$ by showing that $\hat{\gamma_{\textsc{f}}}=\gamma_{\textsc{f}}^*+o_p(1)$, where $\gamma_{\textsc{f}}^*=0_{L+1}$. Second, we establish the asymptotic equivalence between $\hat{\gamma}$ and  $\hat{\tau}$.

\vspace{0.2cm}
\noindent\textit{(i) Convergence of $\hat\gamma_{\textup{\textsc{f}}}$.}  By Lemma \ref{lemma-NeweyMcFadden1994}, it suffices to show that  there exists a function $l_{\infty}(\gamma_{\textsc{f}})$ that satisfies the following three conditions:  (a) $l(\gamma_{\textsc{f}})=l_{\infty}(\gamma_{\textsc{f}})+o_p(1)$  for all $\gamma_{\textsc{f}}\in\mathbb{R}^{L+1}$; (b) $l_{\infty}(\gamma_{\textsc{f}})$ is uniquely maximized at $\gamma_{\textsc{f}}^*$; and (c) $l(\gamma_{\textsc{f}})$ is concave on $\mathbb{R}^{L+1}$. We verify the three conditions sequentially.

Since $\gamma_{\textsc{f}}^*=0_{L+1}$, we have $\exp\{(1, y_i^{\T})\gamma_{\textsc{f}}^*\}=1$, and $\pi(y_i,\gamma_{\textsc{f}})=1/2$ for all $i=1,\ldots,n$. Substituting these results into $l(\gamma_{\textsc{f}})$, $m(\gamma_{\textsc{f}})$, and $H(\gamma_{\textsc{f}})$ yields
\begin{equation*}
\begin{aligned}
l(\gamma_{\textsc{f}}^*)&=-n^{-1}\sum_{i=1}^{n}\log(2)w_i=-\log(2)S_{11}^{w},\\
m(\gamma_{\textsc{f}}^*)&=n^{-1}\sum_{i=1}^{n}w_i(z_i-1/2)(1, y_i^{\T})^{\T}=(S^{w}_{\smash{z1}}-S^{w}_{\smash{11}}/2, S^{w}_{\smash{zy}}-S^{w}_{\smash{1y}}/2)^{\T},\\
H(\gamma_{\textsc{f}}^*)&=-(4n)^{-1}\sum_{i=1}^{n}w_i(1, y_i^{\T})^{\T}(1, y_i^{\T})
=-\frac{1}{4}\begin{pmatrix}
S_{11}^{w} & S_{1y}^{w}\\
S_{y1}^{w} & S_{yy}^{w}
\end{pmatrix}.\\
\end{aligned}
\end{equation*}
The strong LLN implies
\begin{equation*}
\begin{aligned}
l(\gamma_{\textsc{f}}^*)&\rightarrow l_{\infty}(\gamma_{\textsc{f}}^*)=-2\log(2),\\
m(\gamma_{\textsc{f}}^*)&\rightarrow m_{\infty}(\gamma_{\textsc{f}}^*)=\begin{pmatrix}E(WZ)-E(W)/2\\E(WZY)-E(WY)/2 \end{pmatrix},\\
H(\gamma_{\textsc{f}}^*)&\rightarrow H_{\infty}(\gamma_{\textsc{f}}^*)=-\frac{1}{4}\begin{pmatrix}E(W) & E(WY^{\T})\\E(WY) & E(WYY^{\T})\end{pmatrix}<0,
\end{aligned}
\end{equation*}
almost surely. Moreover, the uniform LLN \citep[Lemma 2.4]{Newey1994McFadden} ensures that the derivative
$\partial m(\gamma_{\textsc{f}})/\partial\gamma_{\textsc{f}}=H(\gamma_{\textsc{f}})$ converges uniformly to $H_{\infty}(\gamma_{\textsc{f}})=-E[W\pi(Y,\gamma_{\textsc{f}})\{1-\pi(Y,\gamma_{\textsc{f}})\}(1, Y^{\T})^{\T}(1, Y^{\T})]$ on every compact set  $\Theta\subset\mathbb{R}^{L+1}$. For any  nonzero vector $u\in\mathbb{R}^{L+1}$, we have $u^{\T}H_{\infty}(\gamma_{\textsc{f}})u<0$. Therefore,  $H_{\infty}(\gamma_{\textsc{f}})$ is negative definite.

Let $m_{\ell}(\gamma_{\textsc{f}})=\partial l(\gamma_{\textsc{f}})/\partial\gamma_{\textsc{f}\ell}$. By applying Lemma \ref{lemma-Rudin1976} to each component $f_N=m_{\ell}$, we obtain a limiting function $m_{\infty}(\gamma_{\textsc{f}})$ satisfying
	\begin{equation*}
		m(\gamma_{\textsc{f}})=m_{\infty}(\gamma_{\textsc{f}})+o(1), \quad \partial m_{\infty}(\gamma_{\textsc{f}})/\partial\gamma_{\textsc{f}}=H_{\infty}(\gamma_{\textsc{f}}),
	\end{equation*}
where the convergence is uniform on compact subsets of $\mathbb{R}^{L+1}$. Applying the same lemma to $f_N=l(\gamma_{\textsc{f}})$ yields
	\begin{equation*}
		l(\gamma_{\textsc{f}})=l_{\infty}(\gamma_{\textsc{f}})+o(1), \quad \partial l_{\infty}(\gamma_{\textsc{f}})/\partial\gamma_{\textsc{f}}=m_{\infty}(\gamma_{\textsc{f}}),
	\end{equation*}
for all $\gamma_{\textsc{f}}\in\mathbb{R}^{L+1}$, establishing condition (a).

Because $E(WZ)=1$ and $E(W)=2$, we have $E(WZ)-E(W)/2=0$. In addition, the assumptions $Z\ind\{Y(1),Y(0)\}$ and $\tau=0$ imply $E(WZY)-E(WY)/2=0$.  Consequently, $m_{\infty}(\gamma_{\textsc{f}}^*)=\lim_{n\rightarrow\infty}m(\gamma_{\textsc{f}}^*)=0$. The negative definiteness of $\partial^2 l_{\infty}(\gamma_{\textsc{f}})/\partial\gamma_{\textsc{f}}\partial\gamma_{\textsc{f}}^{\T}=H_{\infty}(\gamma_{\textsc{f}})$ then implies that
$l_{\infty}(\gamma_{\textsc{f}})$ has a unique maximum at $\gamma_{\textsc{f}}^*$, which establishes condition (b).

To verify condition (c), we observe that $(1, y_i^{\T})^{\T}(1, y_i^{\T})\geq0$  and $w_i\pi(y_i,\gamma_{\textsc{f}})\{1-\pi(y_i,\gamma_{\textsc{f}})\}\geq0$. Therefore,  $-n^{-1}\sum_{i=1}^{n}w_i\pi(y_i,\gamma_{\textsc{f}})\{1-\pi(y_i,\gamma_{\textsc{f}})\}(1, y_i^{\T})^{\T}(1, y_i^{\T})\leq0$. Hence, $H(\gamma_{\textsc{f}})\leq0$ and $l(\gamma_{\textsc{f}})$ is concave.

\vspace{0.2cm}
\noindent\textit{(ii) Asymptotic equivalence of $\hat\gamma$ and $\hat{\tau}$.}
Let $m_j(\gamma_{\textsc{f}})=\partial l(\gamma_{\textsc{f}})/\partial\gamma_{\textsc{f}j}$. A Taylor expansion around $\gamma_{\textsc{f}}^*$ gives
$0=m_j(\hat{\gamma_{\textsc{f}}})=m_j(\gamma_{\textsc{f}}^*)+\{H_j(\gamma_{\textsc{f}j}')\}^{\T}(\hat{\gamma_{\textsc{f}}}-\gamma_{\textsc{f}}^*),
$
where $H_j(\gamma_{\textsc{f}})=\partial m_j(\gamma_{\textsc{f}})/\partial\gamma_{\textsc{f}}\in\mathbb{R}^{L+1}$ and $\gamma_{\textsc{f}j}'\in\mathbb{R}^{L+1}$ lies between $\hat{\gamma_{\textsc{f}}}$ and $\gamma_{\textsc{f}}^*$. Since $\hat{\gamma_{\textsc{f}}}=\gamma_{\textsc{f}}^*+o_p(1)$, we have $\gamma_{\textsc{f}j}'=\gamma_{\textsc{f}}^*+o_p(1)$ and $H_j(\gamma_{\textsc{f}j}')=H_j(\gamma_{\textsc{f}}^*)+o_p(1)$.  Let $H'$ denote the matrix whose
$j$th row is  $H_j(\gamma_{\textsc{f}j}')$:
\begin{equation*}
H'=\begin{pmatrix}H_1^{\T}(\gamma_{\textsc{f}1}')\\\vdots\\H_{L+1}^{\T}(\gamma_{\textsc{f}(L+1)}')\end{pmatrix}_{(L+1)\times(L+1)}=H^*+o_p(1).
\end{equation*}
Here, $H^*=H(\gamma_{\textsc{f}}^*)=(H_1(\gamma_{\textsc{f}}^*),\ldots, H_{L+1}(\gamma_{\textsc{f}}^*))^{\T}<0$.  Stacking $m_j(\hat\gamma_{\textsc{f}})$ over $j=\{1,\ldots,L+1\}$ gives
\begin{equation*}
\begin{aligned}
0&=m(\gamma_{\textsc{f}}^*)+H'(\hat\gamma_{\textsc{f}}-\gamma_{\textsc{f}}^*)\\
&=m(\gamma_{\textsc{f}}^*)+\{H^*+o_p(1)\}(\hat\gamma_{\textsc{f}}-\gamma_{\textsc{f}}^*)\\ &=n^{1/2}m(\gamma_{\textsc{f}}^*)+n^{1/2}H^*(\hat\gamma_{\textsc{f}}-\gamma_{\textsc{f}}^*)+o_p\{n^{1/2}(\hat\gamma_{\textsc{f}}-\gamma_{\textsc{f}}^*)\}.
\end{aligned}
\end{equation*}
By the CLT, $n^{1/2}m(\gamma_{\textsc{f}}^*)=O_p(1)$, which implies $n^{1/2}(\hat\gamma_{\textsc{f}}-\gamma_{\textsc{f}}^*)=O_p(1)$. It follows that
\begin{equation*}
\begin{aligned}
n^{1/2}(\hat\gamma_{\textsc{f}}-\gamma_{\textsc{f}}^*)=-(H^*)^{-1}n^{1/2}m(\gamma_{\textsc{f}}^*)+o_p(1),
\end{aligned}
\end{equation*}
where
\begin{equation*}
\begin{aligned}
H^*&=H(\gamma_{\textsc{f}}^*)=-\frac{1}{4}\begin{pmatrix}
S_{11}^{w} & S_{1y}^{w}\\
S_{y1}^{w} & S_{yy}^{w}
\end{pmatrix},\quad
m(\gamma_{\textsc{f}}^*)=(S^{w}_{\smash{z1}}-S^{w}_{\smash{11}}/2, S^{w}_{\smash{zy}}-S^{w}_{\smash{1y}}/2)^{\T}.
		\end{aligned}
	\end{equation*}
The inverse of $H^*$ is
\begin{equation*}
(H^*)^{-1}=
-4
\begin{pmatrix}
(S_{11}^{w})^{-1}
+
(S_{11}^{w})^{-1}S_{1y}^{w}
(\phi_{yy}^{w})^{-1}
S_{y1}^{w}(S_{11}^{w})^{-1}
\quad&
-(S_{11}^{w})^{-1}S_{1y}^{w}(\phi_{yy}^{w})^{-1}
\\[4pt]
-(\phi_{yy}^{w})^{-1}S_{y1}^{w}(S_{11}^{w})^{-1}
\quad&
(\phi_{yy}^{w})^{-1}
\end{pmatrix},
	\end{equation*}
	where $\phi_{yy}^{w}=S^{w}_{\smash{yy}}-S^{w}_{\smash{y1}}(S^{w}_{\smash{11}})^{-1}S^{w}_{\smash{1y}}$. Extracting the components corresponding to $\gamma$ yields
	\begin{equation*}
		\begin{aligned}
			n^{1/2}(\hat{\gamma}-0)&=4n^{1/2}(\phi_{yy}^{w})^{-1}\big\{-S_{y1}^{w}(S_{11}^{w})^{-1}(S^{w}_{\smash{z1}}-S^{w}_{\smash{11}}/2)
+(S^{w}_{\smash{yz}}-S^{w}_{\smash{y1}}/2)\big\}+o_p(1)\\
&=4n^{1/2}(\phi_{yy}^{w})^{-1}\phi_{yz}^{w}+o_p(1)\\
&=4n^{1/2}\phi_{zz}^{w}(\phi_{yy}^{w})^{-1}\hat{\tau}+o_p(1).
		\end{aligned}
	\end{equation*}
This establishes the desired asymptotic equivalence.

\subsection{Proof of Proposition \ref{asyDis-wald-logit-CR-noX}} \label{subsec-propS6-proof}

We first derive the asymptotic distribution of $\hat{\gamma}$ and then use Proposition \ref{Equiv-logit-CR-noX} to obtain the asymptotic distribution of $\hat{\gamma}$. Recall that $\hat{\tau}=(\phi_{zz}^{w})^{-1}\phi_{yz}^{w}$. Under $\tau=0$, we have $\Phi_{\textsc{yz}}^{\textsc{w}}=0$. Therefore, we can write $\hat\tau-\tau$ as
\begin{equation*}
\begin{aligned}
\hat{\tau}-\tau=(\phi_{zz}^{w})^{-1}\big(\phi_{yz}^{w}-\Phi_{\textsc{yz}}^{\textsc{w}}\big).
\end{aligned}
\end{equation*}
Using the arguments analogous to those in the proof of Theorem 1, we obtain
$$n^{1/2}(\phi_{yz}^{w}-\Phi_{\textsc{yz}}^{\textsc{w}})=n^{-1/2}\sum_{i=1}^{n}\big[w_i\{z_i-E(W)^{-1}E(WZ)\}\{y_i-E(W)^{-1}E(WY)\}
-\Phi_{\textsc{yz}}^{\textsc{w}}\big]+o_p(1),$$
and $\phi_{zz}^{w}=\Phi_{\textsc{zz}}^{\textsc{w}}+o_p(1)=1/2+o_p(1)$. Combining the preceding results yields
\begin{equation*}
\begin{aligned}
n^{1/2}(\hat{\tau}-\tau)&=n^{-1/2}(\Phi_{\textsc{zz}}^{\textsc{w}})^{-1}\sum_{i=1}^{n}\big[w_i\{z_i-E(W)^{-1}E(WZ)\}\{y_i-E(W)^{-1}E(WY)\}
-\Phi_{\textsc{yz}}^{\textsc{w}}\big]+o_p(1).
\end{aligned}
\end{equation*}
The CLT then yields
$$n^{1/2}\hat{\tau}\stackrel{d}\rightarrow N\Big(0, (\Phi_{\textsc{zz}}^{\textsc{w}})^{-2}E\big[W^2(Z-1/2)^2\{Y-E(W)^{-1}E(WY)\}\{Y-E(W)^{-1}E(WY)\}^{\T}\big]\Big).$$
Combining this convergence with Proposition \ref{Equiv-logit-CR-noX} and applying Slutsky's theorem gives
$$n^{1/2}\hat{\gamma}\stackrel{d}\rightarrow N\Big(0, 4(\Phi_{\textsc{yy}}^{\textsc{w}})^{-1}E\big[W^2\{Y-E(W)^{-1}E(WY)\}\{Y-E(W)^{-1}E(WY)\}^{\T}\big](\Phi_{\textsc{yy}}^{\textsc{w}})^{-1}\Big).$$
Moreover, the LLN and Slutsky's theorem imply
$$n\hat{V}_{\gamma}=4(\Phi_{\textsc{yy}}^{\textsc{w}})^{-1}E\big[W^2\{Y-E(W)^{-1}E(WY)\}\{Y-E(W)^{-1}E(WY)\}^{\T}\big](\Phi_{\textsc{yy}}^{\textsc{w}})^{-1}+o_p(1).$$
Consequently,
\begin{equation*}
\hat{V}_{\gamma}^{-1/2}\hat{\gamma}\stackrel{d}\rightarrow N(0, I_L), \hspace{0.5cm}
\hat{\gamma}^{\T}\hat{V}_{\gamma}^{-1}\hat{\gamma}\stackrel{d}\rightarrow \chi^2(L).
\end{equation*}

\end{document}